\documentclass[fleqn,usenatbib]{mnras}

\usepackage{newtxtext,newtxmath}

\usepackage[T1]{fontenc}
\usepackage{ae,aecompl}

\usepackage{graphicx}	
\usepackage{amsmath}	
\usepackage{amssymb}	

\usepackage{bm, xcolor}
\usepackage[all]{hypcap}
\usepackage{epsfig}
\usepackage{times}

\DeclareMathOperator\erf{erf}

\newcommand\Msun{\,\rmn{M}_{\sun}}

\newcommand\Gyr{\,\rmn{Gyr}}
\newcommand\Myr{\,\rmn{Myr}}

\newcommand{\pyr}{\,{\rm yr}^{-1}}
\newcommand\pc{\,\rmn{pc}}
\newcommand\kpc{\,\rmn{kpc}}

\newcommand\FeH{\lbrack \rmn{Fe}/\rmn{H} \rbrack}
\newcommand\ZH{\lbrack \rmn{Z}/\rmn{H} \rbrack}

\newcommand{\kms}{\,{\rm km}\,{\rm s}^{-1}}
\newcommand{\numgal}{10}
\newcommand{\numsim}{\sim 200}

\newcommand{\mean}[1]{\overline{#1}}

\newcommand{\pder}[3][]{\frac{\partial^{#1} #2}{\partial {#3}^{#1}}}
\newcommand{\derBr}[3][]{\left( \frac{\rmn{d}^{#1} #2}{\rmn{d} {#3}^{#1}} \right)}

\newcommand{\gadget}         {\textsc{gadget3}}

\newcommand{\K}              {\,{\rm K}}
\newcommand{\cmcubed}              {\,{\rm cm}^{-3}}
\newcommand{\pkpc}                      {\,{\rm pkpc}}

\newcommand{\ckpc}                      {\,{\rm ckpc}}

\newcommand{\Mcstar}{M_\rmn{c,\ast}}
\newcommand{\Mtoomre}{M_\rmn{T}}
\newcommand{\Mgmc}{M_\rmn{GMC}}

\title[The E-MOSAICS project]{The E-MOSAICS Project: simulating the formation and co-evolution of galaxies and their star cluster populations}

\author[J. Pfeffer et al.]{
Joel Pfeffer,$^{1}$\thanks{E-mail: \href{j.l.pfeffer@ljmu.ac.uk}{j.l.pfeffer@ljmu.ac.uk} (JLP); \href{kruijssen@uni-heidelberg.de}{kruijssen@uni-heidelberg.de} (JMDK)}
J. M. Diederik Kruijssen,$^{2}${\color{blue}\footnotemark[1]}
Robert A. Crain$^{1}$
and Nate Bastian$^{1}$
\\
$^{1}$Astrophysics Research Institute, Liverpool John Moores University, 146 Brownlow Hill, Liverpool L3 5RF, UK\\
$^{2}$Astronomisches Rechen-Institut, Zentrum f\"{u}r Astronomie der Universit\"{a}t Heidelberg, M\"{o}nchhofstra\ss e 12-14, 69120 Heidelberg, Germany\\
}

\date{Accepted 2017 November 30. Received 2017 September 18; in original form 2017 August 4}

\pubyear{2017}

\begin{document}
\label{firstpage}
\pagerange{\pageref{firstpage}--\pageref{lastpage}}
\maketitle

\begin{abstract}
We introduce the MOdelling Star cluster population Assembly In Cosmological Simulations within EAGLE (E-MOSAICS) project. E-MOSAICS incorporates models describing the formation, evolution and disruption of star clusters into the EAGLE galaxy formation simulations, enabling the examination of the co-evolution of star clusters and their host galaxies in a fully cosmological context. A fraction of the star formation rate of dense gas is assumed to yield a cluster population; this fraction, and the population's initial properties, are governed by the physical properties of the natal gas. The subsequent evolution and disruption of the entire cluster population is followed accounting for two-body relaxation, stellar evolution, and gravitational shocks induced by the local tidal field. This introductory paper presents a detailed description of the model and initial results from a suite of 10 simulations of $\sim L^\star$ galaxies with disc-like morphologies at $z=0$. The simulations broadly reproduce key observed characteristics of young star clusters and globular clusters (GCs), without invoking separate formation mechanisms for each population. The simulated GCs are the surviving population of massive clusters formed at early epochs ($z\gtrsim1-2$), when the characteristic pressures and surface densities of star-forming gas were significantly higher than observed in local galaxies. We examine the influence of the star formation and assembly histories of galaxies on their cluster populations, finding that (at similar present-day mass) earlier-forming galaxies foster a more massive and disruption-resilient cluster population, while galaxies with late mergers are capable of forming massive clusters even at late cosmic epochs. We find that the phenomenological treatment of interstellar gas in EAGLE precludes the accurate modelling of cluster disruption in low-density environments, but infer that simulations incorporating an explicitly-modelled cold interstellar gas phase will overcome this shortcoming.
\end{abstract}



\defcitealias{S15}{S15}

\section{Introduction}

All galaxies in the local Universe with stellar masses $>10^9~\Msun$ are observed to host globular cluster (GC) populations \citep[for recent reviews, see e.g.][]{Brodie_and_Strader_06, Kruijssen_14}. Even at masses as low as $10^8~\Msun$, the majority of galaxies still contain at least one GC \citep[e.g.][]{Georgiev_et_al_10}. Dwarf galaxies like the Magellanic clouds typically host a few to tens of GCs, the Milky Way (MW) and M31 are known to host a few hundred \citep{Harris_91}, and brightest cluster galaxies can host tens of thousands \citep{Peng_et_al_08, Harris_et_al_17}. The population of galaxies bound to a rich galaxy cluster can host hundreds of thousands of GCs \citep{Alamo-Martinez_et_al_13}. GCs are typically old \citep[ages $>$10 Gyr,][]{Puzia_et_al_05, Strader_et_al_05, Marin-Franch_et_al_09, VandenBerg_et_al_13}, have nearly uniform sizes of a few parsecs \citep{Kundu_and_Whitmore_01, Masters_et_al_10} and the mass distribution of the GC population associated with a galaxy can be reasonably-well approximated by a log-normal function with a characteristic peak mass ($M_\rmn{c,peak} \sim 10^5 \Msun$), which depends weakly on galaxy mass \citep{Harris_91, Jordan_et_al_07_XII}.

Most GCs have ages corresponding to formation times close to the peak of cosmic star formation, and as a consequence GC populations have long been posited as potentially powerful tracers of galaxy formation and assembly \citep[see e.g.~reviews by][]{Harris_91, Brodie_and_Strader_06}.
GC properties broadly correlate with those of their host galaxy, and the observation of bimodal colour distributions (typically interpreted as bimodal metallicity distributions) has fostered the inference of two dominant epochs of star formation within galaxies \citep{Brodie_and_Strader_06}. However, in most cases where the metallicities of GC and field stars can be compared directly, the field stars do not exhibit a bimodal metallicity distribution function \citep[MDF; e.g.][]{Harris_and_Harris_02, Harris_et_al_07, Rejkuba_et_al_11, Lamers_et_al_17}\footnote{\citet{Peacock_et_al_15} recently demonstrated that the stellar halo of NGC 3115 does, however, exhibit a bimodal MDF.}. Moreover, it has become apparent in recent years that galaxies can also exhibit unimodal \citep{Caldwell_et_al_11, Harris_et_al_17} and multimodal GC MDFs \citep[which may also depend on cluster luminosity,][]{Usher_et_al_12}.  The connection between the properties of GCs and these of their host galaxies is therefore not necessarily straightforward; inference of the latter from observation of the former requires a detailed understanding of the co-evolution of GCs and galaxies. Because of this complexity, the promise of using GCs as tracers of galaxy formation remains largely unfulfilled.

The old ages and small sizes of GCs preclude direct, spatially-resolved observations of their formation with current instrumentation. However, the young massive clusters (YMCs) observed to be forming in the local Universe, with masses and densities similar to those of GCs \citep[see e.g.~reviews by][]{Portegies-Zwart_McMillan_and_Gieles_10, Longmore_et_al_14, Kruijssen_14}, are thought to be broadly analogous to proto-GCs. As such, GCs have been interpreted as the surviving population of YMCs that formed in the early Universe. Indeed, star clusters exhibit a continuum of ages between those of YMCs and GCs \citep[e.g.][]{Salaris_Weiss_and_Percival_04, Parisi_et_al_14, Beasley_et_al_15}, and there is a broad range of overlap in their metallicity distributions: low metallicity ($\FeH \sim -1$ dex) YMCs are seen in star-bursting dwarf galaxies \citep[e.g.][]{Ostlin_et_al_07} and examples of YMCs with super-solar metallicity have been observed in the spiral arms of more massive galaxies and in merging pairs \citep[e.g.][]{Gazak_et_al_14}, whereas GCs with super-solar metallicity are ubiquitous in massive galaxies \citep{Harris_and_Harris_02, Usher_et_al_12, Lamers_et_al_17}. Crucially, YMCs are also observed to exhibit similar sizes (few pc) and masses ($\sim 10^3$-$10^8 \Msun$) to GCs \citep{Maraston_et_al_04, Whitmore_et_al_10}.

However, despite exhibiting a similar range of masses, YMCs and GCs populate this range quite differently. The YMC population is well-described by a power-law (index $\sim -2$) cluster mass function with an exponential truncation at high mass \citep[e.g.][]{Larsen_09, Portegies-Zwart_McMillan_and_Gieles_10}, while GCs exhibit a peaked cluster mass function that is relatively insensitive to environmental properties \citep[e.g.~galaxy mass and galactocentric radius;][]{Jordan_et_al_07_XII,Harris_et_al_14}. How the power-law YMC mass function might evolve into the peaked GC mass function through disruption is a topic of energetic debate \citep[see][]{Fall_and_Zhang_01, Vesperini_et_al_03, Elmegreen_10, Kruijssen_15, Gieles_and_Renaud_16}, but a feasible and promising mechanism is dynamical heating by tidal shocks within the interstellar medium (ISM) from which the clusters are born \citep{Elmegreen_10, Kruijssen_15}.

To date, modelling endeavours have largely focused on particular aspects of the problem, such as on cluster formation \citep{Kravtsov_and_Gnedin_05, Katz_and_Ricotti_14, Li_and_Gnedin_14, Mistani_et_al_16, Li_et_al_17} or cluster disruption \citep{Gnedin_and_Ostriker_97, Vesperini_97, Baumgardt_98, Prieto_and_Gnedin_08, Rieder_et_al_13}, on the combination of these mechanisms in idealised galaxy simulations \citep{Kruijssen_et_al_11,Kruijssen_et_al_12}, or on the effects of environment, e.g.~galaxy mergers \citep{Li_MacLow_Klessen_04, Kruijssen_et_al_12, Renaud_and_Gieles_13, Renaud_Bournaud_and_Duc_15} and hierarchical galaxy assembly \citep{Beasley_et_al_02, Bekki_et_al_08, Griffen_et_al_10, Tonini_13, Renaud_Agertz_and_Gieles_17}. In cases where both formation and disruption have been modelled, the environmental dependence of disruption has often been omitted \citep{Muratov_and_Gnedin_10}. With a few exceptions \citep[e.g.][]{Elmegreen_10,Kruijssen_et_al_11,Kruijssen_et_al_12,Kruijssen_15}, most models also neglect the disruptive influence of gas in galaxies. 

This body of work has highlighted a number of important considerations when modelling GC populations. Massive star clusters appear to form within the highest density peaks of the ISM, where star formation efficiencies are greatest \citep{Elmegreen_and_Efremov_97, Elmegreen_08, Kruijssen_12} and the maximum cluster mass-scale is approximately proportional to the Toomre mass in the host galaxy disc \citep[but with important deviations, see][]{Reina-Campos_and_Kruijssen_17}. Tidal shocks in gas-rich environments may be an important means by which the cluster mass function is shaped \citep{Elmegreen_10, Kruijssen_15}, such that survival for a Hubble time likely requires that a cluster migrates away from its natal gaseous disc \citep{Kruijssen_14}, plausibly in response to dynamical heating by minor and/or major galaxy mergers \citep[e.g.][]{Kravtsov_and_Gnedin_05, Kruijssen_15}. Disruption is environmentally dependent \citep[e.g.][]{Baumgardt_and_Makino_03, Gieles_et_al_06, Kruijssen_et_al_11}, rendering a cluster's tidal history markedly sensitive to its environment throughout the formation and assembly of its host galaxy \citep{Prieto_and_Gnedin_08,Kruijssen_et_al_12, Rieder_et_al_13}.

If GCs are the end product of intense star formation episodes, modelling them realistically demands a self-consistent treatment of the formation and subsequent disruption of the entire star cluster population, within the evolving cosmological environment defined by the formation and assembly of the host galaxy. These particularly demanding requirements have thus far obstructed the theoretical understanding of GC formation from keeping pace with discoveries driven by ambitious observational surveys of GCs (e.g.~the ACS Virgo Cluster Survey, \citealt{Cote_et_al_04}; the ACS Fornax Cluster Survey, \citealt{Jordan_et_al_07_I}; SLUGGS, \citealt{Brodie_et_al_14}; NGVS, \citealt{Ferrarese_et_al_12}). None the less, a number of approaches have been deployed to model simultaneously the formation and evolution of star cluster populations in a cosmological context, including analytic methods \citep[e.g.][]{Kruijssen_15}, semi-analytic or sub-grid methods that do not resolve clusters \citep[e.g.][]{Kravtsov_and_Gnedin_05}, and numerical models that directly resolve clusters \citep[e.g.][]{Ricotti_Parry_and_Gnedin_16}. Each method has its own strengths and weaknesses; analytic methods enable a wide range of parameters to be examined but require recourse to restrictive simplifying assumptions. By contrast, direct simulations of cluster formation and evolution in galaxies at extreme resolution may be capable of incorporating disruption self-consistently \citep[e.g.][]{Li_et_al_17,Kim_et_al_17}, but the resulting cluster properties are markedly sensitive to the particular implementation of star formation and stellar feedback adopted. Moreover, these models are generally too computationally expensive to allow a comprehensive examination of the ill-constrained aspects of the model, they are limited to following the evolution of relatively small cosmological volumes, and they can only do so for a brief period of cosmic history.

Here we adopt what we consider to be a well-motivated compromise between these approaches, coupling a semi-analytic model of star cluster formation and disruption to a cosmological hydrodynamical simulation of galaxy formation and evolution. It is a judicious time to adopt such an approach, since the realism of the latter has improved dramatically in recent years, such that they can follow a cosmologically-representative volume (volumes with side $L\sim100$ comoving Mpc) and produce a present-day galaxy population with properties similar to those observed in the local Universe \citep[e.g.][]{Vogelsberger_et_al_14b, S15,Dave_Thompson_and_Hopkins_16,Kaviraj_et_al_17}.

This paper introduces the E-MOSAICS project: MOdelling Star cluster population Assembly In Cosmological Simulations within EAGLE, whereby we couple the semi-analytic {\sc MOSAICS} model of star cluster formation and evolution of  \citet[see Section~\ref{sec:mosaics} below]{Kruijssen_et_al_11, Kruijssen_et_al_12} to the EAGLE simulations of galaxy formation (\citealt{S15}, hereafter S15; \citealt{C15}; see Section~\ref{sec:eagle} below). In brief, this is done by generating a sub-grid population of stellar clusters of which the initial properties are determined by the ambient gas properties associated with each star formation event, and which are calibrated against observations of YMCs in the local Universe. Once formed, clusters undergo mass loss by stellar evolution and dynamical evolution, the latter in response to the evolution of the local tidal field. Our principal aim is to test whether YMC-based cluster formation and disruption models in the context of galaxy formation and assembly are compatible with observations of GCs. If so, the marriage of these models should afford a much more detailed understanding of the formation and co-evolution of galaxies and their GC systems and, by extension, pave the way to unlock the potential of GCs as tracers of galaxy assembly.

To this end, we have conducted and analysed a total of $\numsim$ zoom-in simulations of $\numgal$ MW-like galaxies, with a particular focus on developing a comprehensive understanding of the sensitivity of the resulting cluster populations on the adopted physical models and parameter choices. We do not discuss each of these simulations in detail here, but note that this exploration was a necessity, both for identifying which elements of the model most directly influence the resulting cluster populations, and for identifying the sensitivity of these properties to numerical effects. A single or even a small sample of simulations would have been insufficient to adequately address these factors.

This paper is structured as follows. In Section \ref{sec:methods} we briefly summarize the EAGLE galaxy formation model and describe the MOSAICS implementation of star cluster formation and evolution, and their coupling to the EAGLE model. Section \ref{sec:sims} introduces the $\numgal$ zoom simulations of galaxies, and the suite of simulations incorporating variations of various aspects of the models. In Section \ref{sec:formation_props} we present the results from the cluster formation model and compare the predictions with properties of observed nearby galaxies. In Section \ref{sec:tidal_histories} we investigate the importance of cluster disruption as a function of redshift by comparing the cluster tidal histories in the simulations. In Section \ref{sec:cluster_props} we present the properties of the cluster populations at $z=0$, compare the results with the observed MW GC population, and discuss the origin of the GC mass function. Finally, we summarize our findings in Section \ref{sec:summary}. The paper also includes seven appendices in which the important quantitative tests of the model components are discussed.


\section{Numerical methods} \label{sec:methods}

We incorporate the MOSAICS star cluster formation and evolution model into the EAGLE galaxy formation model. We utilise a sub-grid model where star clusters are attached to the stellar particles formed by the simulation. This is advantageous as it avoids the necessity of adjusting the sub-grid models used by EAGLE, and of recalibrating their parameters (see below). Since the modelling does not (yet) incorporate any back-reaction from the star clusters upon the hydrodynamics, it would in principle be possible to apply all of the MOSAICS models in post-processing to merger trees built from snapshots of an EAGLE volume. However, the temporal resolution required to identify tidal shocks (see Section \ref{sec:clevo}) renders such an approach infeasible, since it would demand storage of $\sim 10^2$ variables associated with $\sim 10^7$ particles, for $>10^4$ output times, requiring (at single precision) $>40$ terabytes per galaxy or $>8$ petabytes for all simulations. We therefore run MOSAICS on-the-fly within EAGLE, and append key variables describing the star cluster populations to regular EAGLE snapshots. We typically output 29 such snapshots per run, requiring storage of 25-300 gigabytes per zoom simulation. The MOSAICS calculations account for only a few per cent of the simulation wallclock time.


\subsection{The EAGLE simulations of galaxy formation}
\label{sec:eagle}

EAGLE \citep[Evolution and Assembly of GaLaxies and their Environments][]{S15, C15} is a suite of hydrodynamical simulations of galaxy formation in the $\Lambda$CDM cosmogony, evolved using a modified version of the $N$-body TreePM smoothed particle hydrodynamics (SPH) code \gadget, \citep[last described by][]{Springel_05}. The key subsequent modifications are to the hydrodynamics algorithm and the time-stepping criteria, and a suite of subgrid routines governing processes that act on scales below the simulation's resolution limit are also included. The updates to the hydrodynamics algorithm, collectively referred to as ``Anarchy'' (see Appendix A of \citetalias{S15}), comprise an implementation of the pressure-entropy formulation of SPH of \citet{Hopkins_13}, an artificial viscosity switch of the form proposed by \citet{Cullen_and_Dehnen_10}, an artificial conduction switch of the form proposed by \citet{Price_08}, the \citet{Wendland_95} $C^2$ smoothing kernel, and the \citet{Durier_and_Dalla_Vecchia_12} time-step limiter. The impact of each of these developments on the EAGLE galaxy population is explored by \citet{Schaller_et_al_15b_short}.

Element-by-element radiative cooling and photoionization heating for 11 species (H, He and 9 metal species) is treated using the scheme of \citet{Wiersma_Schaye_and_Smith_09}, assuming the presence of a spatially-uniform, temporally-evolving radiation field due to the cosmic microwave background and the metagalactic ultraviolet/X-ray background (UVB) from galaxies and quasars, as modelled by \citet{Haardt_and_Madau_01}. This scheme assumes the gas to be optically thin and in ionization equilibrium. Gas with density greater than the metallicity-dependent threshold advocated by \citet{Schaye_04}, and which is within 0.5 decades of a Jeans-limiting temperature floor (see below), is eligible for stochastic conversion to a collisionless stellar particle. The probability of conversion is proportional to the particle's star formation rate (SFR), which is a function of its pressure, such that, by construction, the simulation reproduces the \citet{Kennicutt_review_98} star formation law \citep{Schaye_and_Dalla_Vecchia_08}. Each stellar particle is assumed to represent a simple stellar population with the \citet{Chabrier_03} initial mass function (IMF), and the return of mass and metals from stellar populations to the ISM is implemented with the scheme of \citet{Wiersma_et_al_09}, which tracks the abundances of the same 11 elements considered when computing the radiative cooling and photoionization heating rates. EAGLE also incorporates routines to model the growth of BHs via gas accretion (at the minimum of the Bondi-Hoyle and Eddington rates) and BH-BH mergers \citep{Springel_Di_Matteo_and_Hernquist_05,Rosas_Guevara_et_al_15_short,S15}, and feedback associated with star formation \citep{Dalla_Vecchia_and_Schaye_12} and the growth of BHs \citep{Booth_and_Schaye_09,S15}, via stochastic gas heating. This AGN feedback is implemented as a single heating mode, but nevertheless mimics quiescent `radio-like' and vigorous `quasar-like' AGN modes when the BH accretion rate is a small ($\ll 1$) or large ($\sim 1$) fraction of the Eddington rate, respectively \citep[][]{McCarthy_et_al_11}. 

In general, cosmological simulations lack both the resolution and physics required to model the cold, dense phase of the ISM. Gas is therefore subject to a polytropic temperature floor, $T_{\rm eos}(\rho_{\rm g})$, which corresponds to the equation of state $P_{\rm eos} \propto \rho_{\rm g}^{4/3}$, normalised to $T_{\rm eos} = 8000\K$ at $n_{\rm H} \equiv X_{\rm H,0}\rho/m_{\rm H} = 10^{-1} \cmcubed$, where $X_{\rm H,0}=0.752$ is the hydrogen mass fraction of gas with primordial composition. The exponent of $4/3$ ensures that the Jeans mass, and the ratio of the Jeans length to the SPH kernel support radius, are independent of the density \citep{Schaye_and_Dalla_Vecchia_08}. This is a necessary condition to limit artificial fragmentation. Gas with $\log_{10} T > \log_{10} T_{\rm eos}(\rho_{\rm g}) + 0.5$ is ineligible for star formation, irrespective of its density. 

The resolution and physics limitations of  cosmological simulations currently also precludes the ab-initio calculation of the efficiency of the feedback processes that regulate (and potentially quench) galaxy growth. An effective means of circumventing this problem is to calibrate the subgrid efficiencies of these processes, to ensure that the simulation reproduces appropriate observables. In EAGLE, the subgrid efficiency of AGN feedback is assumed to be constant, and is calibrated to ensure that the simulations reproduce the present-day relation between the mass of central BHs and the stellar mass of their host galaxy \citep[see also][]{Booth_and_Schaye_09}. The subgrid efficiency of feedback associated with star formation is a smoothly-varying function of the metallicity and density of gas local to newly-formed stellar particles, and is calibrated to ensure reproduction of the present-day galaxy stellar mass function, and the size-mass relation of disc galaxies. \citetalias{S15} argue that parameters may need to be recalibrated as the resolution of the simulation is changed; for this reason the parameters adopted for the Reference (`Ref') EAGLE model are slightly different to those that yield the most accurate reproduction of the calibration diagnostics at a factor of 8 (2) better mass (spatial) resolution (the `Recal' model). 

The EAGLE simulations successfully reproduce a broad range of observed galaxy properties and scalings, such as the evolution of the stellar masses \citep{Furlong_et_al_15_short} and sizes \citep{Furlong_et_al_17} of galaxies, their luminosities and colours \citep{Trayford_et_al_15_short}, their cold gas properties \citep{Lagos_et_al_15_short,Lagos_et_al_16,Bahe_et_al_16,Marasco_et_al_16,Crain_et_al_17}, and the properties of circumgalactic and intergalactic absorption systems \citep{Rahmati_et_al_15,Oppenheimer_et_al_16,Rahmati_et_al_16,Turner_et_al_16,Turner_et_al_17}.


\subsection{Star cluster formation and evolution model}
\label{sec:mosaics}

The resolution achieved by the current generation of cosmological simulations of the galaxy population is insufficient to resolve individual star clusters. We therefore use an updated version of the semi-analytic star cluster formation and evolution model for galaxy simulations {\sc MOSAICS} \citep[MOdelling Star cluster population Assembly In Cosmological Simulations,][]{Kruijssen_and_Lamers_08, Kruijssen_09, Kruijssen_et_al_11}. This model has previously been used to study star cluster formation and evolution in isolated disc galaxies and galaxy mergers \citep{Kruijssen_et_al_11, Kruijssen_et_al_12} and has been expanded in this work to include the models of \citet{Kruijssen_12} and \citet{Reina-Campos_and_Kruijssen_17} to account for the environmental dependence of the cluster formation efficiency (CFE) and maximum cluster mass (see below). By using these models, the cluster formation model is `YMC-based', in that the initial cluster populations are set by models which reproduce key properties observed for young stellar clusters (see Section~\ref{sec:clform}).
In {\sc MOSAICS}, the formation of stellar particles in the simulation triggers the formation of a sub-grid population of star clusters. The cluster population is `attached' to the stellar particle, and thus inherits its phase space coordinates and metallicity. Several other properties of the natal cluster population are computed from the properties of particles local to the stellar particle at the instant of its formation. Examples are the CFE and maximum cluster mass, both of which depend on e.g.~the ambient gas pressure, gas volume density, stellar velocity dispersion, and the gas fraction.

\subsubsection{Cluster formation} \label{sec:clform}

Whenever a stellar particle forms in the simulations, a fraction of its mass is used to form a sub-grid cluster population. This mass fraction is set by the CFE \citep[$\Gamma$, i.e.~the fraction of star formation in bound clusters,][]{Bastian_08} which is observed to correlate positively with the surface density of star formation, $\Sigma_\rmn{SFR}$ (see \citealt{Adamo_and_Bastian_15} for a recent review). \citet{Kruijssen_12} present a model that relates $\Gamma$ to the properties of the interstellar medium (ISM). In the model, bound clusters form across the density spectrum of the hierarchically-structured ISM, but most efficiently at the high-density end, where the free-fall times are short and the resulting local star formation efficiencies are high. The CFE is then obtained by integrating over the full density spectrum of the ISM. The fundamental prediction of the model is that the CFE is an increasing function of the turbulent gas pressure, which gives excellent agreement with the observed $\Gamma$--$\Sigma_\rmn{g}$ and $\Gamma$--$\Sigma_\rmn{SFR}$ relations for local galaxies \citep{Kruijssen_12, Adamo_et_al_15, Johnson_et_al_16, Kruijssen_and_Bastian_16}.

We use the `local' formulation of the \citet{Kruijssen_12} model to obtain the CFE based on local quantities (rather than the disc-averaged quantities that are preferable in observational applications of the model), i.e.~$\Gamma (\rho_g, \sigma_\rmn{loc}, c_s)$, where $\rho_g$ is the local gas volume density, $\sigma_\rmn{loc}$ is the one-dimensional gas velocity dispersion and $c_s = 0.3$ km s$^{-1}$ is the thermal sound speed of cold interstellar gas (corresponding to an ideal gas with temperature of $\sim$10 K). Since we do not explicitly model the multiphase ISM, we approximate the unresolved, turbulent velocity dispersion as $\sigma_\rmn{loc} = \sqrt{P_g / \rho_g}$, where $P_g$ is the local gas pressure. 

We exclude the `cruel cradle effect' (i.e.~the tidal disruption of forming clusters by their natal environment, \citealt{Kruijssen_et_al_12b}) when calculating the CFE, because we treat disruption explicitly (see Section~\ref{sec:clevo}). The CFE is therefore specified by the gravitationally bound fraction of star formation based on the local star formation efficiency $\Gamma \left(\rho_g, \sigma_\rmn{loc}, c_s\right) = f_\rmn{bound}$ \citep[equation 26]{Kruijssen_12}. The total mass in field stars represented by a particle of mass $m_*$ is thus $(1-\Gamma)m_*$.

Having assigned the mass fraction of a new-born stellar particle in the form of star clusters, we adopt an initial cluster mass function (ICMF) consistent with observations of young cluster populations in the nearby Universe \citep{Portegies-Zwart_McMillan_and_Gieles_10}. This ICMF is described by a power-law, exponentially truncated at high masses \citep{Schechter_76},
\begin{equation} \label{eq:ICMF}
N \, \rmn{d}M \propto M^{-2} \exp \left(-M/M_\rmn{c,*}\right) \rmn{d}M  ,
\end{equation}
with a minimum and maximum cluster masses of $10^2 \Msun$ and $10^8 \Msun$, respectively. The ICMF truncation mass, $M_\rmn{c,*}$, is assumed to be related to the maximum mass of the molecular clouds from which the clusters form, $\Mgmc$, and the CFE \citep{Kruijssen_14}:
\begin{equation} \label{eq:Mcstar}
M_\rmn{c,*} = \epsilon \Gamma \Mgmc ,
\end{equation}  
where $\epsilon = 0.1$ is the star formation efficiency for an entire molecular cloud \citep*[e.g.][]{Duerr_Imhoff_and_Lada_82, Murray_11}. The observed dynamic range of $\epsilon$ in embedded clusters and nearby molecular clouds covers an order of magnitude around this value ($\epsilon=0.03$--$0.3$, see \citealt{Reina-Campos_and_Kruijssen_17} for a discussion), which is much smaller than the dynamic range of the product $\Gamma \Mgmc$ in our model. Assuming a constant SFE is therefore reasonable. To derive the maximum cloud mass, we adopt the model of \citet{Reina-Campos_and_Kruijssen_17}, which relates $\Mgmc$ to the largest gravitationally unstable mass in a differentially-rotating disc, i.e.~the \citet{Toomre_64} mass $M_\rmn{T}$:
\begin{equation} \label{eq:Mgmc}
\Mgmc = f_\rmn{coll} M_\rmn{T} ,
\end{equation}  
where $f_\rmn{coll}$ is the `Toomre mass collapse fraction'. This fraction reflects the idea that the Toomre mass sets the maximum gas mass which can collapse, but that this mass-scale is not able to collapse into a single object if the stellar feedback timescale is shorter than the collapse timescale. In this `feedback-limited' case, only a fraction $f_\rmn{coll}$ of $M_\rmn{T}$ will be able to collapse before the cloud is disrupted by the onset of feedback. Note that the cloud collapse and feedback time-scales are not resolved by our simulations\footnote{Stellar particles become eligible to trigger stochastic feedback events in the simulation at an age of 30~Myr, corresponding to the maximum lifetime of stars that explode as core collapse supernovae.}, but are evaluated sub-grid according to the \citet{Reina-Campos_and_Kruijssen_17} model. In this model, the fraction of collapsed mass is specified by
\begin{equation} \label{eq:f_coll}
f_\rmn{coll} = \rmn{min} \left( 1 , \frac{ t_\rmn{fb,g} }{ t_\rmn{ff,2D} } \right)^4
\end{equation}
where $t_\rmn{fb,g}$ is the cloud feedback timescale, $t_\rmn{ff,2D} = \sqrt{2 \pi}/\kappa$ is the two-dimensional cloud free-fall time and $\kappa$ is the epicyclic frequency. The quartic exponent arises from the dependence of $M_T$ on $\kappa$ (Eq. \ref{eq:Mtoomre}, see \citealt{Reina-Campos_and_Kruijssen_17} for details).
As in the CFE model, the cloud feedback timescale is expressed as a function of local quantities ($\rho_g$ and $\sigma_\rmn{loc}$):
\begin{equation} \label{eq:t_fb}
t_\rmn{fb,g} = \frac{t_\rmn{sn}}{2} \left( 1 + \sqrt{ 1 + \frac{4 t_\rmn{ff} \sigma_\rmn{loc}^2} {\phi_\rmn{fb} \epsilon_\rmn{ff} t_\rmn{sn}^2} } \right) , 
\end{equation}
where $t_\rmn{ff} = \sqrt{3 \pi / 32 G \rho_g}$ is the gas free-fall time, $\epsilon_\rmn{ff} = 0.012$ is the star formation efficiency per free-fall time \citep{Elmegreen_02}, $t_\rmn{sn} = 3$ Myr is the typical time of the first supernova \citep[e.g.][]{Ekstrom_et_al_12}, and $\phi_\rmn{fb} = 0.16$ cm$^2$ s$^{-3}$ represents the rate at which feedback injects energy into the ISM per unit stellar mass for a simple stellar population with a normal stellar IMF \citep[see Appendix~B of][]{Kruijssen_12}. As shown by \citet[Figure~3]{Reina-Campos_and_Kruijssen_17}, the maximum cloud and cluster mass-scales are feedback-limited in regions of both low shear ($\Omega\lesssim 0.6~\rmn{Myr}^{-1}$) {\it and} low gas pressure or surface density ($\Sigma_\rmn{g}<\Sigma_\rmn{crit}$ with $\Sigma_\rmn{crit}=10^1$--$10^3~\Msun~\rmn{pc}^{-2}$ depending on the local conditions).

The Toomre mass is calculated for each newborn stellar particle according to the largest unstable wavelength (rather than the most unstable wavelength), i.e.:
\begin{equation} \label{eq:Mtoomre}
\Mtoomre = 4 \pi^5 G^2 \frac{\Sigma_\rmn{g}^3}{\kappa^4} ,
\end{equation}
where $\Sigma_\rmn{g}$ is the disc gas surface density local to the particle and $\kappa$ is the epicyclic frequency. The latter is computed from the local tidal tensor (Eq. \ref{eq:tidalTensor}), which enables its definition even in irregular environments such as galaxy mergers. The derivation of $\kappa$ is described in detail in Appendix \ref{app:kappa}. With this formulation of $f_\rmn{coll}$, $t_\rmn{fb,g}$ and $\Mtoomre$, in the feedback-limited regime ($f_\rmn{coll} < 1$) $\Mcstar$ is independent of $\kappa$ since $f_\rmn{coll} \propto \kappa^4$ and $\Mtoomre \propto \kappa^{-4}$. Note that in this `local' formulation, $t_\rmn{fb,g}$ is independent of $\kappa$, which differs from the definition in \citet{Reina-Campos_and_Kruijssen_17} based on global observables, for which the substitution $\sigma=\pi GQ\Sigma/\kappa$ was made. In the feedback-limited regime, we find that $\Mcstar$ scales with the gas pressure as $\log_{10} \Mcstar \propto \gamma_\rmn{EOS}^{3/2} \log_{10} P$ (see Appendix \ref{app:EOS}).

The gas surface density is determined by equating the mid-plane pressure of a hydrostatic equilibrium disc to the gas pressure of parent SPH particle, $P_\rmn{g}$, of the newly-formed stellar particle at the instant of conversion assuming hydrostatic equilibrium \citep[cf.][]{Krumholz_and_McKee_05}: 
\begin{equation} \label{eq:Sigma_g}
\Sigma_\rmn{g} = \left( \frac{2 P_\rmn{g}}{\pi G \phi_P} \right)^{1/2} .
\end{equation}
Here $\phi_P$ is a constant that accounts for the contribution of the gravity of stars to the mid-plane gas pressure, which we write as
\begin{equation} \label{eq:phi_P}
\phi_P = 1 + \frac{\sigma_g}{\sigma_\ast} \left( \frac{1}{f_\rmn{gas}} - 1 \right) ,
\end{equation}
where $f_\rmn{gas} = M_\rmn{g}/(M_\rmn{g}+M_*)$, $\sigma_\rmn{g}$ and $\sigma_*$ are the gas and stellar velocity dispersions, respectively, and $M_\rmn{g}$ and $M_*$ are the gas and stellar masses within the region for which $\phi_P$ is determined.
We calculate $\sigma_g$, $\sigma_\ast$ and $f_\rmn{gas}$ local to the stellar particle within a smoothing kernel with a spatial extent equal to the minimum radius that encloses at least 58 SPH and 48 stellar particles, up to a maximum of 3 times the standard SPH support radius. In cases where the kernel does not enclose 48 stellar particles, we assume $f_\rmn{gas} \simeq 1$ and therefore $\phi_P \simeq 1$. In Appendix \ref{app:Sigma} we demonstrate the accuracy of this method for calculating $\Sigma_\rmn{g}$ and $M_\rmn{T}$.

After defining the mass budget for sub-grid cluster formation and the ICMF according to which clusters are formed, we stochastically draw the cluster masses from the ICMF. The numerical procedure used for generating these cluster masses differs from that used by \citet{Kruijssen_et_al_11}, where cluster masses could not exceed the parent stellar particle mass. That method disfavours the use of high-resolution simulations with low particle masses, because it would impose an undesirable upper limit to the cluster mass. To avoid any implicit limits on the numerical resolution of the simulations, we allow the stochastically-drawn masses of sub-grid clusters to exceed the stellar particle mass. This happens occasionally, but on average the cluster masses are compatible with the ICMF. This is practice is warranted because, under certain conditions, the upper truncation mass of the ICMF can be significantly greater than the baryonic particle mass of our simulations. In principle, our method enables the application of {\sc MOSAICS} in numerical simulations with arbitrarily low particle masses. The approach is effectively identical to that proposed by \citet{Sormani_et_al_17} to assign sub-grid stars to sink particles. 

The number of clusters, $N_\rmn{clust}$, expected to form in a given stellar particle of mass $m_*$ is governed by the ratio of the predicted total mass in clusters to the expected mean cluster mass
\begin{equation} \label{eq:nclust}
N_\rmn{clust} = \frac{\Gamma m_*} {\bar{m_\rmn{c}}} ,
\end{equation}
where the mean cluster mass is calculated by integrating the ICMF
\begin{equation} \label{eq:mcmean}
\bar{m_\rmn{c}} = \int^{10^8 \Msun}_{100 \Msun} M p(M) \, \rmn{d}M ,
\end{equation}
in which $p(M)$ is the normalised probability distribution function corresponding to the ICMF $N(M)$. The actual number of clusters `formed' by the stellar particle is then drawn from a Poisson distribution with mean $N_\rmn{clust}$, and cluster masses are sampled stochastically from the ICMF (Eq. \ref{eq:ICMF}). Clearly, the stochasticity of the method allows some newborn stellar particles to contain no clusters, whereas others contain a total mass in clusters in excess of the particle mass, but on average the drawn mass in clusters is equal to the desired mass, i.e.~$N_\rmn{clust}\bar{m_\rmn{c}}=\Gamma M_*$. To reduce memory requirements, clusters with initial masses below $5 \times 10^3 \Msun$ are discarded, which is warranted because such low-mass clusters are disrupted on short ($\ll\Gyr$) time-scales.

Finally, we assign radii to the clusters, which is necessary for cluster disruption (see below). Young clusters and GCs alike have radii of a few parsecs. YMCs have typical projected radii of $R_\rmn{eff} \sim 2$--$4 \pc$ \citep{Larsen_04,Bastian_et_al_12, Johnson_et_al_12}, with radius increasing with cluster age.
MW GCs have a typical radius $R_\rmn{eff} \sim 3.3 \pc$ \citep{McLaughlin_and_van_der_Marel_05}.
We follow \citet{Larsen_04}, who show that young clusters with masses $M=10^4$--$10^5~\Msun$ have effective radii of $R_\rmn{eff}=2.8$--$3.5 \pc$ and assumed to have a constant half-mass radius of $r_\rmn{h}=4R_\rmn{eff}/3=4 \pc$.
We perform additional simulations with cluster radii of $r_\rmn{h} = 1.5 \pc$ and $r_\rmn{h} = 6 \pc$ to assess how the cluster radius affects the properties of the cluster population (Appendix \ref{app:disruption}). At present, we omit the effects of cluster radius evolution \citep[e.g.][]{Gieles_Heggie_and_Zhao_11}, opting to defer this development to a future study. In Appendix \ref{app:disruption}, we test a simple model for cluster expansion due to stellar mass-loss, finding that simulations with early cluster expansion are nearly indistinguishable from those using constant radii. We note that the combination of relaxation and tidal shocks leads to cluster radii that depend very weakly on mass \citep{Gieles_and_Renaud_16}.

\subsubsection{Cluster evolution} \label{sec:clevo}

In {\sc MOSAICS}, the masses of clusters evolve as a result of stellar evolutionary mass-loss and dynamical processes. Mass-loss from stellar evolution is tracked for each stellar particle by the EAGLE model, based on the implementation of \citet{Wiersma_et_al_09}, using the stellar lifetimes of \citet{Portinari_Chiosi_and_Bressan_98}. At each timestep, the fractional mass-loss of each cluster due to stellar evolution is specified by the ratio of the current and previous mass of the parent stellar particle $m_\rmn{*}/m_\rmn{*,prev}$. For the dynamical evolution, we include mass-loss from both two-body relaxation and tidal shocks. The full derivation of dynamical mass-loss is described by \citet{Kruijssen_et_al_11}; for brevity, we provide here only the key expressions. {\sc MOSAICS} also contains a module that describes the evolution of the stellar content of the clusters (using stellar mass-dependent escape rates from \citealt{Kruijssen_09}), but we omit this part of the model to reduce computational expense. Clusters are evolved down to a minimum mass of $100 \Msun$\footnote{Note that although we only form clusters above masses of $5\times10^3 \Msun$ (which are expected to be disrupted in timescales $\ll\Gyr$), we follow mass-loss down to masses of $100 \Msun$ in order to trace the full disruption of massive clusters.}, after which they are assumed to be fully disrupted. 

The total mass loss rate of a cluster is the sum of the contributions from stellar evolution, two-body relaxation and tidal shocks:
\begin{equation} \label{eq:dmdt}
\derBr{M}{t} = \derBr{M}{t}_\rmn{ev} + \derBr{M}{t}_\rmn{rlx} + \derBr{M}{t}_\rmn{sh} .
\end{equation}
In practice, stellar mass loss is computed after dynamical mass loss such that mass loss is not double-counted. We omit dynamical effects on the cluster induced by stellar evolution \citep[i.e.~extra dynamical mass loss in response to the shrinking of the tidal radius][]{Lamers_Baumgardt_and_Gieles_10}. Dynamical mass loss from a cluster is added to the field star mass budget of the parent stellar particle.

Dynamical mass loss terms are governed by the local tidal field of the parent stellar particle, specified by the tidal field tensor:
\begin{equation} \label{eq:tidalTensor}
T_{ij} = - \frac{ \partial^2 \Phi }{ \partial x_i \partial x_j } ,
\end{equation}
where $\Phi$ is the gravitational potential and $x_i$ is the $i$th component of the coordinate vector.
The tidal tensor is calculated by numerical differentiation (using the forward difference approximation) of the gravitational field with a spatial interval of 1 per cent of the gravitational softening length (which for the simulations presented in this work results in an interval of a few pc). We have verified that our results are insensitive to the exact choice of this length by running simulations with differentiation intervals of 0.5, 5, 20, 50 and 100 per cent of the softening length.

The mass-loss rate from two-body relaxation is determined by the current cluster mass and the tidal field strength $T$:
\begin{equation} \label{eq:dmrlx}
\derBr{M}{t}_\rmn{rlx} = -\frac{\Msun}{\rmn{t}_{0,\sun}} \left(\frac{M}{\Msun}\right)^{1-\gamma} \left(\frac{T} {\rmn{T}_{\sun}} \right)^{1/2},
\end{equation}
where $\gamma$ is the mass dependence of the dissolution time-scale and $\rmn{t}_{0,\sun}$ is the dissolution timescale (which also depends upon $\gamma$) at the solar galactocentric radius with tidal field strength $\rmn{T}_{\sun} \approx 7.01 \times 10^2 \Gyr^{-2}$ \citep{Kruijssen_et_al_11}.
In this work we assume a cluster density profile with King parameter $W_0 = 5$ for which $\gamma=0.62$ \citep{Lamers_et_al_05a} and $\rmn{t}_{0,\sun} = 21.3 \Myr$. $W_0=5$ corresponds to a King concentration $c \simeq 1$ which is found for clusters with masses $\sim 10^5 \Msun$ \citep{King_66, McLaughlin_00}. We also performed simulations adopting $W_0 = 7$ \citep[represented as $\gamma=0.7$ with $\rmn{t}_{0,\sun} = 10.7 \Myr$,][]{Kruijssen_and_Mieske_09}, but found the results are nearly indistinguishable from the fiducial simulations. A mass dependence scaling between $\gamma \approx 0.6$-$0.7$ has been derived from both $N$-body simulations \citep{Baumgardt_and_Makino_03, Gieles_and_Baumgardt_08} and observations \citep{Boutloukos_and_Lamers_03, Lamers_et_al_05a}. \citet{Kruijssen_and_PortegiesZwart_09} also found that $\gamma=0.7$ reproduces the shape of the MW GC mass function when accounting for an evolving $M/L$ ratio due to dynamical evolution. Alternative suggestions for the mass dependence of the dissolution time-scale include a mass-independent mass-loss rate \citep[i.e. $\gamma=1$,][]{Fall_and_Zhang_01, McLaughlin_and_Fall_08}. However, \citet{Gieles_and_Baumgardt_08} found that the fraction of stars lost per relaxation time \citep[assumed to be constant by][]{Fall_and_Zhang_01} depends on the tidal field strength in which case the mass dependence becomes $\gamma=0.65$, consistent with our formulation.
The tidal field strength, $T$, that sets the tidal radius of a cluster is given by $\partial^2 \Phi/\partial r^2 + \Omega^2$ \citep{King_62, Renaud_Gieles_and_Boily_11}. As we show in Appendix \ref{app:tidal_field}, the tidal field strength can be determined from the eigenvalues of the tidal tensor as $T = \rmn{max}(\lambda) + \Omega^2$, where the circular frequency is calculated from the eigenvalues according to Eq. \ref{eq:Omega}. 
We quantify the effect of the inclusion of $\Omega^2$ in Appendix \ref{app:tidal_field}.
If $T < 0$, we assume $(\rmn{d}M / \rmn{d}t)_\rmn{rlx} = 0$. However, fully compressive tidal fields are rare due to the inclusion of the circular frequency term in the tidal field strength.

The mass-loss rate due to tidal shocks in the impulse approximation from the first- and second-order energy terms is given by
\begin{equation} \label{eq:dmsh}
\derBr{M}{t}_\rmn{sh} = - \frac{20.4 \Msun}{\Myr} \left(\frac{r_{\rm h}}{4~\pc}\right)^3 \left(\frac{I_\rmn{tid}}{10^4~\Gyr^{-2}}\right) \left(\frac{\Delta t}{10~\Myr}\right)^{-1} ,
\end{equation}
where $I_\rmn{tid}$ is the tidal heating parameter and $\Delta t$ is time since the previous shock. Note that the constant also depends on $W_0$ (or $\gamma$) and we have again assumed $W_0=5$ \citep[see][]{Kruijssen_et_al_11}.
We write the tidal heating parameter in terms of the tidal tensor \citep{Gnedin_Hernquist_and_Ostriker_99, Prieto_and_Gnedin_08}:
\begin{equation} \label{eq:Itid}
I_\rmn{tid} = \sum_{i,j} \left( \int T_{ij} \, \rmn{d}t \right)^2 A_{\rmn{w},ij} ,
\end{equation}
where $A_{\rmn{w},ij}$ is the Weinberg adiabatic correction \citep{Weinberg_94a,Weinberg_94b,Weinberg_94c} that describes the absorption of energy injection by the adiabatic expansion of the cluster. The integral is performed over the full duration of the shock in each component of the tidal tensor between valid minima which have sufficient contrast with the bounded maximum ($<0.88$ of the maximum, corresponding to a total width equal to $1 \sigma$ in a Gaussian distribution).
The adiabatic correction depends on the time-scale of the shock for the corresponding component of the tidal tensor $\tau_\rmn{ij}^2$ \citep{Gnedin_and_Ostriker_97,Gnedin_Hernquist_and_Ostriker_99}:
\begin{equation} \label{eq:AW}
A_{\rmn{w},ij} = \left(1 + \eta_A \frac{G M}{r_{\rm h}^3} {\tau_\rmn{ij}}^2 \right)^{-3/2} ,
\end{equation}
where $G$ is the gravitational constant and $\eta_A=0.237$ is a constant that weakly depends on the cluster density profile.
For a cluster with a mass of $10^5 \Msun$ and radius of $4 \pc$, tidal shocks with timescales $> 1$ Myr will be absorbed by the cluster expansion captured in the adiabatic correction. Including mass-loss from tidal shocks for massive clusters therefore requires sub-Myr timesteps. Stellar particle timesteps scale as the logarithm of the expansion factor, ensuring smaller physical timesteps at higher redshift when dynamical times are shorter. At $z>1$ all stellar particles have timesteps $<1 \Myr$, with the smallest timesteps being $\sim 0.01 \Myr$, and by $z=0$ the smallest timestep is $\approx 0.5 \Myr$. At the resolution of our simulations (see Section \ref{sec:sims}), tidal shocks caused by encounters with individual particles over the lifetime of a star cluster are not important and would take over 6000 Gyr to disrupt a $10^3 \Msun$ cluster \citep[following the calculation in Section 2.2.4 of][]{Kruijssen_et_al_11}.

The above constitutes a summary of the `on-the-fly', sub-grid model for the disruption of stellar clusters from \citet{Kruijssen_et_al_11}. In combination with the cluster formation model of Section~\ref{sec:clform}, this model is near-exhaustive in the sense that it includes a description of most of the relevant physical processes. One process that we have not discussed, but which may however be important for a subset of the clusters in our simulations, is dynamical friction. We do not model dynamical friction on-the-fly, because stellar particles may host clusters of significantly different masses, resulting in a range of appropriate dynamical friction forces for a single stellar particle. Moreover, the mass of most stellar particles is dominated by the field star fraction. 

We therefore apply an approximate treatment in post-processing as follows. The dynamical friction timescale for a cluster of mass $m_\rmn{c}$ to spiral to the galactic centre is defined \citep{Lacey_and_Cole_93}:
\begin{equation} \label{eq:tdf}
t_\rmn{df} = \frac{f(\epsilon)}{2 B\left(v_c/\sqrt{2}\sigma\right)} \frac{\sqrt{2} \sigma r_c^2 }{ G m_\rmn{c} \ln{\Lambda} } ,
\end{equation}
where $r_c(E)$ is the radius of a circular orbit with the same energy as the actual orbit, $v_c$ is the circular velocity at $r_c$, $\sigma(r_c)$ is the stellar velocity dispersion interior to $r_c$\footnote{For calculating $\sigma(r_c)$ we use either the number of stellar particles interior to $r_c$ or a minimum of 48 particles, with the exception of galaxies with fewer than 48 stellar particles where we use dark matter particles.}, $\ln(\Lambda) = \ln(1+M(r_c)/m_\rmn{c})$ is the Coulomb logarithm with $M(r_c)$ the total mass within $r_c$ and $B(X) \equiv \erf(X) - 2 X \exp(-X^2) / \sqrt{\pi}$. The factor $f(\epsilon) = \epsilon^{0.78}$ \citep{Lacey_and_Cole_93} accounts for the orbital eccentricity, where $\epsilon \equiv J/J_c(E)$ is the circularity parameter (the angular momentum relative to that of a circular orbit with the same energy). Typically $\sqrt{2} \sigma(r_c)/v_c \approx 1.2$ which increases the timescale $t_\rmn{df}$ by a factor $\sim$2 over the standard definition \citep[with $B(1)$ and $\sigma=v_c/\sqrt{2}$,][]{Binney_and_Tremaine_08}.

The dynamical friction timescale for all star clusters is calculated at every snapshot. The current galaxy a stellar particle is bound to at any snapshot is determined by the \textsc{subfind} \citep{Springel_et_al_01, Dolag_et_al_09} algorithm (see following section). Clusters are assumed to be completely removed by dynamical friction (i.e.~we set their mass to zero) at the first snapshot where
\begin{equation}
t_\rmn{df} < t_\rmn{age} ,
\end{equation}
with $t_\rmn{age}$ the age of the cluster.
The method assumes clusters have remained in the current galaxy from birth. This approximation is satisfied by most clusters, as dynamical friction is only important in the central few kiloparsecs where very few clusters have an ex-situ origin.
Though we have assumed that clusters removed by dynamical friction are completely disrupted, in principle such clusters may contribute to the formation of nuclear star clusters \citep[e.g.][]{Capuzzo-Dolcetta_and_Miocchi_08}. We will investigate this effect in future work.


\section{The simulations} \label{sec:sims}

\begin{table*}
\caption{Properties of $L^\star$ galaxies at $z=0$ for the Recal zoom simulations. The columns show (from left to right): simulation name; halo mass; galaxy stellar mass; star forming gas mass; non-star forming gas mass; SFR averaged over 300 Myr; the redshift of the last major merger (stellar mass ratio $M_2/M_1 > 1/4$ where $M_1>M_2$). All baryonic galaxy properties are measured within 30 pkpc.}
\label{tab:sims}
\begin{tabular} {@{}cccccccc@{}}
  \hline
  Name & $\log M_{200}$ & $\log M_*$ & $\log M_\rmn{SF}$ & $\log M_\rmn{NSF}$ & SFR                 & $z_\rmn{MM}$ \\ 
       & [$\Msun$]      & [$\Msun$]  & [$\Msun$]         & [$\Msun$]          & [$\Msun$ yr$^{-1}$] &              \\ 
  \hline
  Gal000 & 11.95 & 10.28 & 9.39 & 10.34 & 0.632 & 1.49 \\
  Gal001 & 12.12 & 10.38 & 9.55 & 11.05 & 0.934 & --   \\ 
  Gal002 & 12.29 & 10.56 & 9.82 & 11.18 & 1.673 & 5.04 \\ 
  Gal003 & 12.18 & 10.42 & 9.80 & 11.05 & 1.810 & 1.49 \\ 
  Gal004 & 12.02 & 10.11 & 9.29 & 10.84 & 0.349 & 2.24 \\ 
  Gal005 & 12.07 & 10.12 & 8.51 & 10.32 & 0.075 & 5.49 \\ 
  Gal006 & 11.85 & 10.16 & 9.74 & 10.76 & 1.049 & --   \\ 
  Gal007 & 11.96 & 10.28 & 9.83 & 10.82 & 1.964 & 2.48 \\ 
  Gal008 & 11.87 & 10.12 & 9.34 & 10.78 & 1.076 & --   \\ 
  Gal009 & 11.87 & 10.16 & 9.62 & 10.52 & 1.356 & 2.24 \\ 
  \hline
\end{tabular}
\end{table*}

\begin{figure*}
  \includegraphics[width=\textwidth]{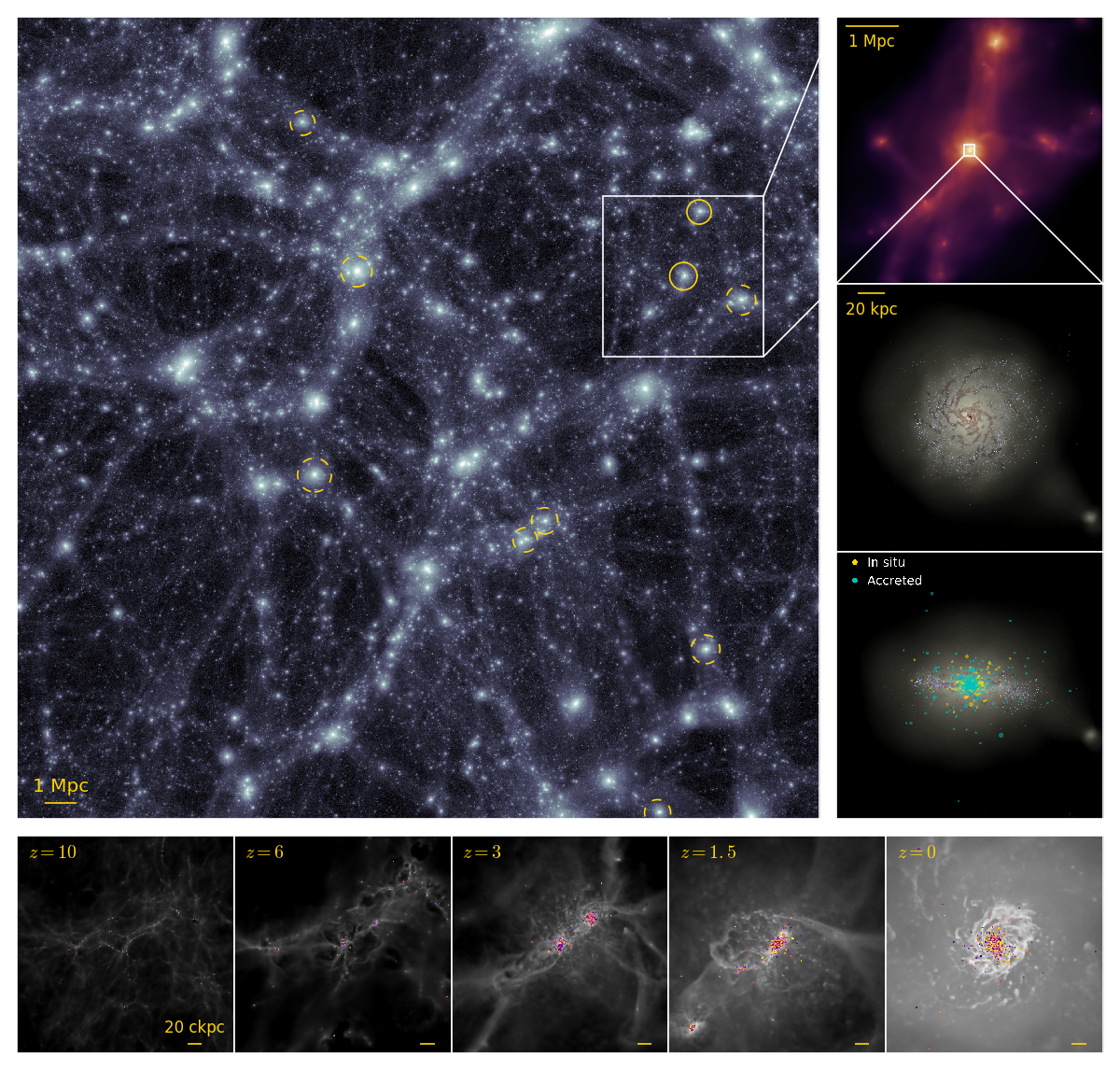}
  \caption{ Visualisation of the E-MOSAICS simulations. The main panel shows the dark matter distribution at $z=0$ from the EAGLE Recal-L025N0752 simulation. Yellow circles highlight the positions of the 10 $L^\star$ galaxies that we have resimulated, where solid lines show the two haloes in the inset on the right. Radii of the circles show the virial radii of the galaxies. The three panels on the right show successive zoom-ins of Gal004: the top panel shows gas density coloured by temperature in the zoom simulation; the lower two panels show mock optical images of face-on and edge-on views of the galaxy (blue for young stars ($<300\Myr$), brown for dense star-forming gas). The bottom panel also shows the locations of massive star clusters ($>5\times10^4 \Msun$) coloured by their formation location (in situ or accreted). The five panels in the bottom row show the formation history of the galaxy and its star cluster population, where grey scale shows the gas surface density and the points show positions of star clusters (with masses $>5\times10^4 \Msun$) coloured by metallicity (yellow for $\ZH=0.5$, blue for $\ZH=-2.5$) and with point area scaling with cluster mass. }

  \label{fig:visualisation}
\end{figure*}

Our focus here is the formation and evolution of GCs in typical $L^\star$ (spiral) galaxies, similar to the MW. We therefore appeal to `zoomed resimulations' \citep[e.g.][]{Katz_and_White_93} in order to follow such environments at high resolution in a computationally efficient fashion. We simulate the evolution of the same set of 10 galaxies studied by \citet{Mateu_et_al_17}, of which the parent volume is the Recal-L025N0752 simulation introduced by S15. This simulation adopts a particle mass that is a factor of 8 lower than the largest volume EAGLE simulation (Ref-L100N1504 in the terminology of S15), and a gravitational softening scale that is a factor of 2 lower. The 10 galaxies were identified as the most disc-dominated examples at $z=0$ within a volume-limited sample of 25 haloes with  total mass $7\times 10^{11} < M_{200}/\Msun < 3 \times 10^{12}$.

Multi-resolution initial conditions for each galaxy were established, such that in each case only the immediate environment of the galaxy's progenitors are followed at high resolution and with hydrodynamics. At $z=0$, the fully-sampled region is roughly spherical, centred on the target galaxy, and has a radius of at least $600~{\rm proper~kpc}$ (hereafter~$\pkpc$). Beyond this region, the large-scale environment is sampled only with collisionless particles, of which the masses increase with distance from the high-resolution region. The zoomed initial conditions were created using the second-order Lagrangian perturbation theory method of \citet{Jenkins_10} and the public Gaussian white noise field \textit{Panphasia} \citep{Jenkins_13}. They adopt the same linear phases\footnote{Descriptors specifying the \textit{Panphasia} linear phases used by each EAGLE volume are given in Table B1 of \citetalias{S15}.} and cosmological parameters as their parent volume, the latter being those specified by \citet{Planck_2014_paperI_short}: $\Omega_{\rm m} =  0.307$, $\Omega_\Lambda = 0.693$, $\Omega_{\rm b} = 0.04825$, $h = 0.6777$ and  $\sigma_8 =  0.8288$. 

Each set of zoom initial conditions was realised at approximately the same resolution as the parent simulation, yielding gas particles with initial masses of approximately $m_{\rm g}=2.25 \times 10^5\Msun$, and high-resolution dark matter particles with masses of approximately $m_{\rm dm}=1.2 \times 10^6\Msun$. The particle masses vary by up to 4 per cent between the runs, as the initial particle load is created by tiling a primitive, periodic cubic glass distribution of $10^3$ particles. The Plummer-equivalent gravitational softening length is fixed in comoving units to $1/25$ of the mean interparticle separation ($1.33~{\rm comoving~kpc}$, hereafter $\ckpc$) until $z=2.8$, and in proper units ($0.35 \pkpc$) thereafter.\footnote{As we show in Section \ref{sec:tidal_histories}, cluster disruption is slightly more efficient prior to $z=2.8$ due to the smaller physical scales of the softening length. However this has the greatest impact at $z>6$ and therefore affects few clusters. At $z=6$ the physical softening length is $1.33 \ckpc = 0.19 \pkpc$, and therefore nearly half the softening length of $0.35 \pkpc$ at $z<2.8$.}
The standard-resolution simulations therefore marginally resolve the Jeans scales at the SF threshold in the warm ($T\simeq 10^4\K$) ISM. The SPH kernel support radius is limited to a minimum of one-tenth of the gravitational softening.

With the above setup, the simulations resolve the formation of galaxies down to stellar masses of $\simeq 2 \times 10^7 \Msun$ with at least 100 stellar particles, and therefore galaxies massive enough to form globular clusters \citep[a similar mass to the Fornax dSph, one of the lowest mass Local Group galaxies with GCs, e.g.][]{Forbes_et_al_00}. We also note that Recal-L025N0752 is a desirable parent volume for these zoom simulations, since the Recal model more accurately reproduces the metallicities of dwarf galaxies than the EAGLE Reference model (see Fig. 13 of S15). This is relevant for modelling low-metallicity GCs.

For each simulation we save 29 snapshots between redshifts 20 and 0, as for the EAGLE simulations. The method for identifying galaxies\footnote{We use the terms galaxy and subhalo interchangeably.} using \textsc{subfind} \citep{Springel_et_al_01, Dolag_et_al_09} is described by \citetalias{S15}.
Briefly, dark matter structures are first identified using the friends-of-friends (FoF) algorithm \citep{DEFW85} with a linking length 0.2 times the mean interparticle separation. Gas, star and black hole particles are associated with the FoF group of their nearest linked dark matter particles. The \textsc{subfind} algorithm then identifies gravitationally bound substructures within the FoF groups. As discussed by \citetalias{S15}, subhaloes separated by less than the stellar half-mass radius of the primary galaxy or 3 pkpc (whichever is smaller), are merged to rectify the occasional misidentification of intra-disc structure as a separate galaxy. We create subhalo merger trees in a similar manner to \citet{Jiang_et_al_14} and \citet{Qu_et_al_17}. Subhaloes are linked between snapshots by searching for the $N_\rmn{link} = \min[100, \max(0.1 N, 10)]$ most bound particles of a subhalo in candidate descendant subhaloes for up to 5 of the following snapshots, where $N$ is the total number of particles in a subhalo. This method can identify a descendant even when most of the outer particles of a subhalo have been stripped away.
Where the $N_\rmn{link}$ particles are spread across multiple subhaloes we rank descendants with a score $\chi = \sum_j \mathcal{R}_j^{-2/3}$, where $\mathcal{R}$ is the binding energy rank of the $N_\rmn{link}$ particles, which ranks the most bound regions most heavily \citep[similar to][]{Boylan-Kolchin_et_al_09}. The subhalo with the largest value of $\chi$ is defined to be the descendant subhalo. This part of the procedure differs from the \citet{Jiang_et_al_14} method and is found to be necessary to determine the main descendant in a very few cases where multiple possible descendants have the same number of $N_\rmn{link}$ particles. The main progenitor branch of a subhalo is chosen as the branch with the highest `branch mass' (the sum of the total subhalo mass for all progenitors on the same branch).

The nature of the E-MOSAICS simulations is visualised in Fig. \ref{fig:visualisation}. The main panel shows the dark matter distribution at $z=0$ in the full $25\times25\times25$ Mpc box from the EAGLE Recal-L025N0752 simulation where yellow circles highlight the positions of the 10 $L^\star$ galaxies that we have resimulated. 
The solid circles in the main panel highlight the two main galaxies in the inset on the right, where the centred galaxy is Gal004. Though Gal008 also appears in the zoom (top of the inset), only Gal004 is free of contaminant, low resolution dark matter particles.
The three panels on the right show successive zoom-ins of Gal004 from our resimulation. The top panel shows a $5\times5\times5$ Mpc region of the zoom-in simulation for which the brightness scales with the logarithm of the gas surface density and the colour scales with the logarithm of the temperature (black for $10$ K, yellow for $10^{5.5}$ K. The bottom two panels show mock optical images of the galaxy within a $150\times150\times150$ kpc box: brightness shows stellar surface density; blue points show young ($<300$ Myr) stars; brown points show star-forming (dense) gas. A dwarf galaxy with tidal tails is clearly visible to the right of the image.
The bottom panel also shows the locations of massive star clusters ($>5\times10^4 \Msun$), split into those with an `in situ' or `accreted' origin (based on the subhalo merger tree and the subhalo the particle was bound to at the last snapshot it was a gas particle). In situ clusters show a very concentrated spatial distribution, with most having galactocentric radii less than 5 kpc. Accreted clusters exhibit a more extended spatial distribution, with radii of up to a few hundreds of kiloparsecs, though most are located within 50 kpc of the galaxy.
The five panels in the bottom row show the formation history of the galaxy and its star cluster population within a $300\times300\times300$ ckpc box. The gas surface density is shown in grey scale. The coloured points show positions of star clusters with masses $>5\times10^4 \Msun$ coloured by metallicity (yellow for $\ZH=0.5$, blue for $\ZH=-2.5$) and with point area scaling with cluster mass.
At high redshift the galaxies undergo a significant number of mergers which redistributes the (mostly) low metallicity clusters that have formed. High metallicity clusters ($\ZH > -0.5$ dex) only form at redshifts $z\lesssim3$, mainly within a few kpc of the galactic centre where galaxy self-enrichment is highest. 
At $z=2.25$ the galaxy undergoes a gas-rich major merger which results in centralised star and stellar cluster formation.

Basic properties of the $\numgal$ $L^\star$ galaxies at $z=0$ are presented in Table \ref{tab:sims}. Following \citet{Qu_et_al_17}, we define major mergers as having a stellar mass ratio $M_2/M_1 > 1/4$ (where $M_1>M_2$). We compare the mass ratio in the three previous snapshots before the merger in order to account for dynamical mass-loss during the merger. 
Two of the galaxies (Gal000 and Gal003, at $z=1.5$) experience a major merger at $z<2$. The region followed with high resolution and hydrodynamics is intentionally kept relatively large, to ensure the targeted galaxy and its progenitors are not contaminated by low-resolution boundary particles at any stage of their evolution. Therefore, the simulations also follow the evolution of `bonus' galaxies that are not satellites of the target galaxy, and many of these are also uncontaminated by boundary particles. The bonus galaxies are mostly sub-$L^\star$ with $M_\ast \sim 10^8$-$10^9 \Msun$, although the Gal000 simulation also contains an uncontaminated elliptical galaxy with $M_{200} = 10^{12.7} \Msun$ and $M_\ast = 10^{10.6} \Msun$, located at a distance of 3 Mpc from the targeted galaxy at $z=0$. Each $L^\star$ galaxy is the most massive galaxy within a distance of 1 Mpc.

\begin{figure}
  \includegraphics[width=84mm]{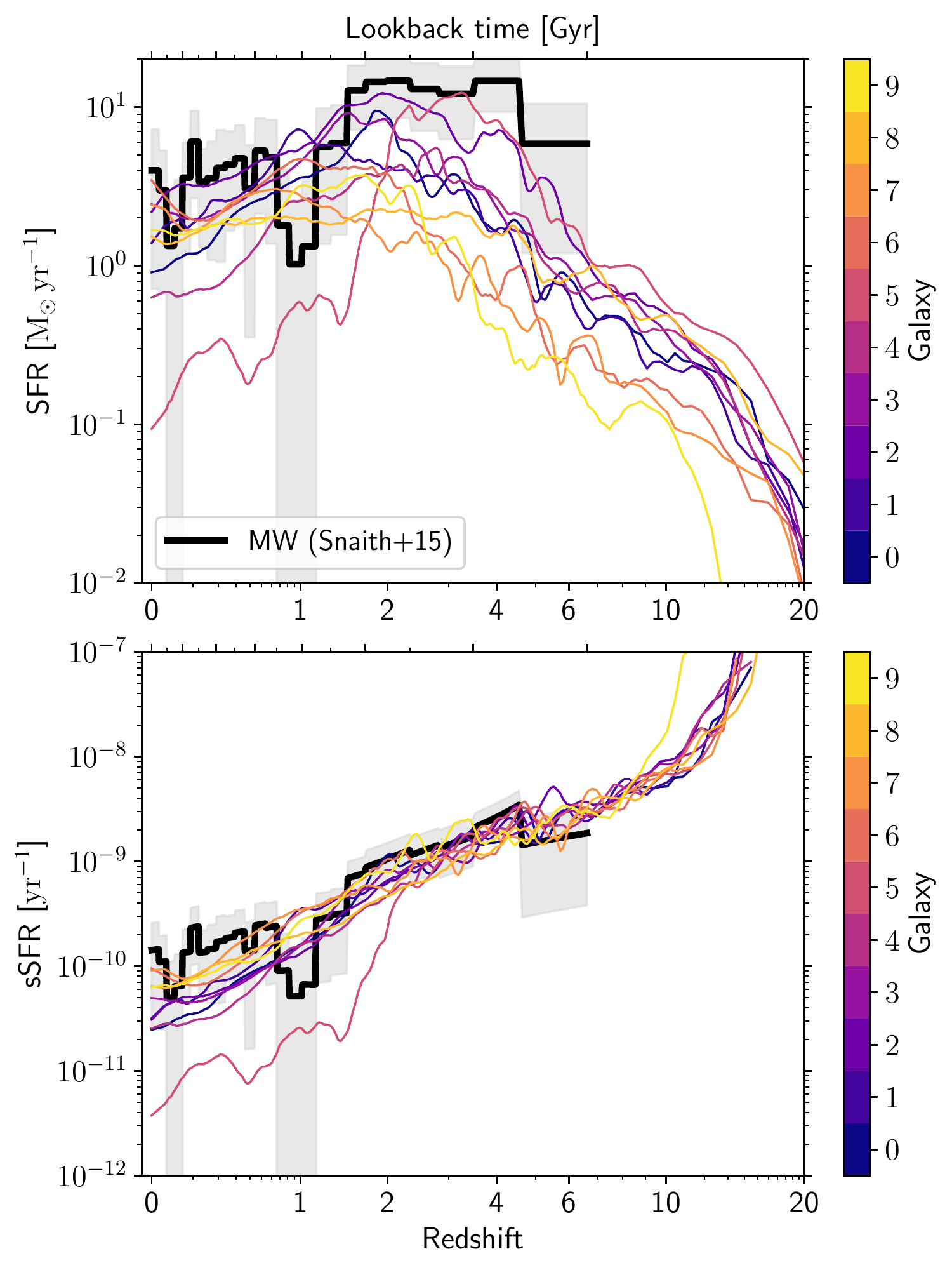}
  \caption{The instantaneous SFR (top) and specific SFR (sSFR; bottom) histories of the $\numgal$ $L^\star$ galaxies comprising our sample. The SFR was calculated in sliding bins of width $\Delta \log(1+z) = 0.05$ (with a minimum physical timescale of $200 \Myr$) using steps of one tenth the bin width. The SFR of each galaxy peaks in the interval $1 < z < 3$. The simulations are in good agreement with the SFR calculated for the MW \citet[solid black line with the grey shaded region showing the standard deviation;][]{Snaith_et_al_14,Snaith_et_al_15}. Though the SFRs at each redshift span more than a decade the sSFRs follow a very tight relation, with the exception of Gal005 which quenches at $z\simeq 2$. Gal004 (our exemplar galaxy below) follows the mean SFR for all galaxies until $z<1$, where it drops to one of the lowest SFRs.}
  \label{fig:SFR}
\end{figure}

The star formation histories of the $\numgal$ targeted galaxies are shown in Fig. \ref{fig:SFR}. The histories are similar and typically reach a peak SFR at redshifts $2 \lesssim z \lesssim 3$. Gal006 and Gal007, however, peak much later at $z<1$. The maximum SFRs achieved are between 2 and 10 $\Msun \pyr$, and the galaxies that peak earlier achieve higher peak SFRs.
For reference, the MW SFR determined from a chemical evolution model by \citet{Snaith_et_al_14,Snaith_et_al_15}, normalised such that the total MW mass at $z=0$ is $5 \times 10^{10} \Msun$ \citep{Bland-Hawthorn_and_Gerhard_16} and accounting for stellar evolution mass-loss, is shown by a solid black line. The grey shaded region shows the standard deviation of the model. We do not show data from $>13 \Gyr$ as the SFR is poorly constrained due to a lack of stars. The simulations are in good agreement with the MW SFR and sSFR. With the exception of the brief dip at $z\approx1$ which is required to fit the $[\rmn{Si/Fe}]$ evolution of MW stars \citep{Snaith_et_al_15}, the MW is consistent with the highest SFRs achieved in the simulated galaxies.
With the exception of Gal005, which appears to be quenched in star formation at $z=0$, the galaxies all follow a very similar trend in specific star formation rate (sSFR). At $z=0$, the SFR does generally not correlate with sSFR, indicating that the galaxies with the highest present-day SFRs are not simply the most massive galaxies.

We have also conducted, in addition to the fiducial models, simulations of all 10 galaxies without cluster formation physics (i.e.~a constant CFE and a power-law ICMF are adopted), and simulations with only one of the models active (i.e.~a variable CFE with a power-law ICMF; a ICMF truncation model with constant CFE) in order to assess the influence of these model components. For Gal004 (chosen simply because the simulation run time was lowest) we also ran a number of simulations to test the influence of the EAGLE sub-grid models on the cluster population, including the use of a constant SF density threshold of $n_\rmn{H} = 0.1$ cm$^{-3}$ (Appendix \ref{app:SFThresh}), different exponents of the polytropic equation of state (isothermal $\gamma_\rmn{EOS}=1$ and adiabatic $\gamma_\rmn{EOS}=5/3$; Appendix \ref{app:EOS}). We have also conducted simulations adopting finer time-stepping  ($1/5$ and $1/10$ the standard timesteps) and differing cluster radii (1.5 and $6 \pc$) to assess the convergence of the cluster disruption rate in the fiducial simulations (Appendix \ref{app:disruption}). In the interest of brevity we confine discussions of the influence of changing these aspects of the model to the appendices.

In total, we have conducted and analysed a total of $\numsim$ zoom-in simulations of the $\numgal$ galaxies listed in Table~\ref{tab:sims}. By varying the adopted physical models and parameter values, we aimed to establish a thorough understanding of how they influence the resulting cluster population. While we do not discuss each of these in detail in this paper, the insights drawn from this comprehensive parameter survey across $\numsim$ simulations have been essential for obtaining the results and conclusions presented in this work. A single or even a handful of simulations is insufficient for isolating which model ingredients are the most important in shaping the modelled cluster populations and for eliminating any numerical effects on the observables of interest. 


\section{Cluster formation properties} \label{sec:formation_props}

In this section we first verify the cluster formation model by comparing the $z=0$ predictions of the model with the properties of observed nearby galaxies and their young cluster populations. We then show results for the predicted cluster formation properties over the full formation history of the galaxies.
In the following sections we exclude the most metal-poor stellar particles ($\ZH<-3$ dex; these are very small in number, see Figures \ref{fig:CFE} and \ref{fig:SFThresh} below) from the analysis because their properties may strongly depend on the treatment of Population III stars, which are not modelled by EAGLE.

\subsection{Cluster formation efficiency}

\begin{figure}
    \includegraphics[width=84mm]{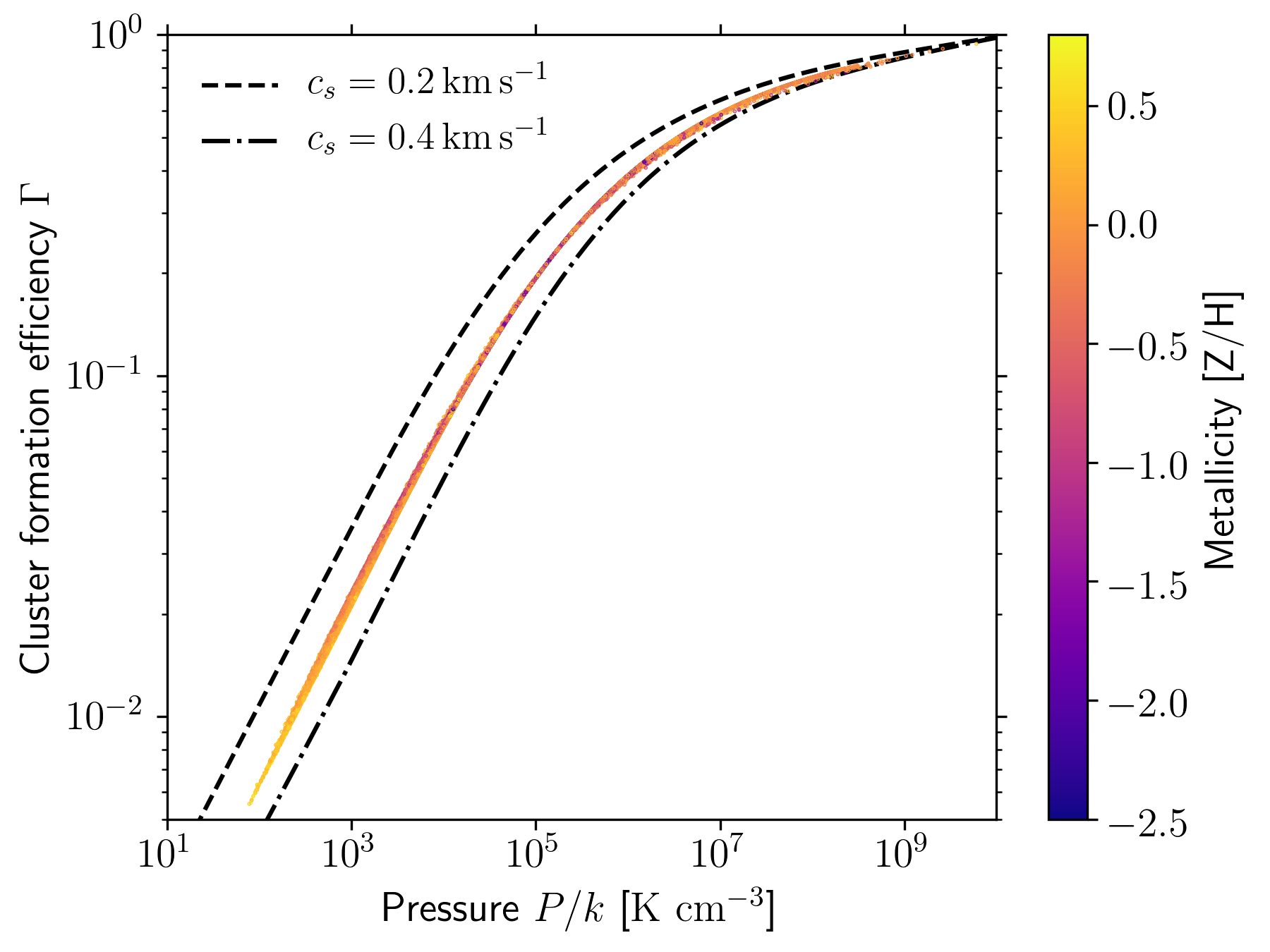}
\includegraphics[width=84mm]{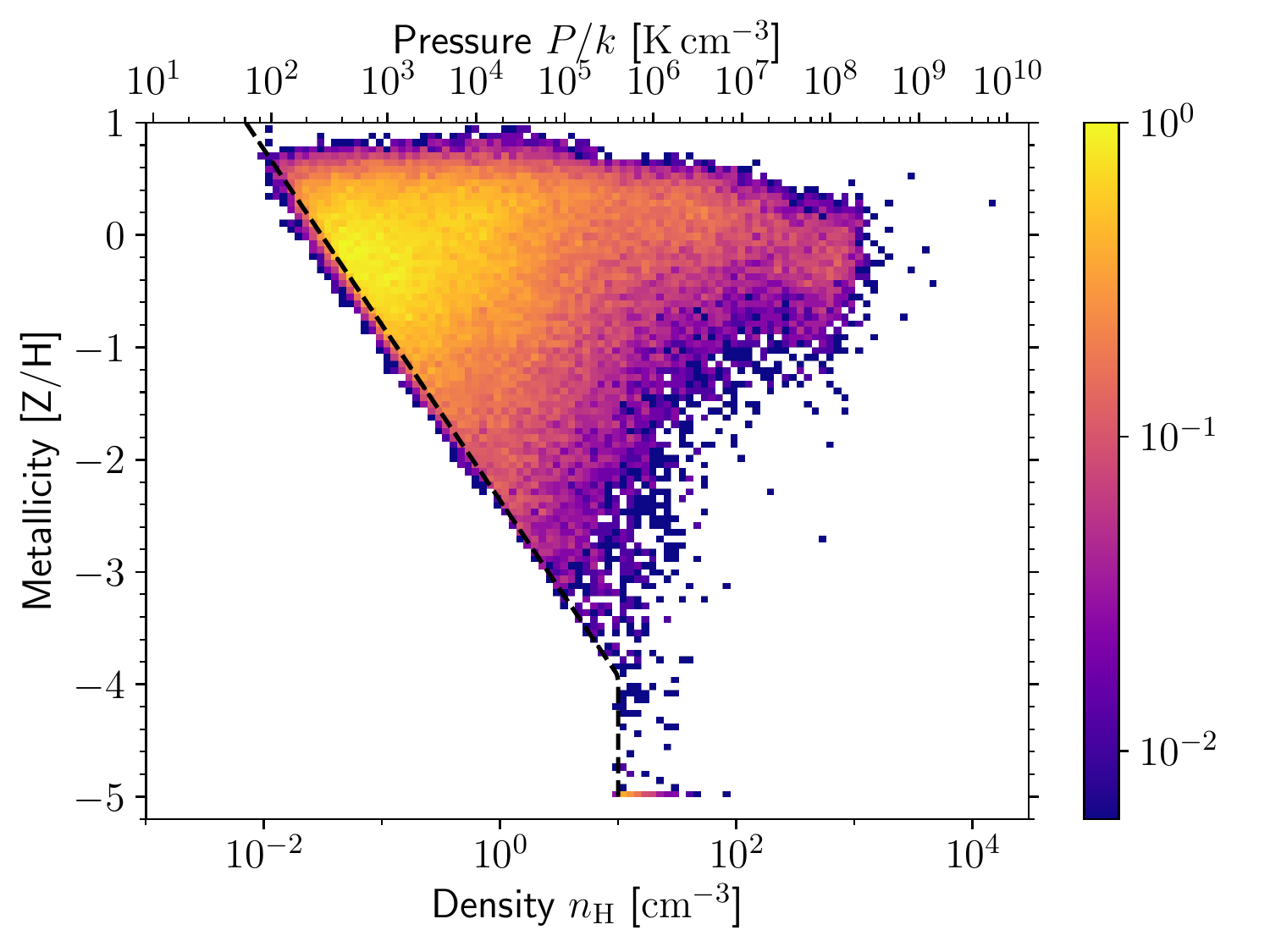}	
  \caption{Upper panel: The CFE, as a function of birth pressure and coloured by metallicity, of all stellar particles within 100 kpc of the centre of mass of Gal004. Dashed and dash-dotted lines show the relation when assuming $c_{\rm s} = 0.2 \kms$ and $c_{\rm s} = 0.4 \kms$ rather than the fiducial $c_{\rm s} = 0.3 \kms$. In our parametrization of the \citet{Kruijssen_12} CFE model, CFE is a monotonic function of pressure. Lower panel: Two-dimensional histogram of the birth pressure and stellar metallicity of the same set of stellar particles. Very low metallicity particles are plotted at a metallicity $\ZH = -5$ dex. The metallicity dependent star formation threshold is shown by the black dashed line. The upper axis shows pressure assuming particles follow the Jeans-limiting polytropic equation of state.}
  \label{fig:CFE}
\end{figure}

The upper panel of Fig. \ref{fig:CFE} shows the CFE as a function of `birth pressure' (i.e.~the gas pressure at the moment the particle was converted from a gas to a stellar particle) for all stellar particles formed in the Gal004 simulation, with points coloured by the metallicity of the star. Birth pressure is the thermodynamic pressure of a cluster's parent gas particle at the instant of conversion, which, as we show below (see Fig. \ref{fig:radial_props}), is a reasonable approximation of the pressure of cold gas in observed galaxies. 

In the \citet{Kruijssen_12} model, $\Gamma$ (as a function of $\rho_{\rm g}$, $\sigma_\rmn{loc} = \sqrt{P_{\rm g}/\rho_{\rm g}}$ and a constant sound speed $c_{\rm s} = 0.3$ km s$^{-1}$) depends almost entirely on the birth pressure. Since we adopt a fixed sound speed for the putative cold ISM phase, $c_{\rm s}$ acts primarily as a normalization for $\Gamma$. Increasing (decreasing) $c_{\rm s}$ by 0.1 km s$^{-1}$ changes $\Gamma$ by a factor 0.7 (1.7) at $P/k = 100 \K \cmcubed$, and by less than 10 per cent for $P/k > 10^7 \K \cmcubed$. 

For the range of birth pressures realised by clusters in Gal004, the CFE varies between $\Gamma \sim 0.01$ and unity. Formation efficiencies of approximately 1 per cent or lower are achieved only for stellar particles with super-solar metallicity (and at redshifts $z < 1$, see Fig. \ref{fig:particles} below), for which the density threshold for star formation in EAGLE is $\lesssim 0.03 \cmcubed$. The lower panel of Fig. \ref{fig:CFE} shows a two-dimensional histogram of the birth density-metallicity plane. Birth density can be connected uniquely to the birth pressure subject to the approximation that stars are born on the polytropic Jeans-limiting equation of state.\footnote{Recall from Section~\ref{sec:eagle} that gas with density greater than the density threshold for star formation is in fact eligible for star formation at temperatures up to 0.5 dex higher than those set by the equation of state.} We use this approximation to draw the upper $x$-axis on the plot, thus visualising the connection between $\Gamma$, $Z$ and $n_{\rm H}$. The appearance of this plot is similar for all of our $\numgal$ simulated galaxies, though the peak metallicity and birth pressure, and the fraction of high-pressure star formation ($P>10^6 \K \cmcubed$), differ slightly in each case.

\subsection{Radial distributions at $z=0$}

\begin{figure*}
  \includegraphics[width=\textwidth]{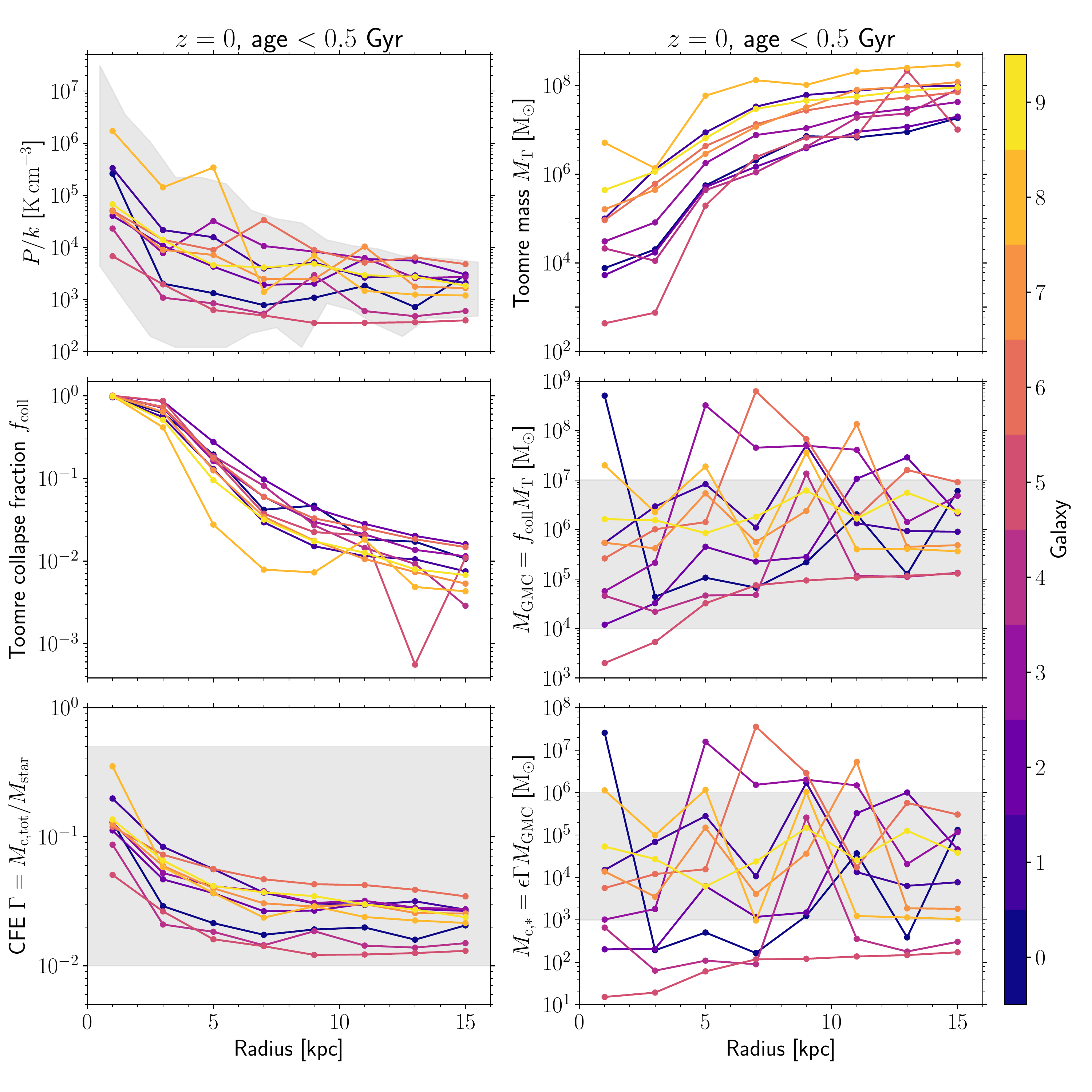}
  \caption{Radial distributions of the cluster formation properties for young disc stars (ages less than 0.5 Gyr) in all $\numgal$ galaxies at redshift $z=0$. The lines for each galaxy show the mean value as a function of radius for all panels, with the exception of $\Mtoomre$ which shows the median since it is particularly susceptible to outliers. In the top left panel the shaded region shows the pressure distributions of observed disc galaxies from \citet[assuming hydrostatic equilibrium, see text]{Leroy_et_al_08}. The shaded regions in the middle right and bottom panels show the observed ranges of $\Mgmc$, CFE and $\Mcstar$ in nearby galaxies (see text for references). Although the pressure, and therefore also the CFE, show a decreasing trend with radius, $\Mgmc$ and $\Mcstar$ show nearly flat trends with radius at $z=0$ due to the decreasing Toomre mass collapse fraction with increasing radius, indicating that these maximum mass scales become increasingly limited by stellar feedback.}
  \label{fig:radial_props}
\end{figure*}

Fig. \ref{fig:radial_props} shows the radial distributions of the stellar particle birth pressure and the cluster formation properties $\Mtoomre$, $f_\rmn{coll}$, $\Mgmc$, CFE ($\Gamma$) and $\Mcstar$ for all $\numgal$ $L^\star$ galaxies at $z=0$ for stars younger than 0.5 Gyr. 
The mean birth pressure for stellar particles (top left panel) show a strong trend with radius for all galaxies. Pressure peaks at the galactic centre and decreases until a radius of $\sim8 \kpc$, at which point pressure becomes approximately constant with radius. However, the distributions show large variation between galaxies. In particular Gal005 (the quenched galaxy with the lowest sSFR, see Fig. \ref{fig:SFR}) shows the lowest star birth pressures, while Gal008 shows the highest pressures as a result of very central star formation.
As a verification of the star-forming gas pressure of galaxies in the EAGLE model, since this variable underpins much of the cluster formation model, in Fig. \ref{fig:radial_props} we also compare the pressure distributions to estimated values for nearby disc galaxies from the sample of \citet[where we include only those galaxies with CO measurements]{Leroy_et_al_08}. The galaxies have stellar masses in the range $10^{10}$-$10^{11} \Msun$, similar to the range of stellar masses for our simulated galaxies (Table~\ref{tab:sims}).
Total cold gas surface density for the observed galaxies is calculated as the sum of the $\rmn{H_I}$ and $\rmn{H_2}$ surface densities: $\Sigma_{\rm g} = \Sigma_\rmn{HI} + \Sigma_\rmn{H_2}$. Gas surface density is then converted to pressure assuming $P = \pi \phi_P G \Sigma_{\rm g}^2 / 2$ \citep[with $\phi_P = 3$,][]{Krumholz_and_McKee_05}. The full range of pressures for galaxies in this sample are shown as the grey range in the figure.
Overall, the range of pressures shows very good correspondence between the simulated and observed galaxies. The observed galaxies show a very similar trend of decreasing pressure with radius to the simulated galaxies and the scatter for both sets of galaxies is similar over the full radial range shown. This indicates that the simulated galaxies provide realistic initial conditions for cluster formation at $z=0$.

We now focus on the cluster formation properties in Fig. \ref{fig:radial_props}.
The top right panel shows the Toomre mass, $\Mtoomre$. Recall from Eq. \ref{eq:Mtoomre} that $\Mtoomre$ is a function of the gas surface density $\Sigma_{\rm g}$ (itself calculated from the star-forming gas pressure) and the epicyclic frequency $\kappa$.
Although $\Sigma_{\rm g}$ decreases with radius in the galaxies (see Appendix~\ref{app:Sigma}), $\Mtoomre$ increases with radius for all galaxies due to decreasing $\kappa$, reaching a maximum of $\sim 10^8 \Msun$ beyond $\sim$10 kpc. The smallest radial bins for $\Mtoomre$ show a larger range (4 dex) of values for the galaxies than the largest radial bins (1 dex). This is due to the large range of gas surface densities $\Sigma_g$ of the galaxies (mainly through $\phi_P$ and, hence, the gas fraction $f_\rmn{gas}$, since $\Mtoomre$ does not directly correspond to $P$ in the first panel). The galaxy with the lowest $\Mtoomre$ at nearly all radii, Gal005, also has the lowest star-forming gas pressure and SFR at $z=0$ of all galaxies in our sample. 
As we show in Appendix \ref{app:Sigma}, our `particle-centric' calculation of $\Mtoomre$ underestimates the true value by $\sim0.5$ dex at small radii due to the underestimation of $\Sigma_{\rm g}$ following this approach. Specifically, we approximate the mid-plane gas pressure $P_\rmn{mp}$ through the local gas pressure of the star-forming particle, which in general underestimates the true pressure in the mid-plane $P_\rmn{mp}$ due to the vertical offsets of particles from the mid-plane. Additionally, the scale heights of disc galaxies in EAGLE are too large by a factor of $\sim$2, which also results in lower gas pressures (though this also affects $\phi_P$, making the quantitative effect uncertain).

Although the Toomre mass governs the maximum possible mass that may collapse, given an infinite timescale, it makes no statement on the actual mass in a given area that will collapse into star-forming molecular clouds.
To determine the maximum masses of molecular clouds, $\Mgmc$, we calculate the fraction of $\Mtoomre$ which can collapse before stellar feedback destroys the cloud \citep{Reina-Campos_and_Kruijssen_17}.
The Toomre collapse fraction $f_\rmn{coll}$ (middle left panel) shows the opposite trend with radius to $\Mtoomre$, having a maximum of unity at the smallest radial bins and reaching a minimum beyond 10 kpc. Therefore, within $\sim 2$ kpc, $\Mgmc$ (middle right panel) is limited by the Toomre mass and is feedback-limited beyond this radius. 
The combination of $\Mtoomre$ and $f_\rmn{coll}$ results in $\Mgmc$ being approximately independent of galactocentric radius, though with significant scatter for some galaxies (e.g.~Gal004 at 9 kpc). The typical values for the galaxies ranges between $10^5$ and $10^7 \Msun$ and is in good agreement with observed molecular clouds in MW, M31, and M83 \citep[$\sim10^4$-$10^7 \Msun$, shown as the grey shaded region;][Schruba et al.~in prep.]{Heyer_et_al_09, Freeman_et_al_17, Johnson_et_al_17}. 

The bottom left panel shows the mean CFE for all galaxies.
The characteristic shape of the CFE distributions, with $\Gamma \gtrsim 10$ per cent within $\simeq 2$ kpc and a few per cent farther out, is similar to the observed distributions of nearby disc galaxies \citep{Silva-Villa_et_al_13, Johnson_et_al_16}. The global CFE at $z=0$ of the $\numgal$ galaxies span a range from 1.5 (Gal005) to 30 (Gal008) per cent, covering a similar range to that of observed galaxies \citep[$\approx$1-50 per cent, shown as the grey shaded region; e.g.][]{Adamo_et_al_11,Adamo_et_al_15,Johnson_et_al_16}.
Given that the CFE is a function of pressure in our model (Fig. \ref{fig:CFE}), the similarity of the pressure (top left) and CFE panels is expected.

Because the ICMF truncation mass, $\Mcstar$ (bottom right panel), is linearly proportional to both $\Gamma$ and $\Mgmc$, the radial profiles are approximately flat with galactocentric radius. For most galaxies $\Mcstar$ is in good agreement with observed galaxies, being in the range $10^3$-$10^6 \Msun$ \citep[shown as the grey shaded region]{Johnson_et_al_17}. The galaxies with the lowest $\Mcstar$ are also the galaxies with the lowest gas pressures and star formation rates at $z=0$ ($<1$ $\Msun \, \rmn{yr}^{-1}$).
Again, we see a strong correlation between $\Mcstar$ and pressure (which we demonstrate directly in Appendix \ref{app:EOS}), except where the epicyclic frequency $\kappa$ is highest in the galaxies (the inner few kpc).
For some galaxies (Gal004 and Gal005), the mean $\Mcstar$ is below the minimum mass for cluster formation ($100 \Msun$). However for nearly all points in the figure the maximum $\Mcstar > 100 \Msun$, meaning some clusters are still expected to form and an `observed' $\Mcstar$ for the galaxies may be higher than the mean shown here. 

The results from Fig. \ref{fig:radial_props} demonstrate the good correspondence at low redshift between the galaxy and cluster formation properties realised by the E-MOSAICS model, and those observed in nearby disc galaxies. They verify the ability of the model to predict cluster formation properties from local gas and dynamical properties in the simulated galaxies. We therefore now turn to the application of the model over the full galaxy formation history in the simulations, and discuss the resulting predictions for GC population properties.

\subsection{Redshift evolution of the cluster formation physics}

\begin{figure*}
  \includegraphics[width=\textwidth]{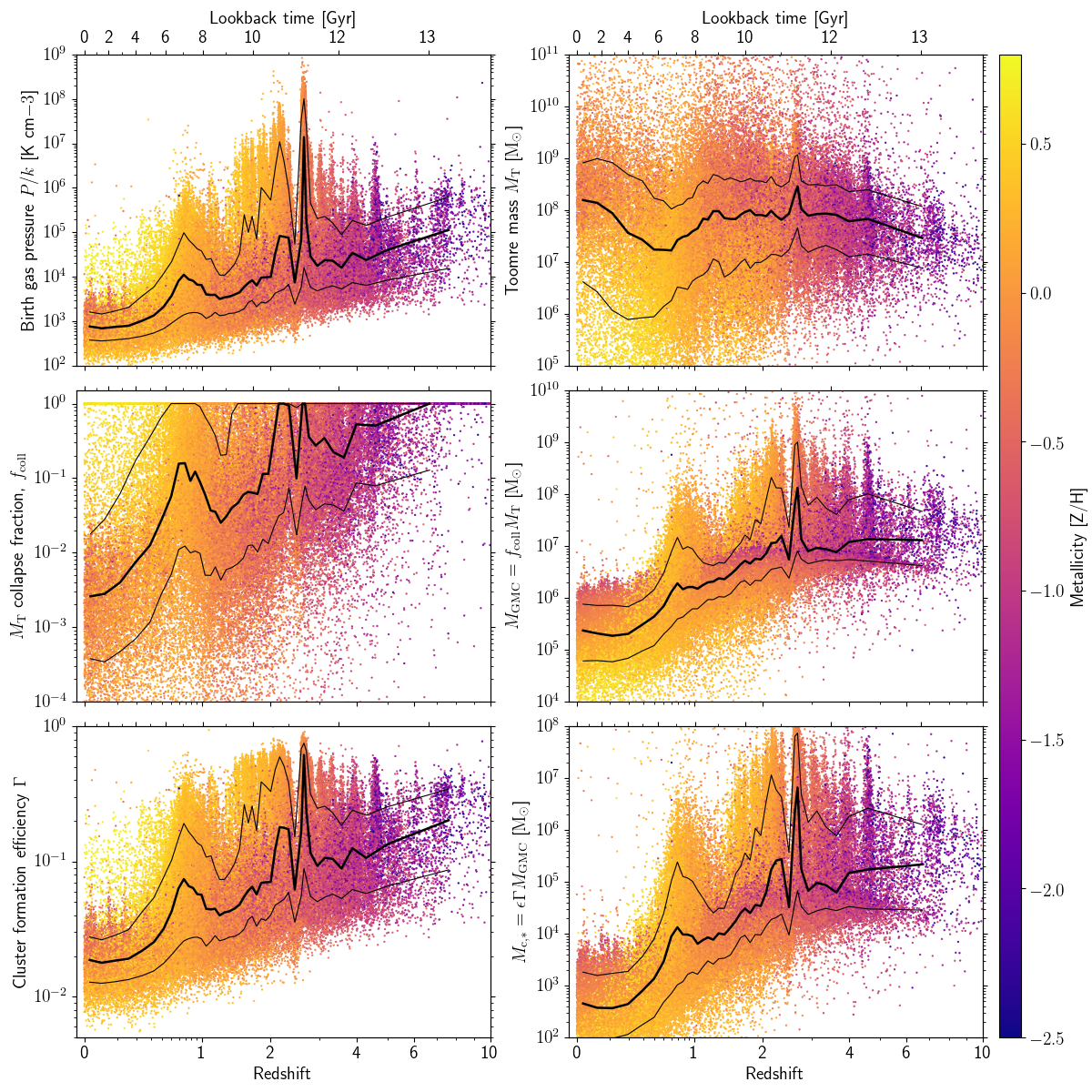}
  \caption{Cluster formation properties for all stellar particles within 100 kpc of Gal004 at $z=0$. Points are coloured by the metallicity of the parent stellar particle hosting the cluster population. The thick (thin) black lines show the median (standard deviation) as a function of redshift. The cluster formation properties for this galaxy decline at low redshifts due to falling pressures of star formation, but with peaks $z=2.5$ (at the same redshift as the SFR, Fig. \ref{fig:SFR}) and $z=0.7$ due to increased pressures from gas-rich galaxy mergers (Fig. \ref{fig:trees}).}
  \label{fig:particles}
\end{figure*}

Fig. \ref{fig:particles} shows the cluster formation properties (panels as per Fig. \ref{fig:radial_props}) as a function of redshift for Gal004, with the points coloured by the metallicity of the parent stellar particle. 
The birth pressure of clusters declines with advancing cosmic time and exhibits a peak at the same time as the SFR, at $z \approx 2.5$ for this particular example (see Fig. \ref{fig:SFR}). This galaxy exhibits two major peaks in birth pressure, at $z\approx2.5$ and $z\approx0.7$, corresponding to gas-rich mergers (see also the galaxy merger tree in Fig. \ref{fig:trees} below) that foster higher birth pressures, but without triggering major episodes of star formation. 
The trend for birth pressure to decline with redshift is driven in part by the metallicity-dependent density threshold for star formation (see Fig. \ref{fig:CFE}) implemented in the EAGLE simulations which also affects the birth pressure through the EOS. The threshold is motivated by the onset of the thermogravitational collapse of warm, photoionized interstellar gas into a cold, dense phase, which is expected to occur at lower densities and pressures in metal-rich gas \citep{Schaye_04}. 
We have re-run Gal004 (Appendix \ref{app:SFThresh}), adopting instead a constant density threshold for star formation of $n_\rmn{H} = 0.1$ cm$^{-3}$, and find a nearly constant median birth pressure of $10^4 \K \cmcubed$ for $z>0.5$ and $2\times10^3 \K \cmcubed$ for $z<0.5$, where the drop at low redshift is the result of high metallicity gas being more able to cool to the $8000\K$ temperature floor. This change in the star formation threshold most strongly affects the birth pressures for low metallicity stars at $z>3$ and decreases the median birth pressure by a factor of 10 at $z>6$. We discuss the main affects of this change on the cluster properties in Appendix \ref{app:SFThresh}.

The median $\Mtoomre$ shows a very weak trend with redshift, increasing from $10^8 \Msun$ at $z=6$ to $2 \times 10^8 \Msun$ at $z=0$. $\Mtoomre$ shows a slight peak at $z=2.5$, corresponding to the peak in birth pressure. However, while the pressure changes by 3 dex, the median $\Mtoomre$ changes by only 0.5 dex. This is a consequence of the star formation being very centralised and therefore strongly limited by the epicyclic frequency $\kappa$, which is also indicated by the median $f_\rmn{coll}=1$ at this time. 
The second major peak in birth pressure, at $z=0.7$, doesn't foster an increase in $\Mtoomre$, because the birth pressures (and therefore gas surface densities) are significantly less elevated than during their peak at $z=2.5$. 
The median $f_\rmn{coll}$ is less than unity for almost the entire formation history. The periods where $f_\rmn{coll}=1$ correspond to centralised star formation, within $\sim$1-2 kpc of the galactic centre. A collapse fraction close to unity also occurs when $\Mcstar$ is maximal, indicating that $\kappa$ plays an important role in governing the maximum mass with which clusters can form (see also Fig \ref{fig:MToomre}). However, clusters born close to the centres of galaxies are particularly susceptible to dynamical friction, and may rapidly merge into the galactic centre unless they are heated away from the galactic centre by mergers.

At early times, corresponding to redshifts $z>3$, $\Mgmc$, $\Gamma$ and $\Mcstar$ (middle right and bottom panels in Fig. \ref{fig:particles}) are relatively constant for metal-poor ($\ZH \lesssim -1$) stars. For redshifts $z \lesssim 2$, $\Gamma$ and $\Mgmc$ decline as a consequence of the decreasing characteristic star formation pressures, and at $z \lesssim 0.5$ the typical CFE has declined to only a few per cent. At late times, a small fraction of stars form with $\Gamma > 0.1$ at small galactocentric radii ($r<1$ kpc), owing to their high gas birth pressures. 
The evolution of $\Gamma$ and $\Mgmc$ with redshift, acting in concert, result in the truncation mass $\Mcstar$ attaining a broad maximum between redshifts 1.5 and 5 for this galaxy. This is similar to the inferred ages of MW GCs \citep[e.g.][]{Dotter_Sarajedini_and_Anderson_11}. The contrast between the redshift-dependencies of $\Mtoomre$ and $\Mcstar$ confirms the conclusion of \citet{Reina-Campos_and_Kruijssen_17} that the decrease of the maximum cloud and cluster masses with cosmic time is driven by a transition between physical regimes. At high redshift, cloud and cluster masses are mostly limited by Coriolis and centrifugal forces, whereas at low redshift, they are mostly limited by stellar feedback preventing the Toomre-limited volume to collapse into a single unit. This allows the more prevalent formation of massive stellar clusters in high-redshift environments than in low-redshift galaxies.

\subsection{Galaxy to galaxy diversity of the evolving cluster populations}
\label{sec:diversity}

\begin{figure*}
  \includegraphics[width=\textwidth]{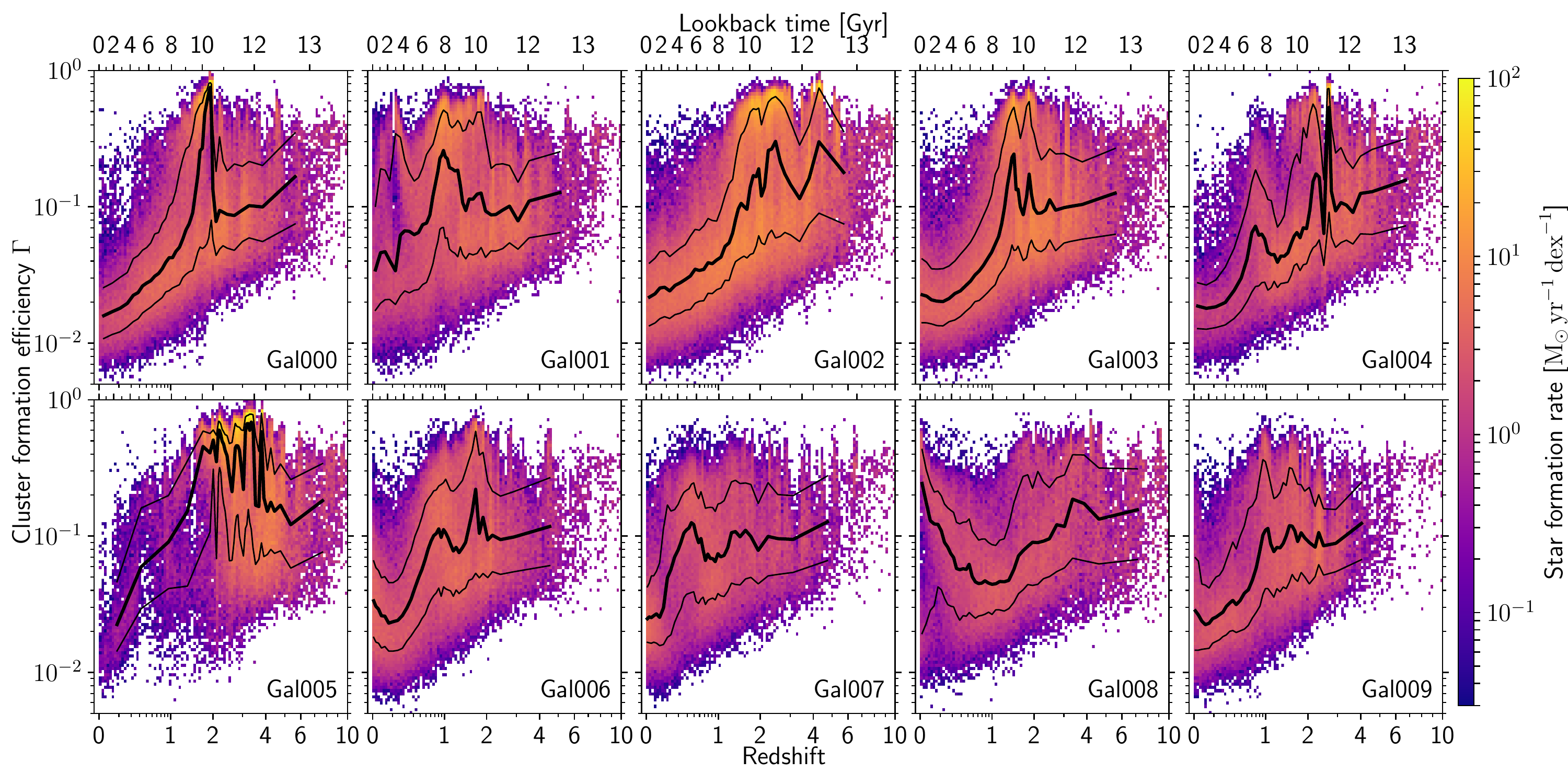}
  \caption{Two-dimensional histogram (log-scale) for the CFE ($\Gamma$) as a function of redshift for all $\numgal$ galaxies. The colour scale is identical for all galaxies and shows the star formation rate in each bin. The thick (thin) black lines show the median (standard deviation) as a function of redshift. The CFE generally peaks between $z=1$-$4$, though with significant deviation between galaxies, and at $z=0$ most galaxies have a median CFE of a few percent. The sharp drop in CFE for Gal005 after $z=2$ occurs due to quenching of star-formation in the galaxy by AGN feedback, while the increase at $z=0$ for Gal008 occurs due to very central, high pressure star formation.}
  \label{fig:all_CFE}
\end{figure*}

\begin{figure*}
  \includegraphics[width=\textwidth]{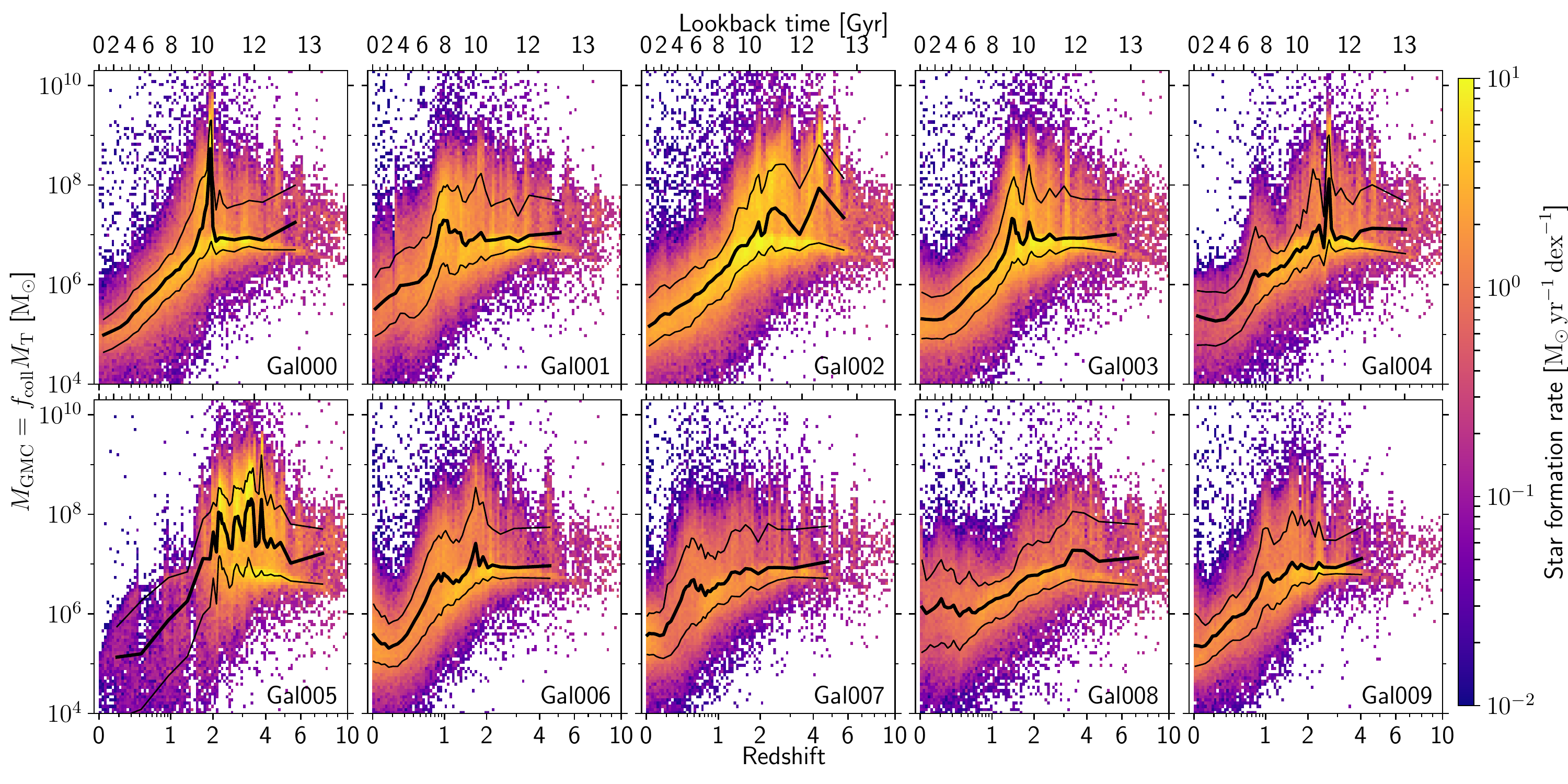}
  \caption{Two-dimensional histogram (log-scale) for the molecular cloud mass, $\Mgmc$, as a function of redshift of the $\numgal$ simulated galaxies. The colour scale is identical for all galaxies and shows the star formation rate in each bin. The thick (thin) black lines show the median (standard deviation) as a function of redshift. In the case of feedback limited GMC masses ($f_\rmn{coll}<1$), $\Mgmc$ scales with the gas pressure, while for $f_\rmn{coll}=1$, the GMC masses become limited by the epicyclic frequency $\kappa$. $\Mgmc$ peaks at the same redshifts as the CFE (Fig. \ref{fig:all_CFE}) since both variables scale with the gas pressure distribution of star formation. However in the case of very central star formation the correspondence between CFE and $\Mgmc$ may deviate as masses become limited by $\kappa$. Specifically, for Gal008 at low redshifts, the CFE increases due to increase stellar birth pressures while $\Mgmc$ remains constant since the masses are $\kappa$-limited.}
  \label{fig:Mgmc}
\end{figure*}

\begin{figure*}
  \includegraphics[width=\textwidth]{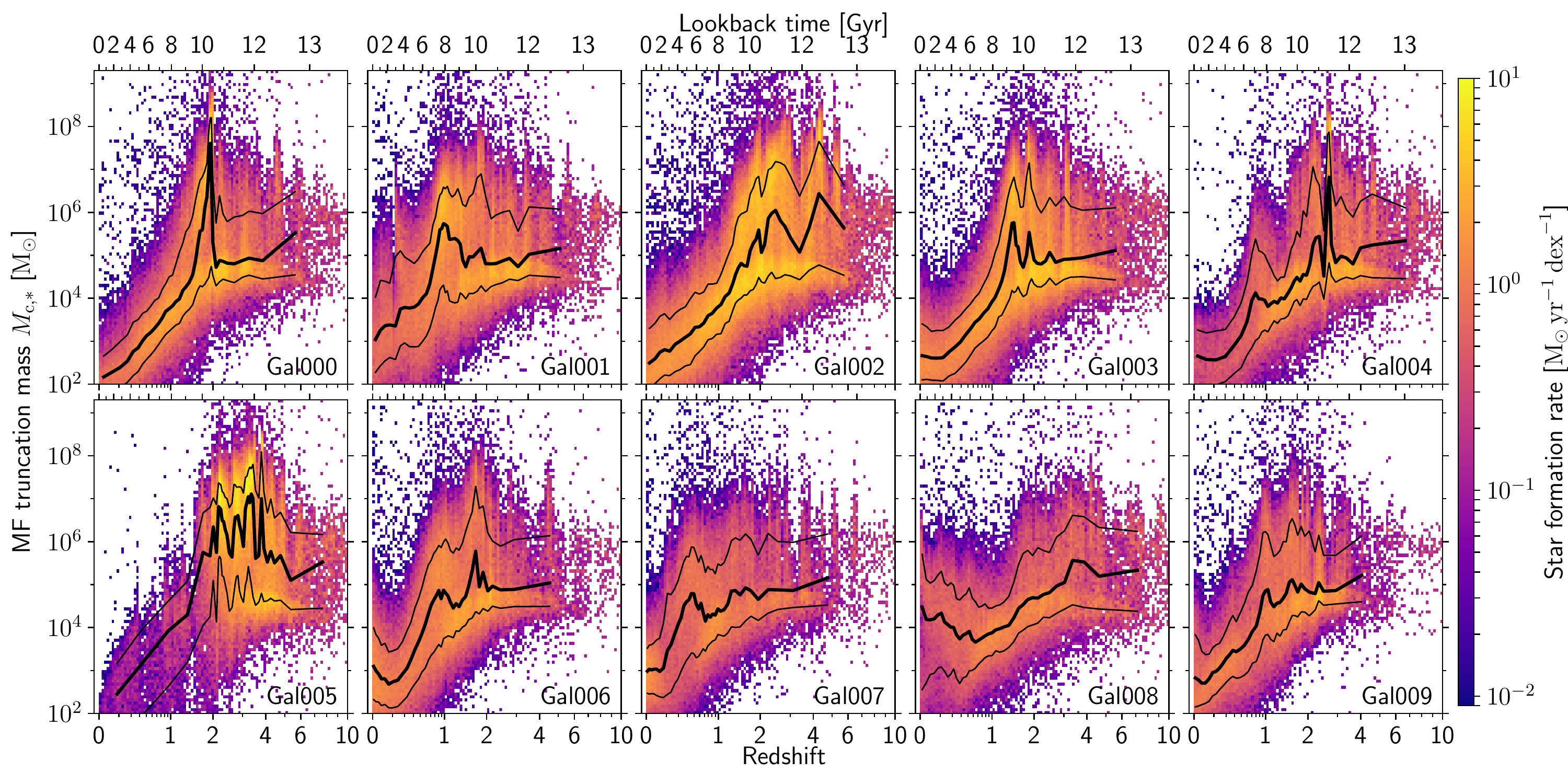}
  \caption{ Two-dimensional histogram (log-scale) for the ICMF truncation mass, $\Mcstar$, as a function of redshift for all $\numgal$ galaxies. The colour scale is identical for all galaxies and shows the star formation rate in each bin. The thick (thin) black lines show the median (standard deviation) as a function of redshift. Since $\Mcstar$ is a linear combination of CFE and $\Mgmc$, $\Mcstar$ shows very similar redshift evolution to $\Mgmc$ in Fig. \ref{fig:Mgmc}, peaking at redshifts $z=1$-$4$. $\Mcstar$ typically peaks at $\sim10^8 \Msun$ for the bulk of the population, with some limited scatter to higher values due to very high pressure particles.}
  \label{fig:Mcstar}
\end{figure*}

\begin{figure*}
  \includegraphics[width=\textwidth]{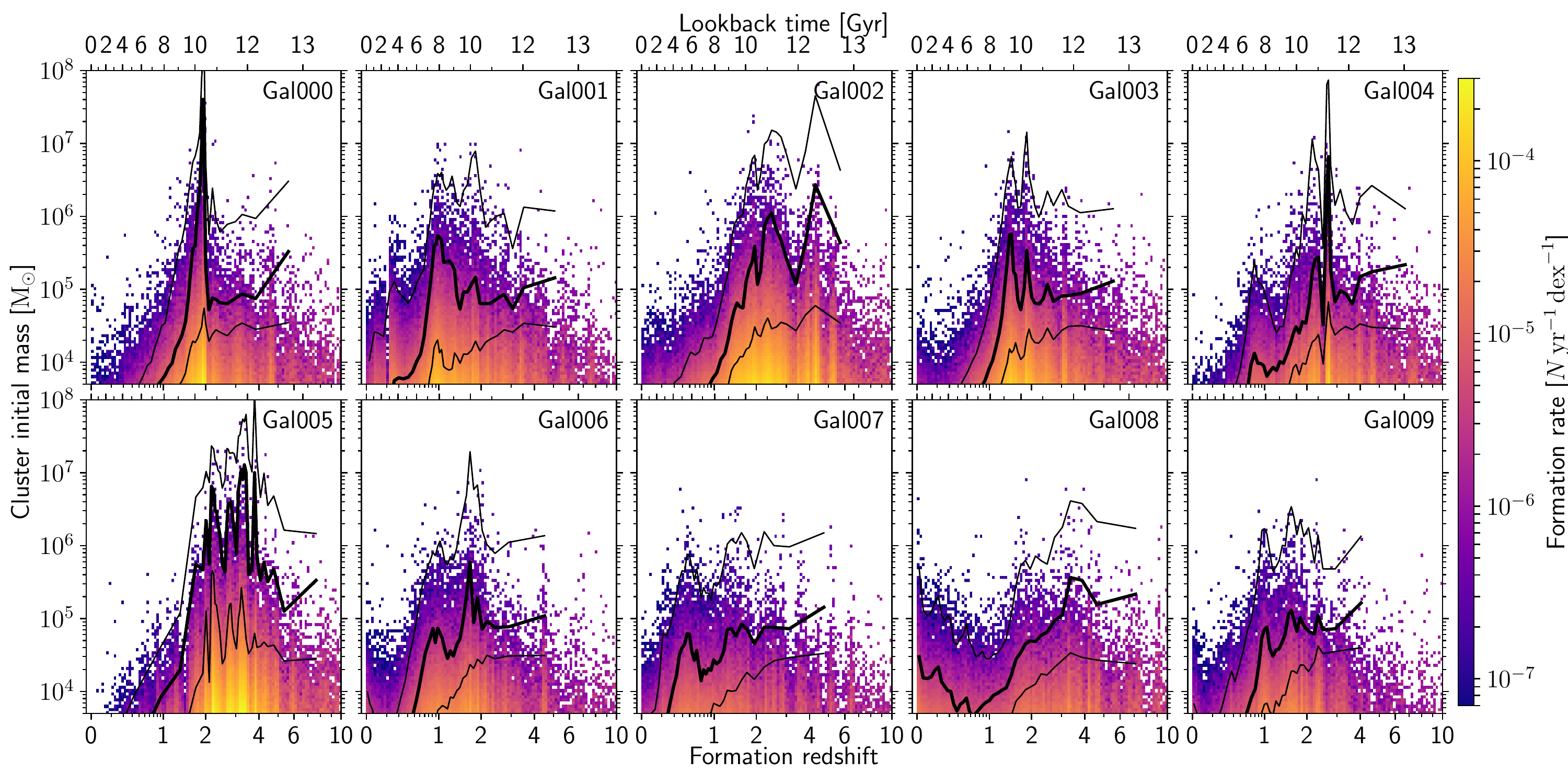}
  \caption{Cluster initial masses with median and standard deviations of $\Mcstar$ overplotted. The colour scale is identical for all galaxies and shows the number of star clusters per two-dimensional bin. Note the minimum cluster mass limit at $5 \times 10^3 \Msun$. At $z \lesssim 4$, cluster masses are strongly limited by $\Mcstar$, while at early times few clusters are formed. The most massive cluster formed for each galaxy typically has a mass $\sim10^7 \Msun$, however these clusters will be particularly susceptible to disruption by dynamical friction-driven inspiral to the galactic centre.}

  \label{fig:Mcinit}
\end{figure*}

To illustrate the degree of variation as a function of the assembly and environment history of the galaxy sample, Figs. \ref{fig:all_CFE} to \ref{fig:Mcinit} show the cluster formation properties for all $\numgal$ of our simulated galaxies. The CFE (Fig. \ref{fig:all_CFE}) tends to peak in the redshift interval $1 \lesssim z \lesssim 3$ for most galaxies. This epoch $1 \lesssim z \lesssim 3$ broadly coincides with the peak of star formation (Fig. \ref{fig:SFR}). However, as noted in the specific case of Gal004 in Fig. \ref{fig:particles}, the CFE does not follow directly from the SFR, but from the gas pressure. In the case of Gal008 the CFE peaks at $z=0$ and $z=4$, while the SFR has remained almost constant over this redshift range. The $z=0$ peak is caused by the majority of star formation taking place in a high-pressure disc within 3 kpc of the galactic centre (Fig. \ref{fig:radial_props}). 
Gal005, the quenched galaxy at $z=0$ (Fig. \ref{fig:SFR}), shows an increasing median CFE from 20 per cent at $z=6$ up to 70 per cent at $z=2$, at which point the CFE declines rapidly to 2 per cent at $z=0$. This decline in CFE at $z=2$ also coincides with the rapid drop in SFR of the galaxy at the same epoch due to quenching by AGN feedback.

The molecular cloud mass and ICMF truncation mass (Figs. \ref{fig:Mgmc} and \ref{fig:Mcstar}) also peak at a similar epoch to the CFE. Since both CFE and $\Mgmc$ scale with pressure, this is not unexpected. 
These figures also highlight the diversity of cluster formation in galaxies of the same mass range. 
In general, the galaxies peak in their cluster formation properties between redshifts 1 and 4, though the exact epoch of differs between galaxies. In particular, some galaxies peak in cluster formation early in their formation history (Gal008 at $z=3.5$), some later (Gal001 at $z=1$) and some have broad peaks over a long timescale (Gal005 from $z=2$-$4$, Gal007 from $z=0.5$-$4$). We will explore in Section~\ref{sec:assembly} to what extent this diversity depends on the galaxy assembly history.

The combination of CFE and $\Mgmc$ places important limits on when massive star clusters may form in MW-like galaxies.
Fig. \ref{fig:Mcinit} shows the initial masses of clusters as a function of redshift. The majority of the galaxies form their most massive clusters prior to $z\sim1$, and at $z \sim 0$ very few clusters are born with masses $> 10^5 \Msun$. Consequently, such galaxies typically host only an old population of massive clusters.
The figure also shows the mean and standard deviation of $\Mcstar$, enabling comparison with the initial cluster masses.
At $z \lesssim 4$, $\Mcstar$ is the key factor governing the upper envelope of the cluster mass distribution. By contrast, at $z \gtrsim 4$ fewer clusters are born, even though $\Mcstar$ can remain high, implying that the upper envelope of the mass distribution in Fig.~\ref{fig:Mcinit} is shaped by small-number stochastic sampling. Therefore, $\Mcstar$ plays a smaller role in governing cluster masses at early times, and it is clear that the redshift evolution of the maximum cluster mass does not simply follow from the SFR.

To summarise the above findings, the YMC-based cluster formation model predicts that, on average, the massive clusters in MW-like galaxies that survive to the present day (i.e.~GCs) should be predominantly old, with mean formation redshifts of $z\sim2$ (Reina-Campos et al.~in prep.). 
This follows from the evolution of the star formation birth pressures with redshift, which typically declines to low pressures in extended star-forming discs at $z=0$. The model also predicts that few MW-like galaxies should be forming massive clusters at $z=0$, because the CFE and ICMF truncation mass ($\Mcstar$) are much lower than required for the formation of such clusters. These predictions are in good agreement with observations of star clusters in they MW and M31 \citep[e.g.][]{Dotter_Sarajedini_and_Anderson_11, Caldwell_et_al_11}.

\subsection{Cluster formation throughout galaxy assembly}
\label{sec:assembly}

\begin{figure*}
  \includegraphics[width=0.48\textwidth]{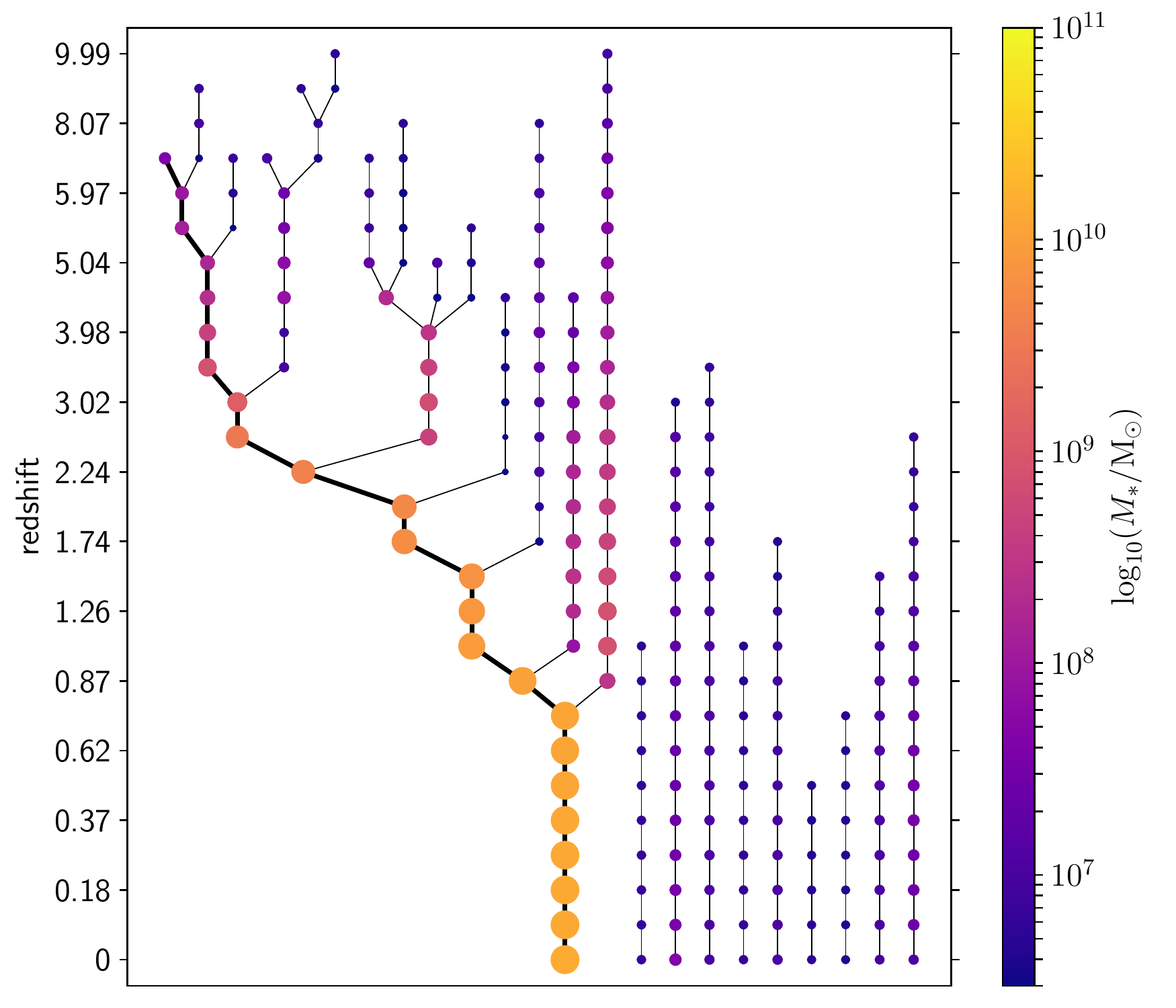}
  \includegraphics[width=0.48\textwidth]{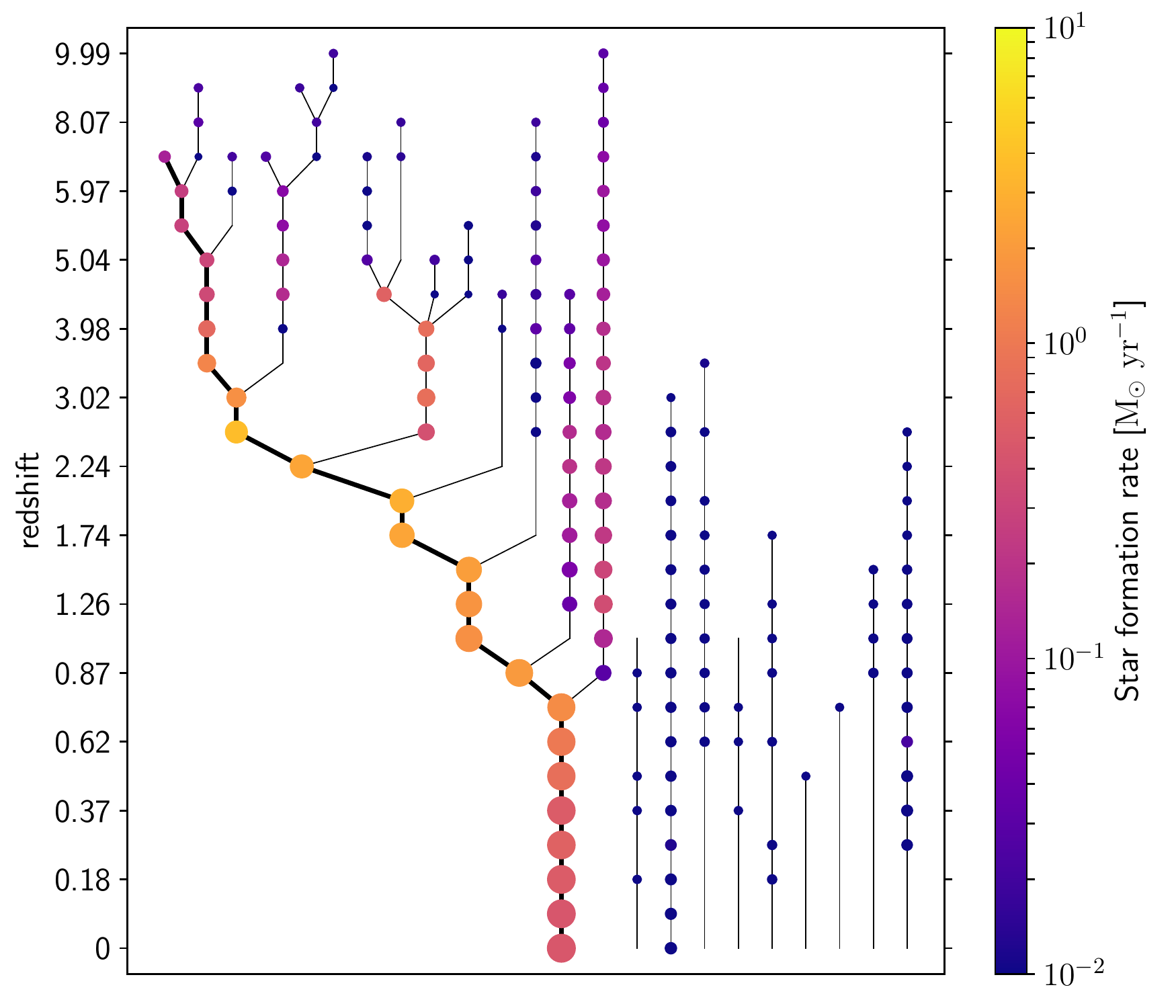}
  \includegraphics[width=0.48\textwidth]{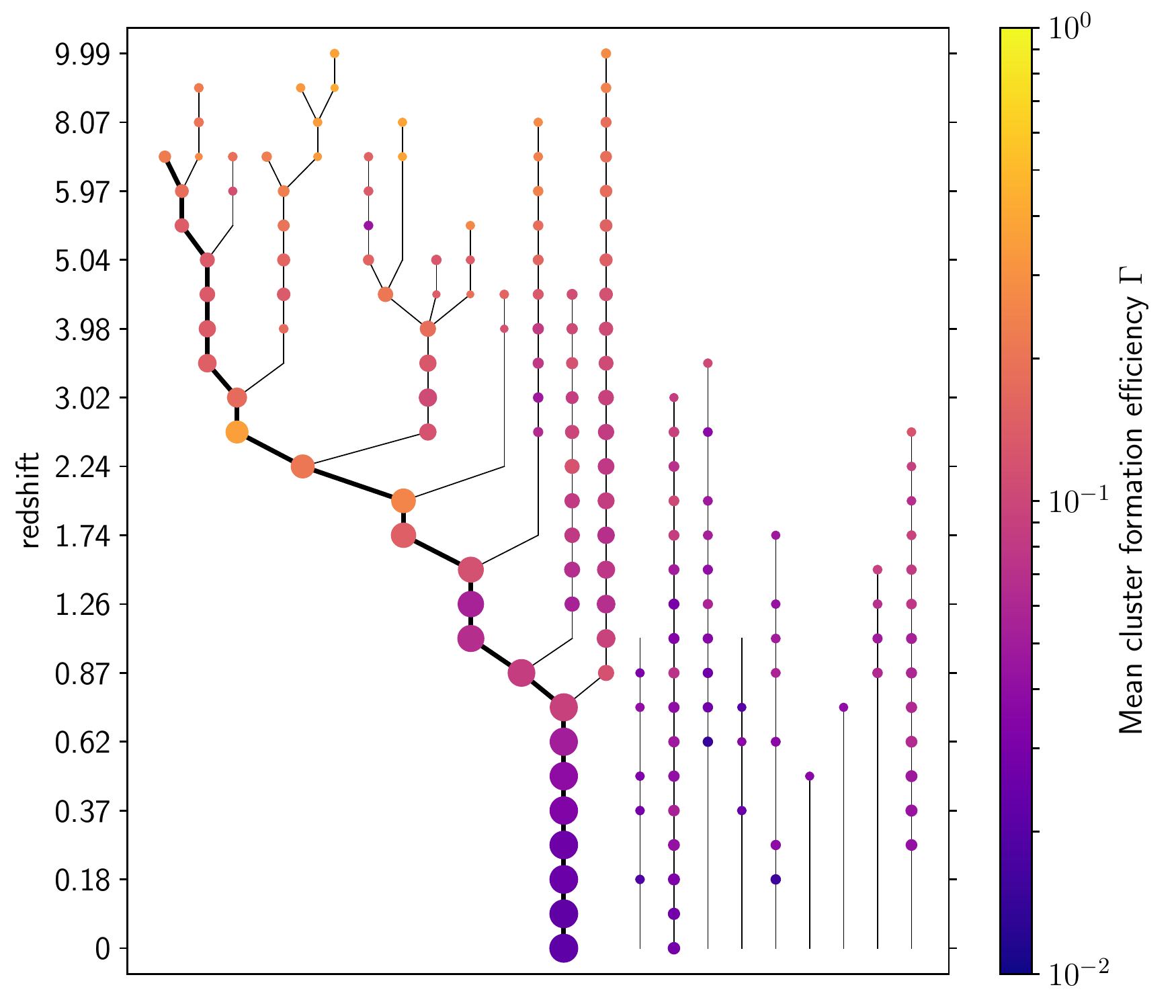}
  \includegraphics[width=0.48\textwidth]{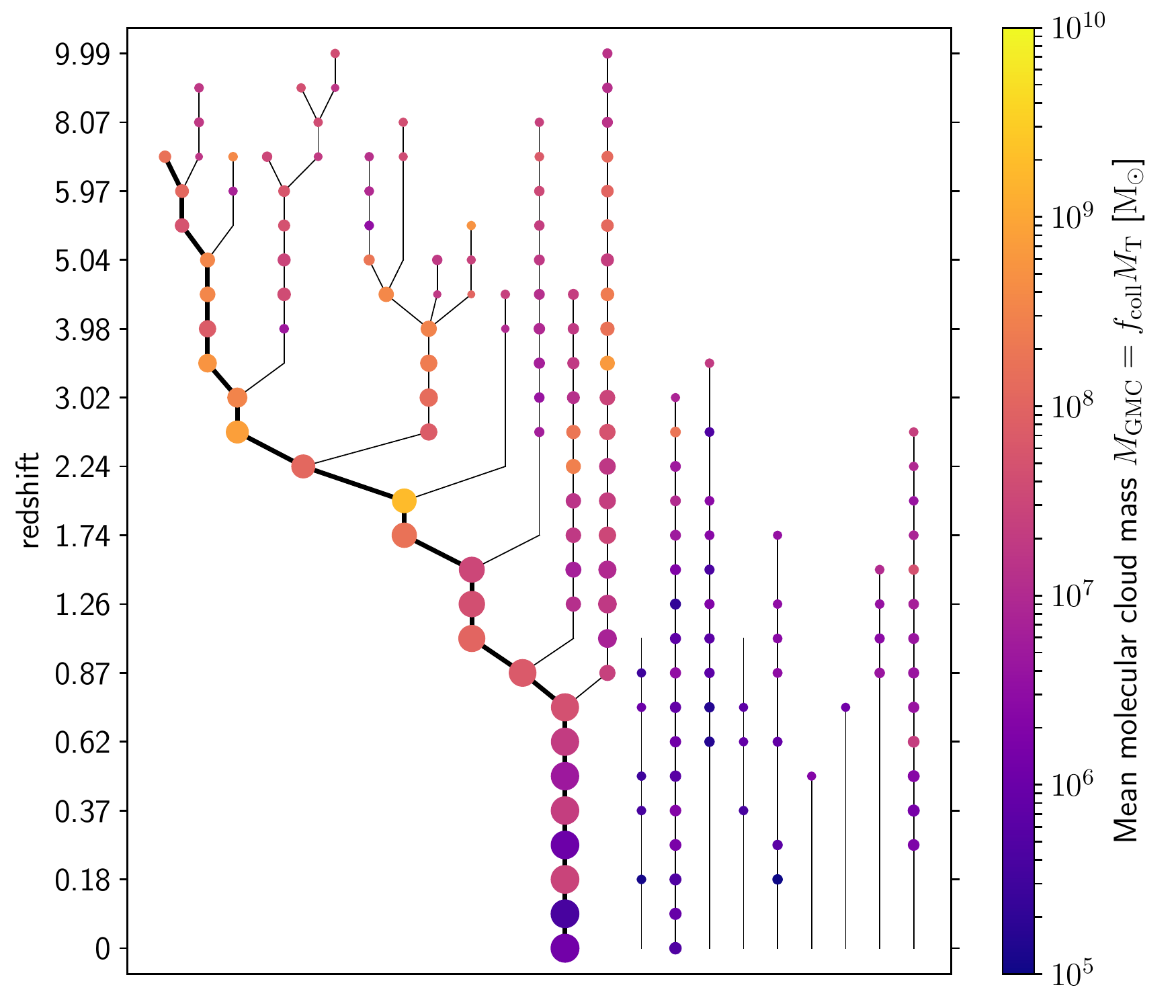}
  \includegraphics[width=0.48\textwidth]{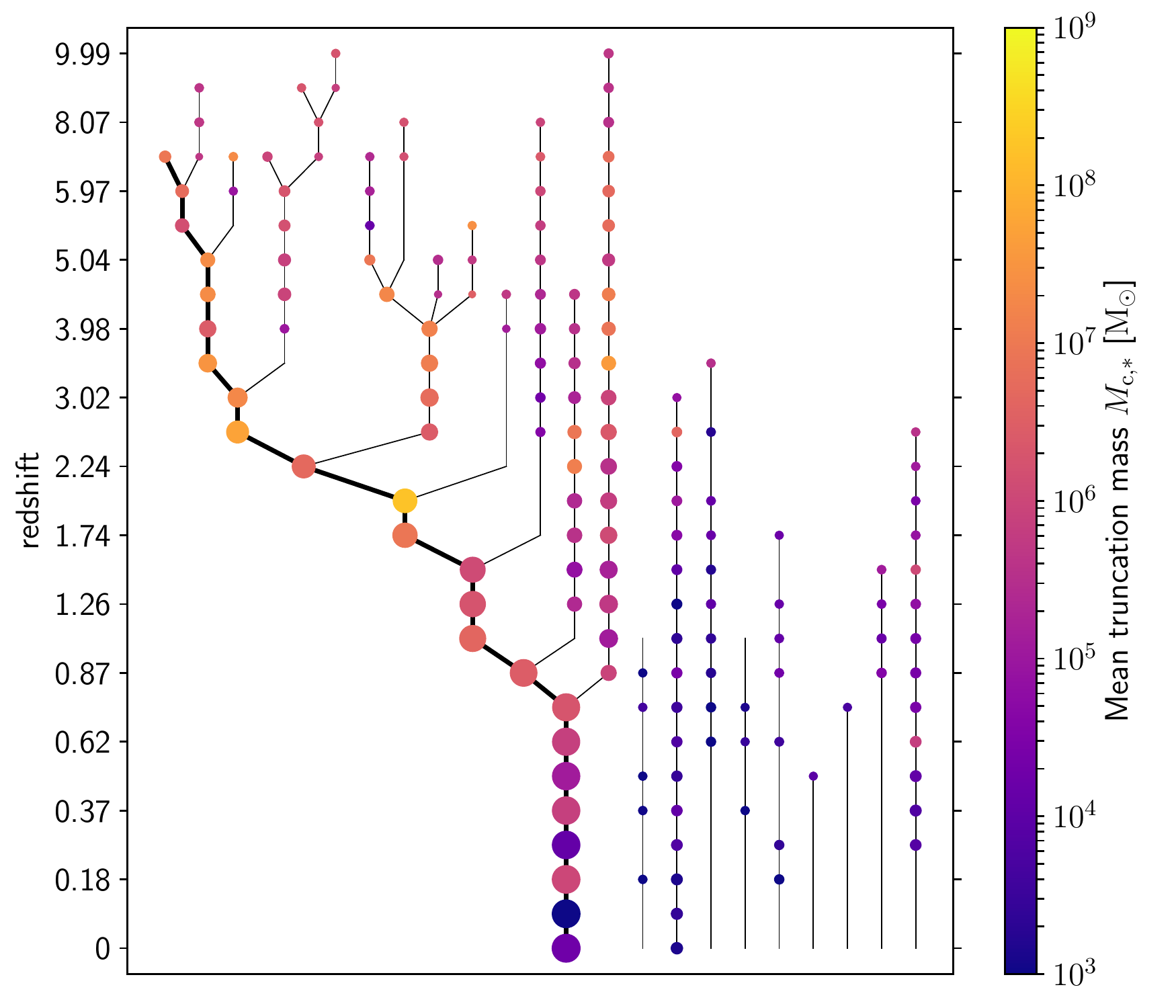}
  \includegraphics[width=0.48\textwidth]{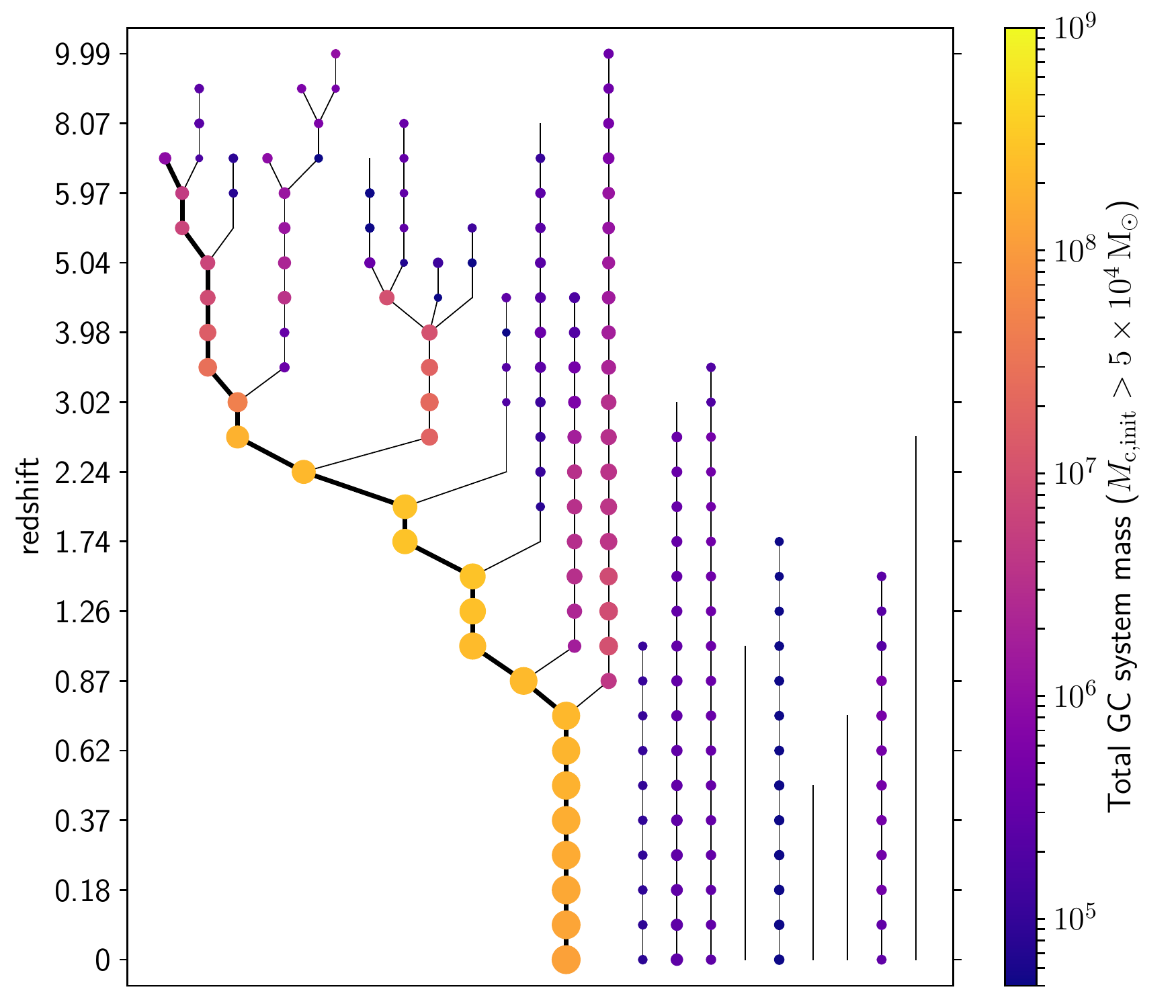}
  \caption{Galaxy merger tree for Gal004 and its satellite population coloured by (from left to right, top to bottom) galaxy stellar mass, star formation rate, cluster formation efficiency, molecular cloud mass, ICMF truncation mass and total GC system mass (including surviving clusters with initial masses $M_\rmn{c,init}>10^5 \Msun$ and metallicities $\ZH>-3$). The cluster formation properties (cluster formation efficiency, molecular cloud mass, ICMF truncation mass) show the mean value computed for stars younger than 300 Myr at each epoch along the tree. Only galaxies with stellar mass $M_\ast > 20 m_{\rm g}$ (where $m_{\rm g}$ is the initial gas particle mass) are shown and the point area for each galaxy is proportional to galaxy stellar mass. The thick line shows the main branch of the central galaxy. Note that not all galaxies in the tree have points for all snapshots due to lack of star formation at that epoch.}
  \label{fig:trees}
\end{figure*}

To further understand the environmental dependence of cluster formation, we now investigate how cluster formation properties vary throughout the galaxy assembly process.
Fig. \ref{fig:trees} shows the galaxy merger tree of Gal004 and its satellite galaxies (i.e.~associated with the same FoF group) coloured by galaxy properties ($M_\ast$, SFR, top row) and cluster population properties (CFE, $\Mgmc$, $\Mcstar$, GC system mass, middle and bottom rows). The SFR and cluster formation properties are calculated for stars younger than 300 Myr in each progenitor and at each epoch along the tree. This timescale is sufficiently long that the properties for the satellite galaxies are not significantly affected by poor particle sampling. The satellite galaxies remaining at $z=0$ reside at radii between 120 and 320 kpc, with respect to the central galaxy, and have stellar masses between $10^6$ and $10^8 \Msun$, typical of dwarf spheroidal galaxies \citep[e.g.][]{McConnachie_12}. 

The top left panel shows the galaxy merger tree coloured by galaxy stellar mass. The merger rate is highest at high redshifts ($z>4$). The main galaxy undergoes major mergers at $z=3$ and $z=2.25$, and accretes two $M_\ast = 10^9 \Msun$ galaxies at $z\approx0.8$. The satellite galaxies do not undergo any mergers with structures comprising 20 or more particles.
The top right panel shows the galaxy merger tree coloured by SFR. The absence of a point for galaxies in the figure  indicates an absence of star formation at this epoch. The SFR peaks at $z=2.5$ for Gal004 (Fig. \ref{fig:SFR}) during a gas-rich galaxy merger (see the middle lower panel of Fig. \ref{fig:visualisation}), though the SFR remains high ($>2 \Msun \pyr$) for this galaxy between redshifts 3.5 and 1.
Only one of the satellites (the second in the figure) is still star-forming at $z=0$ (within 300 Myr).

The CFE and $\Mgmc$ (and therefore $\Mcstar$) reach their highest values roughly co-temporally with the peak of star formation in the main galaxy branch (at $z=2.5$ and $z=2$, respectively), and at high redshift ($z \gtrsim 4$) for the progenitor galaxies that merge onto the main branch. Along the main galaxy branch, the CFE decreases as the galaxy's stellar mass grows, until an episode of very centralised star formation takes place at $z \simeq 2.5$ when the galaxy mass is $\simeq 3 \times 10^9 \Msun$ and the CFE peaks. The CFE then continues the declining trend until $z=0$, punctuated by a brief period of elevation in response to (merger-induced) elevated star formation pressures at $z \approx 0.8$. The trend is similar for $\Mgmc$ (middle right panel) and $\Mcstar$ (bottom left panel) since both also correlate with gas pressure. 
Though of similar stellar mass, galaxies that will become satellites of the central galaxy at $z=0$ show significantly different cluster formation properties than those that merge with the central galaxy.
For a fixed galaxy stellar mass, CFE and $\Mgmc$ are higher at earlier times. This is due to a combination of declining gas accretion rates (resulting in lower peak pressures) towards later times, and a tendency for star formation at late times to occur at larger galactocentric radii, where the pressure is markedly lower than in galactic centres (Fig.~\ref{fig:radial_props}; see also Fig.~8 of \citealt{C15}), resulting in low CFEs \citep[see Fig.~\ref{fig:CFE} and][]{Kruijssen_12} and causing $\Mcstar$ to become feedback-limited \citep{Reina-Campos_and_Kruijssen_17}.

\begin{figure*}
  \includegraphics[width=\textwidth]{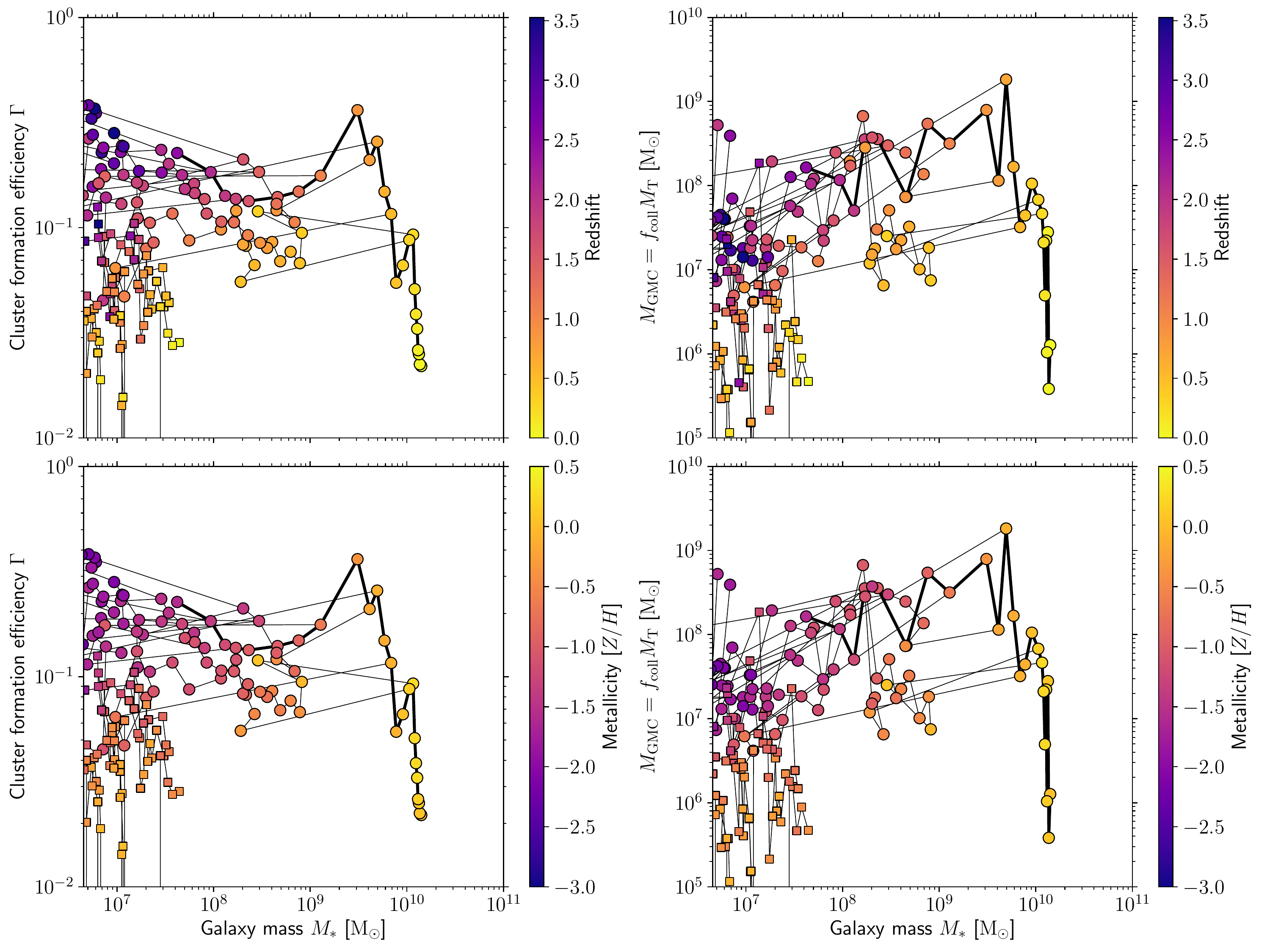}
  \caption{Cluster formation properties as a function of galaxy mass for all progenitors in the Gal004 merger tree (Fig. \ref{fig:trees}) with $\rmn{SFR}>0$, coloured by redshift (top panels) and metallicity (bottom panels). The filled circles show the merger tree of the central galaxy, while filled squares show the satellite galaxies. Lines connect galaxies as in the merger tree, with the thick line showing the main branch of the central galaxy. For a given galaxy mass, galaxies formed at earlier times and with lower metallicities form clusters more efficiently and from higher-mass GMCs than those formed later.}
  \label{fig:CFE-galMass}
\end{figure*}

We show this more directly in Fig. \ref{fig:CFE-galMass}, where we compare the CFE and $\Mgmc$ as a function of galaxy mass, with galaxies connected as per the merger tree. Galaxies in the merger tree of the central galaxy are shown as large filled circles, while satellite galaxies at $z=0$ are shown as small filled squares.
At a fixed galaxy stellar mass, the CFE is highest for early formation times and low metallicities. 
During the assembly of the main galaxy, the mean CFE remains relatively constant between 10 and 20 per cent until the galaxy reaches a mass $>6\times10^9 \Msun$ (about half its final stellar mass) at $z\approx1.5$. The CFE reaches a peak of 30 per cent at a galaxy mass of $3\times 10^9 \Msun$ due to very central, high-pressure star formation. From this time onwards the CFE drops to a few per cent at $z=0$, with a brief increase to $\Gamma=10$ per cent during the accretion of two gas-rich dwarf galaxies at $z\approx0.8$. 
For the present day satellite galaxies, the CFE remains less than 10 per cent over nearly the entire formation history of the galaxies.

$\Mgmc$ shows a similar trend to the CFE, being highest at early formation times and low metallicities for a given galaxy mass. However during the assembly of the main galaxy $\Mgmc$ shows a steady increase from $\sim10^8 \Msun$ at early times to $\sim10^9 \Msun$ at $z\approx2$ near the peak of star formation (see also Fig. \ref{fig:particles}). After this point $\Mgmc$ significantly drops to a mean of $\sim10^6 \Msun$ in the central galaxy at $z=0$. While the satellites that end up being accreted by the central galaxy commonly reach $\Mgmc\sim10^8 \Msun$, the satellite galaxies at $z=0$ have $\Mgmc < 10^7 \Msun$ over nearly their entire formation histories.

These results imply that galaxies with earlier and more rapid formation have more abundant star cluster populations that extend to higher cluster masses than those with late and more extended formation histories. This is caused by the differing birth pressure distributions of star formation in these cases, with star formation occurring at higher gas pressures in the early Universe than at low redshift \citep[see also][]{Mistani_et_al_16}. 

Returning briefly to Fig. \ref{fig:trees}, the bottom right panel shows the instantaneous (i.e.~including mass-loss) total mass of massive clusters for the galaxies in the merger tree (birth masses larger than $10^5 \Msun$). 
Galaxies in the merger tree may lose their cluster population through cluster disruption, dynamical friction (which is applied at each snapshot) or the stripping of clusters during the merging process.
The GC population of the main galaxy branch is largely in place by $z=2$ (i.e.~just after the peak of cluster formation, Fig. \ref{fig:Mcinit}). Comparison with the top left panel of Fig.~\ref{fig:trees} shows that most of the GC system forms before the stellar mass of the galaxy.
From redshift 3 to 0.75, a number of galaxies merge into the main galaxy which also contribute their GC systems. As visualised in the bottom panels of Fig. \ref{fig:visualisation} (particularly the $z=1.5$ panel), these mergers result in clusters being redistributed from the star-forming disc of the galaxy into the `halo' of the galaxy, or otherwise into orbits no longer coinciding with that of the dense star-forming gas in the galaxies which may disrupt clusters by tidal shocks (see below). 
Such a process is thought to be necessary for the survival of clusters to the present time \citep{Kruijssen_15}.
For this particular galaxy, accreted galaxies contribute 11 per cent of the mass of the cluster population and 13 per cent of stellar mass.
At $z=0$, nearly half of the satellite galaxies have at least one massive cluster. All satellite galaxies formed their clusters very early in their formation history, when $\Mcstar$ and the CFE were highest. Two other satellite galaxies (the 7th and 9th in the figure) also formed massive clusters but these were removed by the dynamical friction calculation at the first snapshot for which the galaxies appear in the figure. 

In summary, we see that there exists a close link between cluster formation and galaxy assembly history. This connection opens up the potential of tracing galaxy formation and assembly histories using the observed GC population. We plan to address this aspect in more detail in a companion paper \citep{Kruijssen_et_al_18}.


\section{Tidal histories} \label{sec:tidal_histories}

\begin{figure*}
  \includegraphics[width=0.47\textwidth]{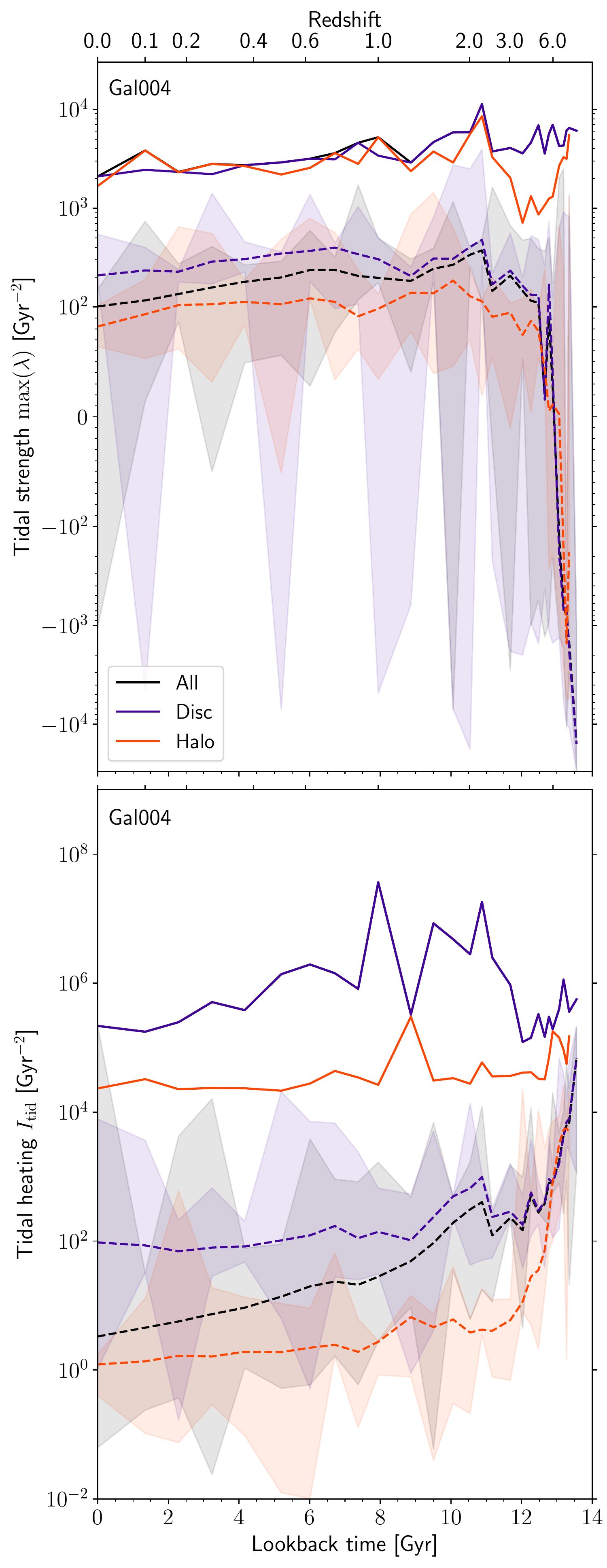}
  \includegraphics[width=0.47\textwidth]{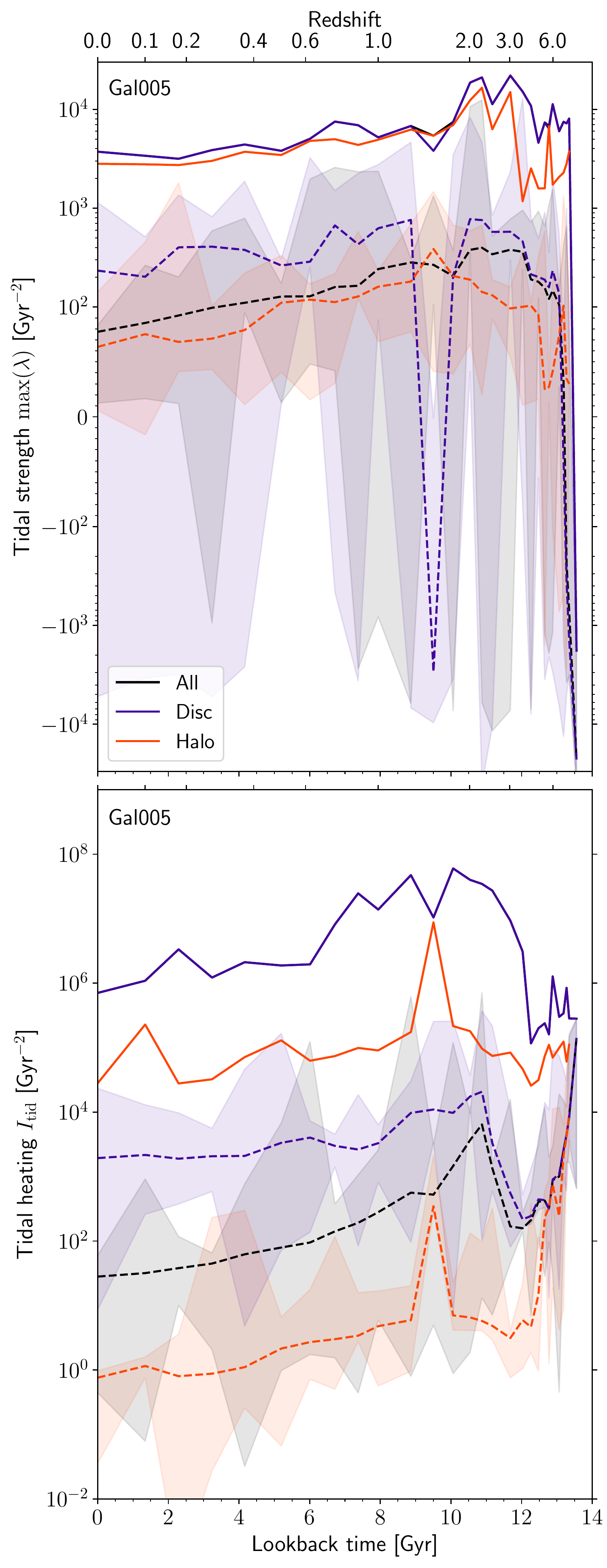}
  \caption{Tidal field strengths (upper panels, reflecting disruption by two-body relaxation) and tidal heating parameters (lower panels, reflecting disruption by tidal shocks) at each snapshot as a function of time for stellar particles hosting star clusters in Gal004 (left panels) and Gal005 (right panels). For the upper panels, scales are linear between $10^2$ and $-10^2 \Gyr^{-2}$ and logarithmic otherwise. Solid lines show the maximum values, dashed lines show the median and shaded areas show the $1\sigma$ scatter about the median. Stellar particles are divided into `disc' (star-forming regions) and `halo' (non star-forming regions) by the current gas smoothing length at $h_\rmn{sml} < 1 \kpc$ and $h_\rmn{sml} > 1 \kpc$, respectively, based on the maximum smoothing length of star-forming gas particles. Disk particles show significantly higher tidal field strengths (for median values) and tidal heating parameters (both median and maximum) than halo particles, showing that cluster disruption is significantly more effective in regions of high gas density.}
  \label{fig:tid_stren}
\end{figure*}

Before discussing the properties of the present-day cluster populations (i.e.~including disruption, see Section \ref{sec:cluster_props} below), we first investigate the efficiency of cluster disruption (via the tidal field strength and tidal heating parameter) as a function of environment and cosmic time.
As we will show directly in Section \ref{sec:cluster_props}, mass-loss by tidal shocks is a strong function of the ambient gas density that a cluster experiences after formation. Therefore the evolution of the gas density and pressure with redshift has strong implications not only for cluster formation (previous section), but also for cluster mass-loss.

In Fig. \ref{fig:tid_stren} we show at each snapshot the maximum (solid lines), median (dotted lines) and $1\sigma$ scatter about the median (shaded regions) of the tidal strength, $\max (\lambda)$ (upper panels), and tidal heating parameter $I_\rmn{tid}$ (lower panels), for stellar particles in Gal004 (left panels) and Gal005 (right panels). Particles are chosen to be within $100 \kpc$ of the galaxy at $z=0$.
We show $T = \max (\lambda)$, rather than $T = \max (\lambda) + \Omega^2$ as we define in Section \ref{sec:clevo}, to be consistent with previous studies.
Very negative values of $\max (\lambda)$ imply very central particles (see Fig. \ref{fig:tidal_field}), whilst the maximum value occurs at $\approx2 \kpc$ at $z=0$ for Gal004. The radius of the maximum depends on the mass distribution in the galaxy and will therefore occur at somewhat different radii in different galaxies.
The tidal heating parameter, $I_\rmn{tid}$ (Eq. \ref{eq:Itid}), is the integral of the tidal field strength throughout the duration of a shock. As a result, the saved tidal heating parameters in a given snapshot reflect the integrated tidal heating during a shock so far, at some intermediate stage during the shock. As the value for $I_\rmn{tid}$ increases over the course of a tidal shock, particles are statistically unlikely to be at peak $I_\rmn{tid}$ at the moment a snapshot was saved. However, a change of the median or maximum tidal heating parameter across the population still tracks relevant macroscopic changes in the tidal field properties. In this figure, we also neglect the adiabatic correction, $A_{\rm w}$, which depends on the individual cluster properties and is only applied at the completion of a tidal shock.
Particles are divided into `disc' and `halo' particles by the current gas smoothing length of the stellar particles\footnote{The SPH kernel is used for distributing stellar mass-loss by star particles to the neighbouring SPH particles. Since there is a nearly monotonic relation between $h_\rmn{sml}$ and $n_\rmn{H}$ this division could be made in either quantity. However, the current value, is only saved in the snapshots for $h_\rmn{sml}$. For $h_\rmn{sml} = 1 \kpc$, $n_\rmn{H} \approx 4 \times 10^{-3} \cmcubed$.} at $h_\rmn{sml} = 1 \kpc$, based on the maximum smoothing length of star-forming gas particles. Star particles in environments with lower smoothing lengths therefore reside in star-forming regions, whereas star particles in environments with higher smoothing lengths reside in non-star-forming regions. Therefore this division in smoothing length separates particles (approximately) into disc and halo particles.
As the tidal field strength and tidal heating parameter are only calculated in the simulation for stellar particles that contain star clusters, the particles represented in this figure have been selected self-consistently by the physics of both cluster formation and disruption included in the simulations. Therefore, it is important to consider both effects when interpreting the figure.

At lookback times $>13 \Gyr$, the median tidal field strengths (upper panels) for both galaxies are strongly negative, indicative of very central star formation in the first galaxies to form at this epoch. As star formation becomes more spatially extended in the galaxies, the median tidal field strengths increase to positive values at lookback times $<13 \Gyr$.
The peak of both the median and maximum tidal field strength occurs near $z=2$ for both galaxies. However, from this time onwards, the evolution of the median and maximum tidal field strengths is mild -- they decrease by only 0.5~dex to their values at $z=0$. Therefore, after the initial rapid evolution of the galaxy, the \textit{typical} tidal field strength experienced by clusters is relatively constant over the lifetime of the galaxy, though at any epoch clusters in a galaxy will experience a wide range of environments (as shown by the $1\sigma$ regions).

The peak for the tidal field strengths occurs at a similar epoch to the peak of star formation for the galaxies (Fig. \ref{fig:SFR}) for which star formation occurs at small galactocentric radii (as indicated by high birth pressures, Fig. \ref{fig:particles}). At lookback times $>9 \Gyr$, the median tidal field strength for all particles traces that of the disc particles, indicating most clusters exist at that time in star-forming regions.
For lookback times of $<9$ Gyr, the median tidal strength tends more towards that of the halo cluster population as, in general, only the clusters that migrate out of star-forming regions avoid disruption and survive until $z=0$. There is a large scatter about the median tidal field strength, reflecting the diversity of environments that clusters experience in any given galaxy. The excursions of the $1\sigma$ regions to very negative values, particularly for the disc particles, correspond to episodes of very central star formation within the galaxy (as shown in Fig. \ref{fig:tidal_field} very negative values for $\max(\lambda)$ only occur at the galactic centre).

The tidal field strength governs the cluster mass-loss rate by evaporation. However, as we will show in Section \ref{sec:GCMF} (see Table \ref{tab:P-disruption}), evaporation contributes significantly less to cluster mass-loss than tidal shock heating. The diagnostic relevant for assessing the impact of tidal shocks is the tidal heating parameter, $I_\rmn{tid}$, of which we show the evolution in the lower panels of Fig. \ref{fig:tid_stren}. As with the tidal field strength, the median $I_\rmn{tid}$ of all particles broadly traces that of young clusters in the disc at lookback times $>9 \Gyr$, whereas it tends towards evolved clusters in the halo at later times. The maximum $I_\rmn{tid}$ experienced by halo clusters is typically 1-2 orders of magnitude lower than by disc clusters, with a maximum of nearly 4 dex difference for Gal005 at a lookback time of $\approx11 \Gyr$. This highlights that disruption by tidal shocks is much more effective in regions of high gas density than in more diffuse environments (we quantify this further in Section \ref{sec:cluster_props}), and indicates that the migration of clusters away from high density regions is a necessity for long term cluster survival \citep{Kruijssen_15}.

The median tidal heating first peaks at lookback times of $>13 \Gyr$. This is due to the high absolute values of the corresponding tidal field strengths at these epochs, which is caused by very central star formation (recall the very negative tidal field strengths).
The median $I_\rmn{tid}$ peaks a second time near the peak SFR, when gas densities are highest. The maximum $I_{\rm tid}$ follows suit and experiences its first (and typically only) major peak at the same lookback time, showing that the strongest tidal shocks occur when most of the stars are being formed. Interestingly, the median $I_\rmn{tid}$ peaks later for halo clusters than for disc clusters (by $2 \Gyr$ in Gal004, $1.5 \Gyr$ in Gal005), which we interpret as the effect of surviving clusters migrating from dense star-forming regions into more diffuse environments (since clusters migrating from high density regions will initially have a higher $I_\rmn{tid}$ than the median for halo particles).

\begin{figure*}
  \includegraphics[width=\textwidth]{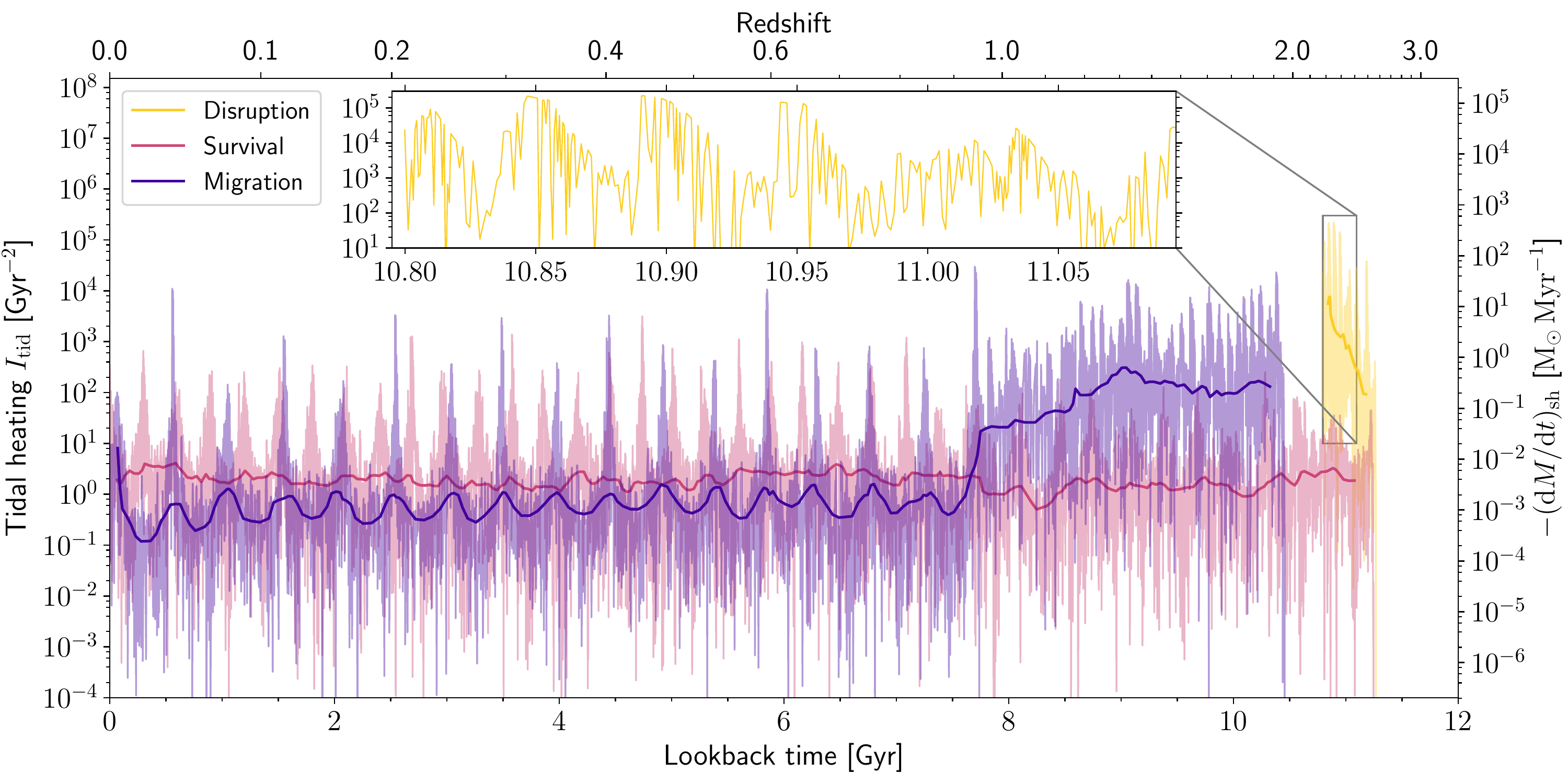}
  \caption{ Temporal evolution of the tidal heating parameter at the particle timestep level for three example cases in Gal004. The three examples were chosen to have formed star clusters with masses of $2\times10^4 \Msun$ and to show typical evolutions of each case. The solid lines shows the running median for each particle. The inset highlights a $300 \Myr$ period of the `disruption' example to show the individual timesteps for the particle. For reference, the right-hand axis shows the (approximate) mass-loss rate corresponding to $I_\rmn{tid}$ from Eq. \ref{eq:dmsh}, assuming $r_{\rm h} = 4 \pc$ and $\Delta t = 10 \Myr$. In the `disruption' example, the cluster is formed in a region of high tidal heating and is disrupted over a short timescale ($0.5 \Gyr$). In the `survival' example, the cluster is formed in a region with low tidal heating and survives until $z=0$, with a final mass of $3000 \Msun$. In the `migration' example, the cluster forms in a region of high tidal heating and migrates to a region with low tidal heating through a minor galaxy merger at $z=0.9$, enabling the cluster to survive to $z=0$ with a final mass of $3200 \Msun$. The regular modulations at $<7.5 \Gyr$ are due to the non-circular orbit of the particle.}
  \label{fig:heating_ts}
\end{figure*}

To investigate particle tidal histories in more detail, in Fig. \ref{fig:heating_ts} we show the tidal heating parameter at the timestep level for three example cases in Gal004. For this purpose, we re-ran the simulation of Gal004 in order to output the tidal heating parameter at all timesteps for all star particles currently hosting star clusters. By inspecting the tidal histories of a large number of particles (each of which formed a $2\times 10^4 \Msun$ cluster) we classified the evolutions into three broad classes: cluster disruption, where clusters are disrupted before $z=0$ due to high tidal heating; cluster survival, where clusters survive to $z=0$ due to low tidal heating; and cluster migration, where clusters migrate from regions of high tidal heating to regions of low tidal heating, enabling their survival to $z=0$. Each of these cases is represented in Fig. \ref{fig:heating_ts} with particles that show typical evolutions for each class. In the survival and migration cases, each cluster survives to $z=0$ with a final mass of $\approx 3000 \Msun$. For the disruption and survival cases, the clusters were chosen to form at similar times. However one cluster forms in a region of high tidal heating and is disrupted within $0.5 \Gyr$, while the other evolves in a region of low tidal heating to $z=0$. In the case of migration, the cluster forms in a region of high tidal heating but migrates at $z\approx0.9$ in to a low tidal heating region which enables the cluster to survive to $z=0$. This cluster is formed in a galaxy with a stellar mass of $2\times10^8 \Msun$ which merges into the main galaxy at $z=0.9$ (see the galaxy merger tree in Fig. \ref{fig:trees}), highlighting the importance of galaxy mergers for cluster migration and survival \citep[as had been suggested by analytical models, cf.][]{Kruijssen_15}.

Inspection of the evolution of the tidal field strength and tidal heating parameter for Gal004 and Gal005 in Fig. \ref{fig:tid_stren} reveals significant differences, in spite of the two sharing similar present day masses and morphologies. The differences are most clearly seen in the disc cluster population, with Gal005 achieving significantly greater values of both quantities. This is largely a reflection of the galaxies' differing gas density distributions and star formation rates, and underlines the significant diversity of cosmic environments that have been presented to star clusters in galaxies that, by many measures, are similar at the present day.

\begin{figure*}
  \includegraphics[width=\textwidth]{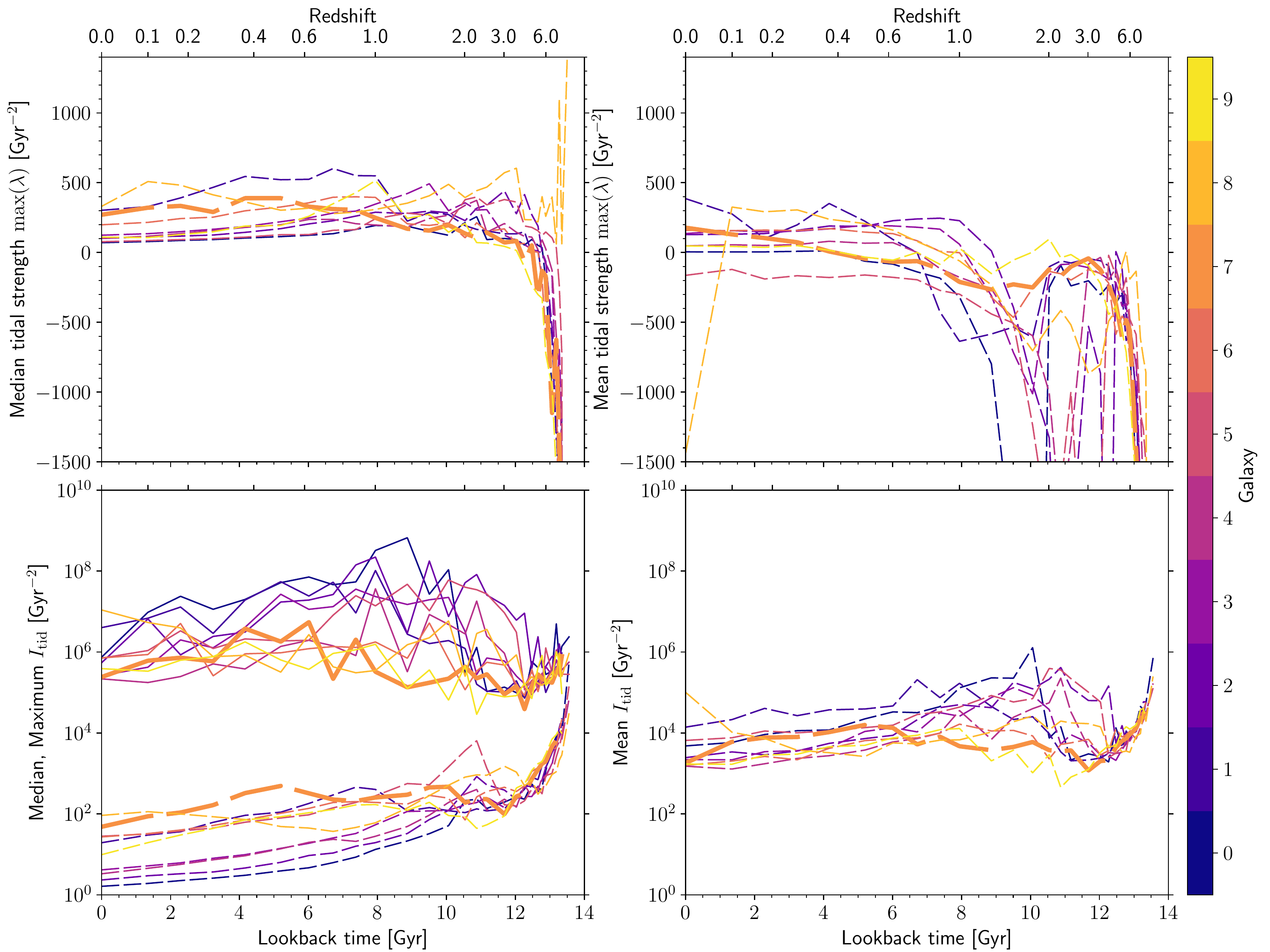}
  \caption{Median (left panels) and mean (right panels) tidal strength ($\rmn{max}(\lambda)$, upper panels) and tidal heating parameter ($I_\rmn{tid}$, lower panels) as a function of time for the cluster-bearing particle populations in all $\numgal$ galaxies. For reference in the discussion, Gal007 is highlighted with a thicker linewidth. The bottom left panel also shows the maximum $I_\rmn{tid}$ (solid lines) in addition to the median values (dashed lines). The median and mean values show little correlation between them for both tidal strength and $I_\rmn{tid}$. For most galaxies, the median tidal strength (top left) remains nearly constant for $z<3$. Negative values for the  mean tidal strength (top right) indicate very central cluster formation episodes, occurring for all galaxies at $z>6$, for most galaxies at $1<z<4$ during the peak of cluster formation, and again at $z=0$ for Gal008. The mean tidal strength for Gal000 remains negative over its entire history due to very central star formation occurring at $z \approx 2$. The median $I_\rmn{tid}$ generally decreases over time as clusters are disrupted or migrate out of regions of high gas density. However the maximum and mean $I_\rmn{tid}$ are more constant over time, with the highest values attained during periods of highest gas density. Galaxies with higher peak gas densities and more central star formation are therefore more elevated in $I_\rmn{tid}$ (e.g.~Gal000).}
  \label{fig:mean_tid_stren}
\end{figure*}

To illustrate the influence of the diversity in tidal histories more clearly, we show in Fig. \ref{fig:mean_tid_stren} the median and mean values of the tidal field strength and tidal heating parameter of the cluster-bearing particle populations in all $\numgal$ simulated galaxies. 
In spite of the similarities between the galaxies at the present day (Table~\ref{tab:sims}), both the median and mean tidal field strengths show a large range between galaxies, which reflects differences between the galaxies in terms of both the cluster radial distributions and the underlying galaxy mass profiles. 
The median tidal field strengths (top left panel) generally show a similar trend between the galaxies, being negative at early times ($z>5$) and then flattening out to nearly constant values for $z<4$.
However, the mean tidal strength shows significantly different evolution with redshift, bearing little correspondence to the median. In particular, Gal000 shows negative values for the mean tidal strength for the entire history of the galaxy while the median remains positive after the rapid evolution at early times. 
A significant deviation between the mean and median tidal field strengths also occurs for nearly all galaxies between $z=1$--$4$, during which the mean tidal field strength experiences a (negative) minimum, driven by intense central star formation, after the initial increase to $\max(\lambda)\approx0 \Gyr^2$ at $z\sim5$.
The time for the mean to return to positive values (which is of the order a few Gyr) reflects the timescale for cluster disruption and (mostly) migration away from the galactic centre, such that cluster-bearing particles that recently formed near the galactic centre are no longer weighted strongly in the mean.
A negative mean tidal field strength also occurs at $<1 \Gyr$ for Gal008, which is undergoing a central starburst near $z=0$ (see the discussion of the CFE for Gal008 in Section~\ref{sec:diversity}).
In summary, the mean tidal field strength appears to be a measure of how central the cluster formation is in the galaxy, with very negative values indicating very central star and stellar cluster formation. 

A similar figure was presented by \citet[see their fig. 10]{Renaud_Agertz_and_Gieles_17}, who found the mean tidal strength was nearly constant at $z>1.2$, increased from $z=1.2$ to $z=0.6$ and reached a maximum at $z=0.5$ (the final time in the simulation).
This differs significantly from the time evolution found in our simulations which generally do not have a constant mean tidal strength until $z<1$.
However, the subset of particles for which tidal tensors are calculated differs significantly between the two studies. 
In their work, a random subset of stellar particles was chosen for which to output tidal tensors and study the tidal histories. 
In our simulations, tidal tensors are calculated only for the particles that at that time contain at least one star cluster, based on the physically-motivated cluster formation and disruption model used in this work. This modifies the particle distribution in a number of ways:
\begin{enumerate}
\item
Cluster formation occurs mainly in stellar particles with high birth pressures, which emphasises particles at early formation times and in galactic centres, where cluster formation is more efficient.
\item
Tidal shocks are most effective at high gas densities and rapidly removes low-mass clusters (see following section), which preferentially leaves particles that formed massive clusters, again gearing the sample towards star particles that formed in high-pressure environments.
\item
Newly formed stellar particles hosting only low-mass clusters may dominate the particle numbers prior to the disruption of low-mass clusters.
\end{enumerate}
The differing subsets of particles therefore impedes direct comparison between the studies.
However, the median and mean tidal strength for Gal007 (highlighted in Fig. \ref{fig:mean_tid_stren}) has a qualitatively similar evolution with time to the figure in \citet{Renaud_Agertz_and_Gieles_17}, being low at redshifts $z>1$ and increasing from $z=1$ to the present time. Gal007 peaks in SFR at a lookback time of $6 \Gyr$ (Fig. \ref{fig:SFR}), significantly later than most of the galaxies in our sample, consistent with the peak in the median tidal strength. In contrast, galaxies with an earlier peak SFR generally have a near constant mean tidal strength at $z<1$. This diversity between galaxies, even within a narrow range of present-day masses, highlights the necessity of considering a sample of galaxy simulations, since no two undergo an identical formation and assembly history.

Finally, in the bottom panels of Fig. \ref{fig:mean_tid_stren} we compare the median, mean and maximum tidal heating parameters for all galaxies. Again, the median and mean $I_\rmn{tid}$ show significant differences between them, and a high median value does not necessarily correspond to a high mean value. As discussed for Fig. \ref{fig:tid_stren}, the median $I_\rmn{tid}$ traces star-forming regions at high redshifts ($z\gtrsim1$) and non star-forming regions at low redshifts ($z\lesssim 1$) and shows an overall trend of a decreasing tidal shock intensity with time. 
However, though the maximum and mean $I_\rmn{tid}$ typically peak between $0.8<z<2$, they do not significantly differ between low ($z<0.5$) and high ($z>2$) redshifts. For some galaxies (Gal007 and Gal009) the maximum and mean $I_\rmn{tid}$ are nearly constant over the full simulation (though short timescale peaks may have been missed between snapshots).
The peaks in the mean $I_\rmn{tid}$ correlate with peaks in birth pressures and are therefore more representative of values that the young cluster population experiences at each epoch (which is weighted higher in the mean through the large dynamic range). The relatively flat behaviour of the mean $I_{\rm tid}$ as a function of lookback time therefore indicates that the conditions favourable to cluster formation drive efficient cluster disruption independently of the redshift. However, the general trend remains that the mean $I_\rmn{tid}$ achieves higher values during the peak of cluster formation at redshifts $z>1$. As a result, most cluster disruption by tidal shocks takes place at high redshifts, when the gas pressures and densities are the highest.


\section{Properties of the present day cluster populations} \label{sec:cluster_props}

In this section, we discuss the properties of the star cluster population of each simulated galaxy at $z=0$, after cluster mass-loss (stellar evolution, tidal shocks and two-body relaxation) and destruction by dynamical friction have been included, and compare the cluster populations to the MW GCs. We also discuss the origin of the GC mass function and the necessary modelling requirements for simulations aiming to study its origin.

\subsection{Cluster masses}

\begin{figure*}
  \includegraphics[width=\textwidth]{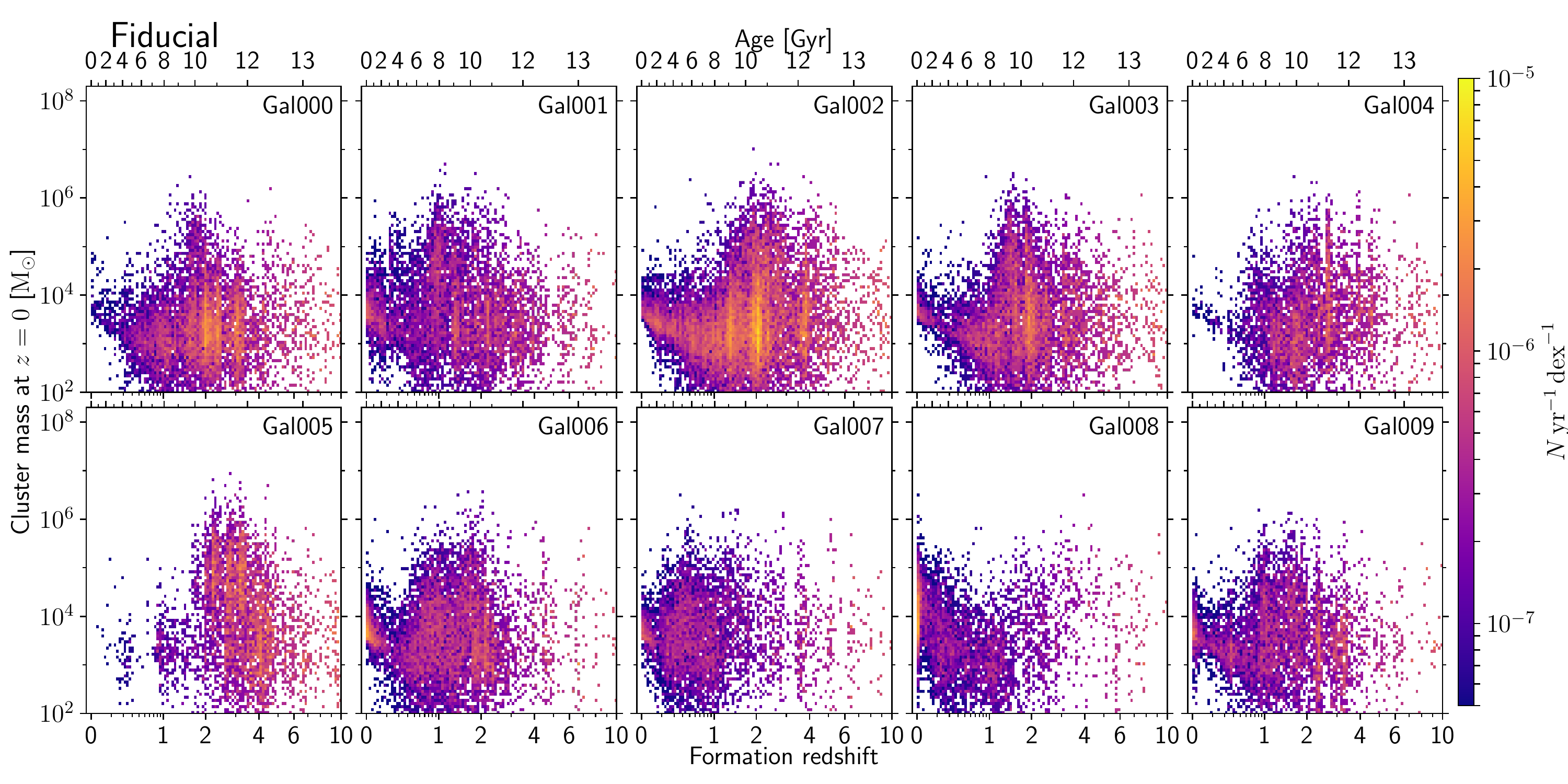}
  \includegraphics[width=\textwidth]{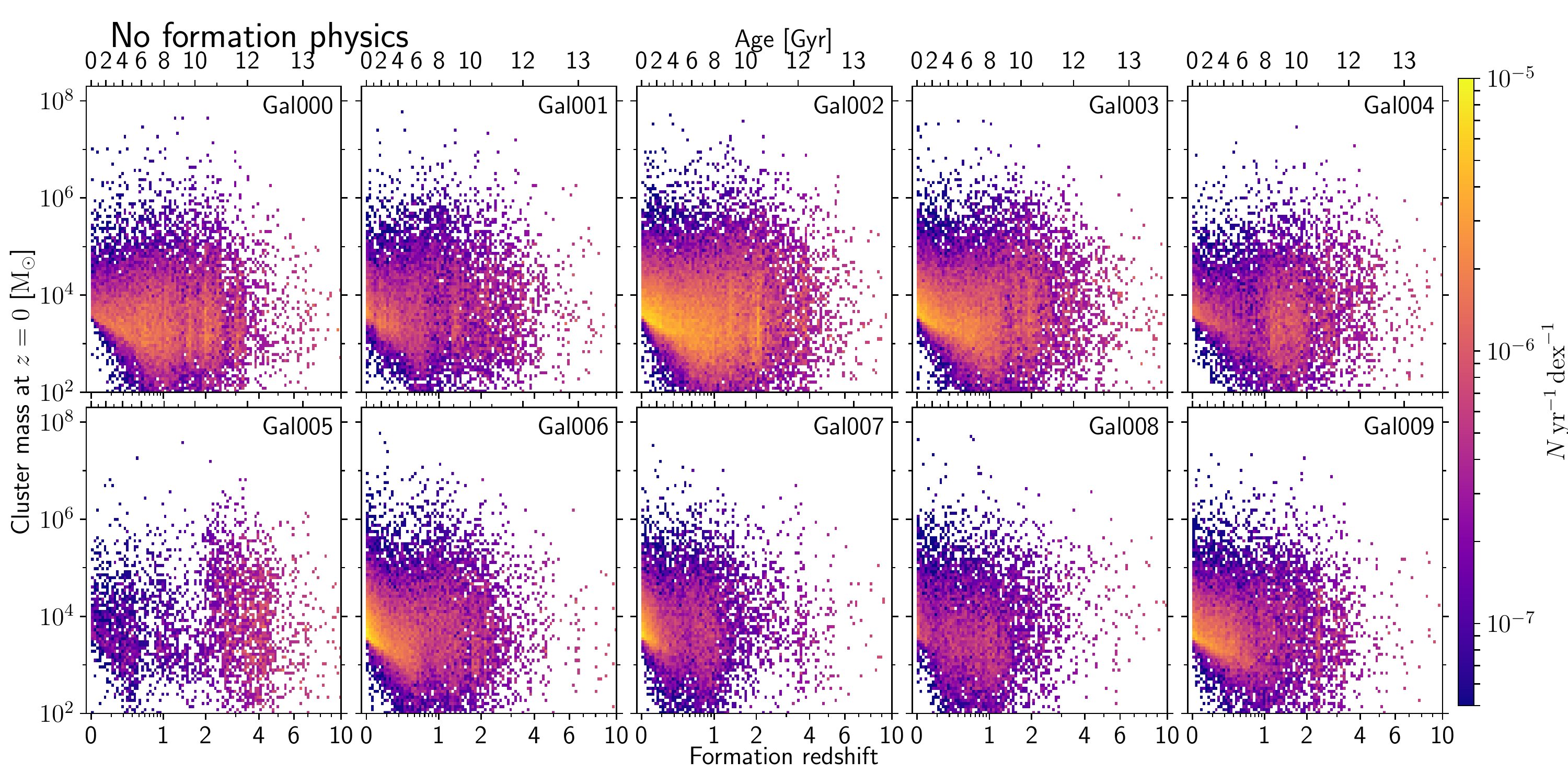}
  \caption{ Two-dimensional histogram of final ($z=0$) cluster masses as a function of time for all $\numgal$ galaxies. The colour scale is identical for all galaxies and shows the number of star clusters per two-dimensional bin. Upper: Evolved cluster masses for the fiducial simulations. This figure corresponds to Fig. \ref{fig:Mcinit} after all cluster mass-loss has been included (stellar evolution, tidal shocks, evaporation and dynamical friction). Most low-mass clusters have been completely disrupted, mainly by tidal shock heating (see Section \ref{sec:GCMF}). The few massive clusters in each galaxy are typically removed by dynamical friction. Lower: Evolved cluster masses in the simulations with no cluster formation physics, i.e.~a constant cluster formation efficiency $\Gamma=0.1$ and an infinite cluster truncation mass (i.e.~power-law mass function). In contrast to the fiducial model, here the cluster formation rate traces the SFR and the formation of massive clusters ($>10^5 \Msun$) no longer peaks at old ages.}
  \label{fig:Mcfin}
\end{figure*}

The upper panel of Fig. \ref{fig:Mcfin} shows the final cluster masses (at $z=0$) as a function of cluster formation redshift for the fiducial runs. Compared to the initial cluster masses (Fig. \ref{fig:Mcinit}) most low-mass clusters ($5\times 10^3 \Msun$) have been completely disrupted. This is particularly evident for Gal008 at formation times $z>3$ where nearly all of the clusters with masses $<10^4 \Msun$ have been disrupted.
Disruption of low mass clusters can also be seen with the lower cluster mass decreasing with increasing redshift, from the initial mass limit of $5 \times 10^3 \Msun$ to the limit of $100 \Msun$ (e.g.~for Gal002).
At the high-mass end, the most massive clusters formed in the galaxy are typically removed by dynamical friction, which is most effective for high cluster masses and small galactocentric radii.

In the lower panel of Fig. \ref{fig:Mcfin} we show the final cluster masses for the simulations with a constant CFE ($\Gamma = 0.1$) and no upper truncation to the mass function, i.e.~equivalent to a simple particle tagging method. The galaxies differ slightly between the fiducial and `no formation physics' runs because of stochasticity in star and cluster formation, however the star formation histories are generally very similar. Note that the value of the constant CFE is not particularly important, since it merely represents a simple scaling of the total number of clusters.
Here, the formation of massive clusters no longer peaks at $z\sim2$ as in the fiducial runs. This demonstrates that the age distribution of GCs in the fiducial runs is not just a consequence of the star formation history, but is also influenced by the physically-motivated cluster formation model. Simple particle tagging would therefore be unable to simultaneously reproduce the GC population and young clusters in $z=0$ MW-like galaxies. Either they would predict present-day young clusters that are too massive, or high-redshift GCs that are not massive enough. Though we do not explicitly demonstrate it here, we also note that cluster disruption by tidal shocks is less effective when omitting cluster formation physics (in particular the environmentally-dependent CFE model), because clusters are then no longer predominantly formed at the peaks of the density distributions.

\subsection{The GC mass function} \label{sec:GCMF}

\begin{figure*}
  \includegraphics[width=\textwidth]{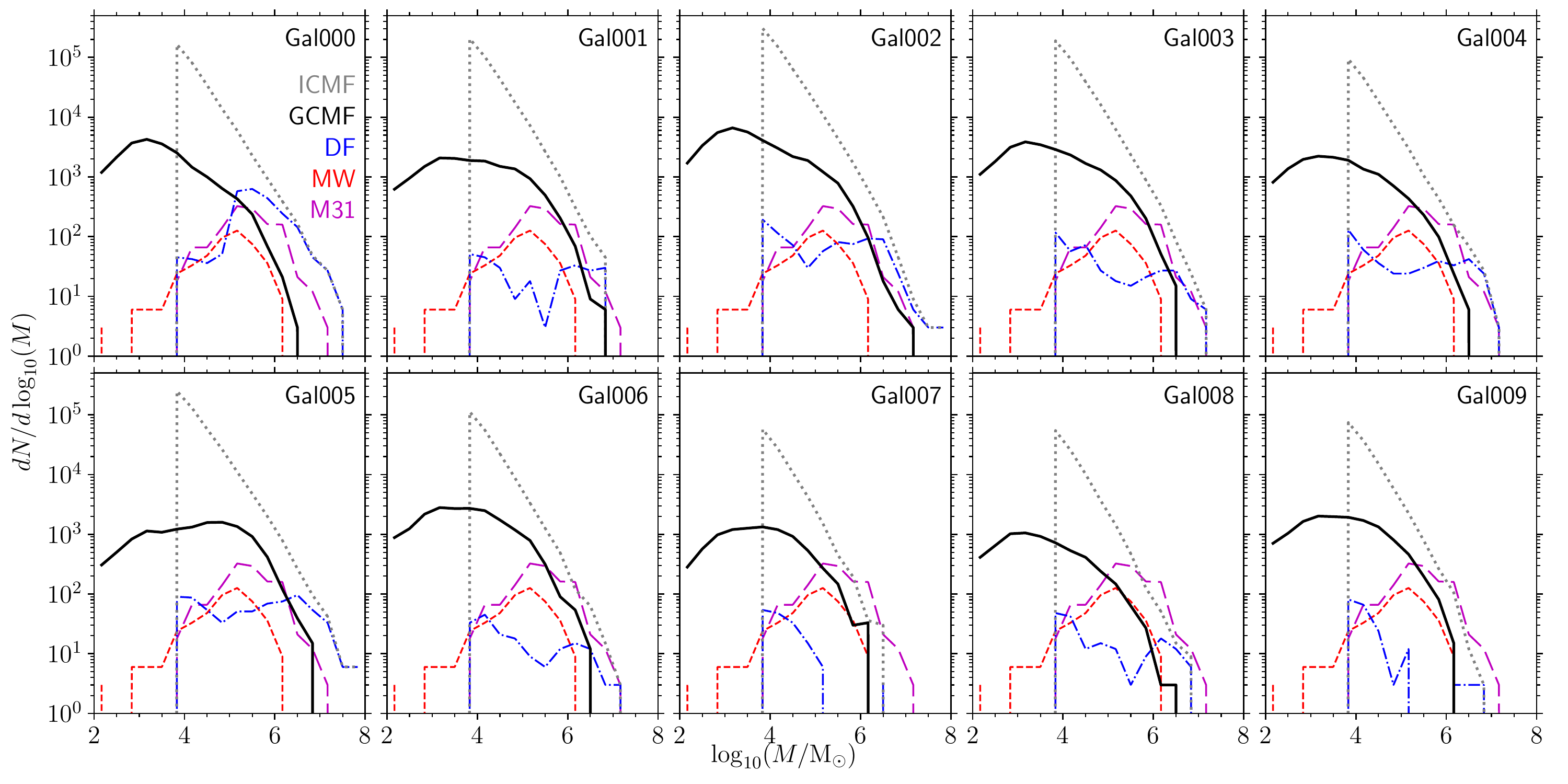}
  \caption{ Globular cluster mass function (GCMF; for clusters older than 6 Gyr at $z=0$) for the simulated $L^\star$ galaxies. Solid black lines show the final ($z=0$) mass function, grey dotted lines show the ICMF, blue dash-dotted lines show the initial masses of clusters removed by dynamical friction, red dashed lines show the MW GC mass function \citep{Harris_96} and magenta long-dashed lines show the M31 GC mass function \citep{Caldwell_et_al_11, Huxor_et_al_14}. The high mass end of the predicted cluster mass functions are generally in good agreement with the MW and M31 GC mass functions. Low mass clusters are much more abundant than observed, likely due to the lack of an explicit model for the cold ISM phase and therefore insufficient disruption by tidal shocks (see the text). }
  \label{fig:GCMF}
\end{figure*}

This section focuses on the mechanisms shaping the $z=0$ cluster mass function in our model, and explores whether the simulations are able to reproduce the observed mass distribution. Fig.~\ref{fig:GCMF} shows the cluster mass function for the 10 fiducial simulations, and compares with the MW and M31 GC mass functions (GCMFs). The age cut for clusters in the simulations at $6 \Gyr$ is motivated by the late peak in cluster formation for some galaxies (e.g. Gal001 and Gal007, see Fig. \ref{fig:Mcfin}). For the MW GC masses we use the catalogue from \citet[2010 edition]{Harris_96} and convert luminosity to mass assuming $M/L_V = 1.7 \, \rmn{M/L}_{\sun}$ \citep[the mean for MW clusters][]{McLaughlin_and_van_der_Marel_05}. For M31 GC masses we combine the catalogues of \citet[using the given masses]{Caldwell_et_al_11} and \citet[again assuming $M/L_V = 1.7 \, \rmn{M/L}_{\sun}$, e.g.~\citealt{Strader_et_al_11}]{Huxor_et_al_14}. 

The high-mass end of the simulated GCMFs ($> 10^{5.5} \Msun$) are in good agreement with the observed MW and M31 GCMFs, with these two cases approximately bracketing the range of distributions found in the simulations (with the MW at the low number end, M31 at the high number end). The majority of distributions more closely resemble the MW GCMF, however Gal002 and Gal005, the two galaxies with the highest SFR at early times ($z>2$, see Fig \ref{fig:SFR}), have significantly more clusters than the MW at the high-mass end and more closely match the observed mass function of M31 \citep[which has nearly three times more clusters than the MW,][]{Huxor_et_al_14}. The most massive clusters hosted by our simulated galaxies are between $10^6$ and $10^7 \Msun$, consistent with what is observed in the MW and M31, as well as the Virgo cluster \citep{Jordan_et_al_07_XII}. 
We have included in the observed mass functions clusters such as $\omega$ Cen, which likely has a nuclear star cluster origin \citep{Lee_et_al_99}. However the contribution of stripped nuclear clusters to the GC mass function in MW-mass galaxies is expected to account for $<4$ clusters \citep{Pfeffer_et_al_14} and is therefore negligible.

The blue dash-dotted lines in the figure show the initial masses of clusters older than 6 Gyr that are removed by dynamical friction. Dynamical friction is most effective at removing high-mass clusters, for which the mass ratio between the cluster and the enclosed mass of the galaxy is highest, or clusters at very small galactocentric distances. This effect of affecting the most massive clusters is enhanced further due to the cluster formation model: the most massive clusters form at the highest gas pressures (due to the dependence of CFE and $\Mcstar$ on pressure), which occur at the galactic centre during the peak of cluster formation (see Fig.~\ref{fig:particles} and discussion in text). Gal000 has more clusters removed than other galaxies due to very central cluster formation with high CFE ($\sim$80 per cent, see Fig.~\ref{fig:all_CFE}) occurring at $z\approx1.7$, making dynamical friction highly effective at removing clusters. For this galaxy most clusters are removed from the central 3 kpc. 

Despite the good agreement with observed galaxies at the high-mass end of the GCMF, all 10 simulations produce too many low mass clusters, by a factor of $2$--$10$ at $10^5 \Msun$ and a factor of $10$--$100$ at $10^4 \Msun$. The observed MW cluster mass function is likely incomplete below $10^4 \Msun$ and one should additionally account for old open clusters, such as NGC 6791 \citep[$\sim 5 \times 10^4 \Msun$,][]{Platais_et_al_11}, since they would not be excluded from the mass function in the simulations. However, incompleteness does not account for the discrepancy between the observed and simulated mass functions, because the MW mass function is likely complete at $10^5 \Msun$, given that the peak mass is similar to the near-universal peak mass observed in extragalactic GC populations \citep[e.g.][]{Jordan_et_al_07_XII}.

We posit that the simulations do not adequately disrupt low-mass clusters through tidal shocks induced by interaction with the cold and substructured ISM. Theoretical studies \citep{Gieles_et_al_06,Elmegreen_and_Hunter_10,Kruijssen_et_al_11} predict, and observational campaigns \citep{Bastian_et_al_12, Miholics_et_al_17} reveal, a strong correlation between cluster lifetimes and the properties of the cold ISM that is indicative of cluster destruction by shocks. As reiterated throughout this work, this process is expected to be particularly important for shaping the GC population \citep{Elmegreen_10,Kruijssen_15}.
The failure to disrupt clusters through tidal shocks in our simulations is most likely a consequence of the EAGLE model not incorporating an explicit model of the cold, dense phase of the ISM, which is predicted to contribute the vast majority of the disruptive power in real galaxies. Therefore, stars and stellar clusters form in the simulations from gas particles of which the density and temperature are characteristic of photoionized interstellar gas ($T \sim 10^4$ K, $n_{\rm H} \sim 0.1-1.0$ cm$^{-3}$). Consequently, the ISM of EAGLE galaxies is considerably smoother than that of real galaxies and, since the Jeans length of such gas is $\sim 1 \pkpc$, the stellar discs of EAGLE galaxies are also thicker than observed by a factor of $\simeq$2. These shortcomings of the galaxy formation model reduce the impact of cluster mass loss from tidal shocks induced by the dense ISM.

\begin{figure}
  \includegraphics[width=84mm]{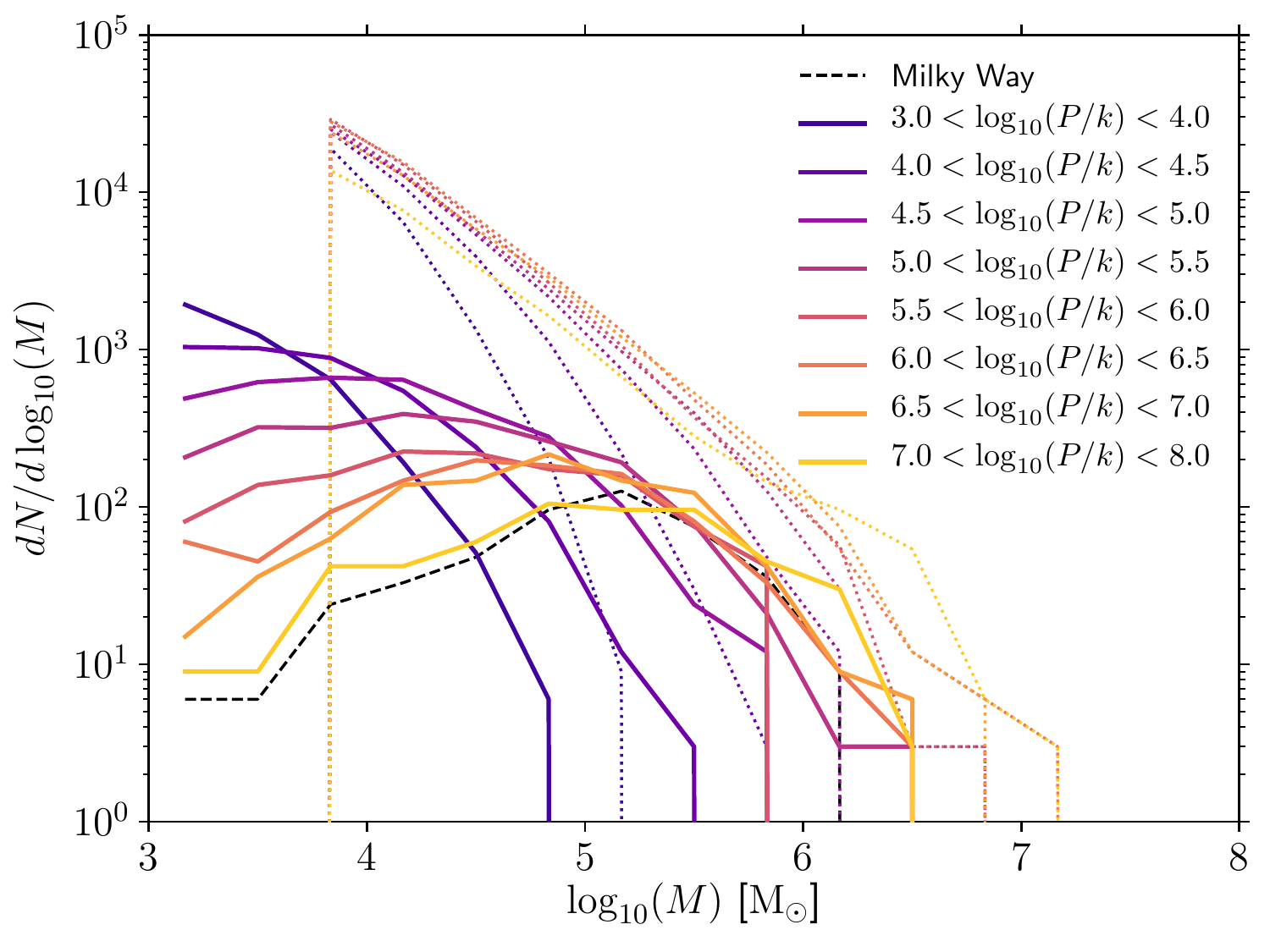}
  \caption{Initial (dotted) and $z=0$ (solid) cluster mass functions for clusters older than 6 Gyr as a function of cluster birth pressure (units $\K \cmcubed$) for Gal003. The dashed line shows the MW GC mass function \citep{Harris_96} for comparison. Disruption of low mass clusters is significantly more effective at high birth pressures and reflects the increasing contribution of tidal shocks to cluster mass-loss at high gas pressures/densities (Table \ref{tab:P-disruption}).}
  \label{fig:MF_Pressure}
\end{figure}

\begin{table}
\caption{Fraction of dynamical mass-loss (i.e.~excluding stellar evolution mass-loss and dynamical friction) due to tidal shocks for clusters in Fig. \ref{fig:MF_Pressure}.}
\label{tab:P-disruption}
\centering
\begin{tabular} {@{}ccc@{}}
  \hline
  Birth pressure           & Birth density & Shock mass-loss \\
  $\log_{10}$(K cm$^{-3}$) & $\log_{10}$(cm$^{-3}$) & fraction \\
  \hline
  3.0 - 4.0  &  $-0.95$ - $-0.2 $  &  0.30 \\
  4.0 - 4.5  &  $-0.2 $ - $ 0.05$  &  0.41 \\
  4.5 - 5.0  &  $ 0.05$ - $ 0.5 $  &  0.50 \\
  5.0 - 5.5  &  $ 0.5 $ - $ 0.9 $  &  0.59 \\
  5.5 - 6.0  &  $ 0.9 $ - $ 1.3 $  &  0.67 \\
  6.0 - 6.5  &  $ 1.3 $ - $ 1.65$  &  0.70 \\
  6.5 - 7.0  &  $ 1.65$ - $ 2.0 $  &  0.69 \\
  7.0 - 8.0  &  $ 2.0 $ - $ 2.8 $  &  0.66 \\
  \hline
\end{tabular}
\end{table}

To illustrate the influence of the ambient ISM properties on cluster evolution, Fig. \ref{fig:MF_Pressure} shows the initial (dotted lines) and $z=0$ (solid lines) mass functions of clusters formed within Gal003, binned by their birth pressure. We adopt Gal003 as the exemplar in this case because its clusters exhibit a relatively broad distribution of birth pressures.\footnote{Note that the dependence of $\Mcstar$ on pressure results in the high-mass end of the ICMF appearing to be steeper at lower birth pressure.} The figure shows that significantly fewer low-mass clusters survive when formed from gas at high pressures, illustrating the strong dependence of cluster disruption on the birth pressure (or, equivalently, birth density). This reflects the fact that the birth density is representative of the ambient ISM densities that clusters experience in their early lives (see Appendix \ref{app:env_density}), and thus also how efficiently they get disrupted by tidal perturbations from the ISM.

Table \ref{tab:P-disruption} shows the total fraction of dynamical mass-loss due to tidal shocks for the initial cluster population (i.e. including those completely disrupted) in the birth pressure bins used in Fig. \ref{fig:MF_Pressure}. The fraction of tidal shock-induced mass-loss increases with birth pressure -- at pressures $> 10^6 \K \cmcubed$, tidal shocks account for 75 per cent of the dynamical mass loss. 
The fact that the tidal shock mass-loss fraction does not increase in the final pressure bin ($10^7$-$10^8$ $\K \cmcubed$) is caused by the larger fraction of high-mass clusters ($>10^{5.5} \Msun$) for which mass-loss by tidal shocks is minimal. 
The increase of tidal shock-driven mass-loss with density implies rapidly changing tidal fields at the highest densities, which in turn reflects an increased degree of substructure in the ISM. The monotonic nature of this increase also suggests that disruption will further increase as higher densities can be realised with more detailed ISM models (up to some limit, imposed by the density distribution function).
Our finding that the majority of cluster disruption in the simulations takes place due to tidal shocks is consistent with previous estimates \citep{Lamers_and_Gieles_06,Kruijssen_et_al_11}.

Returning to Fig.~\ref{fig:MF_Pressure}, we see that at low birth pressures ($10^{3-4}$ K cm$^{-3}$), clusters experience little mass loss other than that due to stellar evolution. Therefore, they retain a power law mass function from formation to the present day. Clusters formed from gas particles at higher pressure experience stronger tidal disruption, such that the mass function of clusters formed from gas with pressure greater than $10^6$ K cm$^{-3}$ evolves into a peaked (i.e.~close to log-normal) distribution by the present day. As we found in Section \ref{sec:tidal_histories}, disruption by tidal shocks is most effective at redshifts $z>1$ (although the peak epoch varies from galaxy to galaxy), meaning that the peaked mass functions were in place soon after cluster formation. This concurs with the predictions of \citet{Kruijssen_15}.
The mass functions at the highest birth pressures ($P/k > 10^7$ K cm$^{-3}$) peak at a mass and number density that is similar to the MW GCMF. Encouragingly, birth pressures of $\sim 10^7$ K cm$^{-3}$ are similar to those proposed by \citet{Elmegreen_and_Efremov_97} for GC formation. They are also similar to those observed in star-forming galaxies at high redshift \citep[e.g.][]{Swinbank_et_al_11}. Therefore, even though our simulations do not explicitly model the cold phase of the ISM (which will be addressed in future work), they indicate that the evolving cosmic environments experienced by young clusters induce tidal-shocks sufficiently strong to shape a power-law initial mass function into the observed log-normal GCMF, as predicted by \citet[][]{Elmegreen_10} and \citet{Kruijssen_15}.

\subsection{Radial cluster properties}

Finally, we briefly investigate the radial distributions of cluster metallicities and masses at $z=0$. A more detailed analysis of predicted radial distributions and metallicities will be addressed in future work. As in the previous sections, we find that these two observables again highlight the necessity of including physically-motivated formation models when studying star cluster populations.

\begin{figure*}
  \includegraphics[width=84mm]{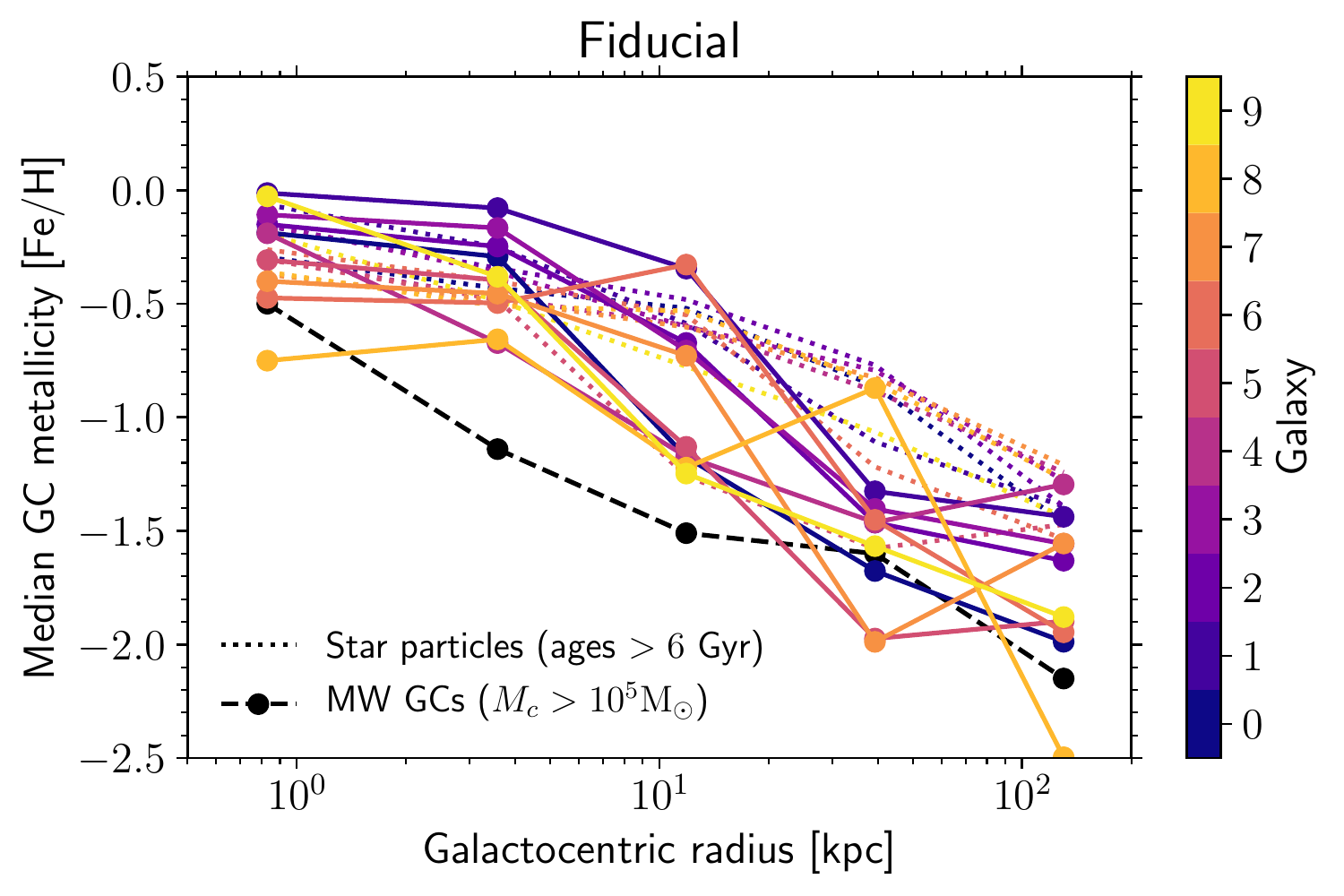}
  \includegraphics[width=84mm]{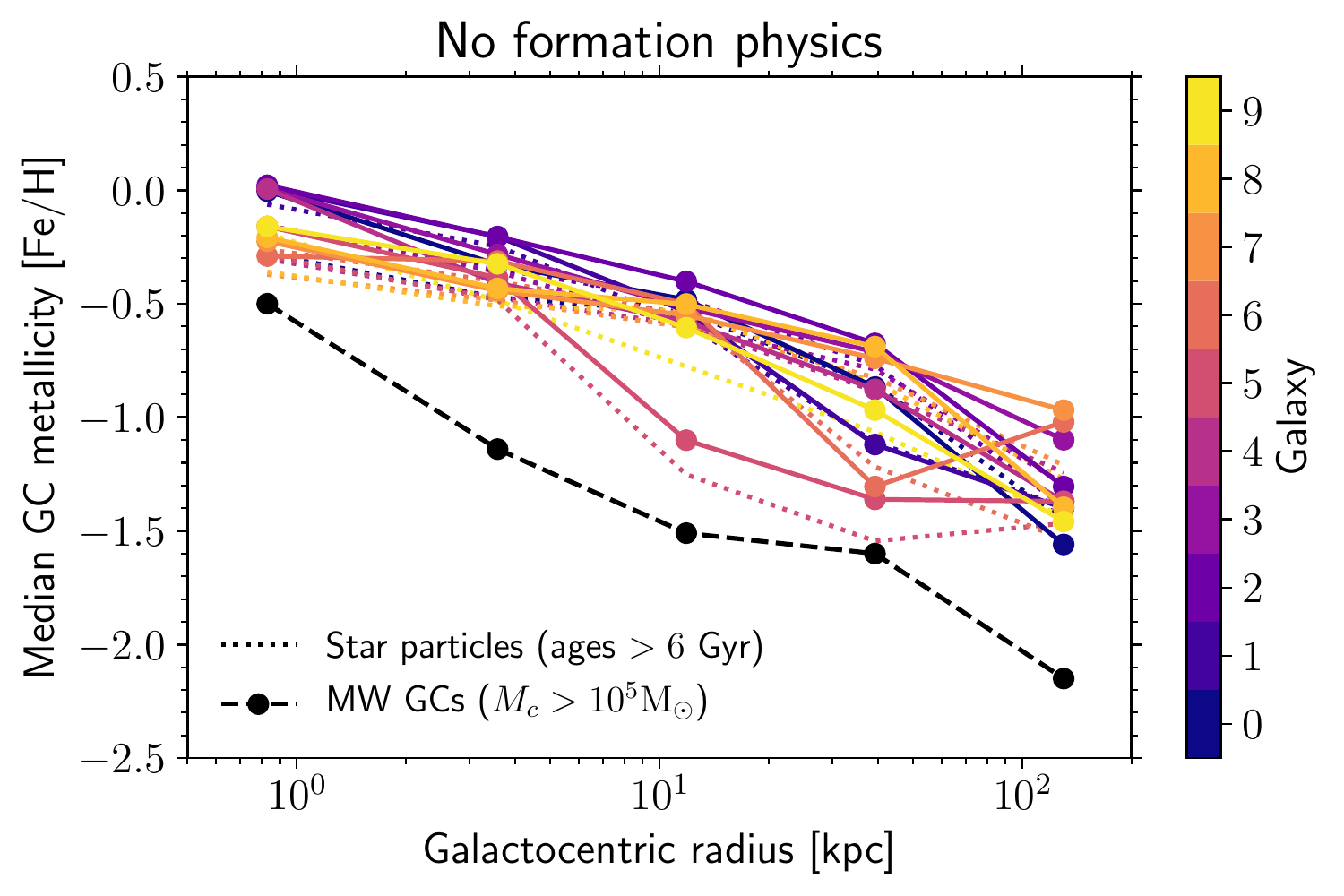}
  \caption{Median cluster metallicity (for cluster masses $>10^{5} \Msun$) as a function of galactocentric radius for all $\numgal$ galaxies and cluster ages greater than 6 Gyr. Dashed black lines show MW GCs \citep{Harris_96} with $M/L_V = 1.7 \, \rmn{M/L}_{\sun}$. Dotted lines show the metallicity relations for field stars in the simulated galaxies with the same age limit as the clusters. The left panel shows the results for the fiducial runs, while the right panel shows the results for the simulations omitting cluster formation physics, i.e.~adopting a constant CFE ($\Gamma=0.1$) and power-law mass function. }
  \label{fig:radial-props}
\end{figure*}

Fig. \ref{fig:radial-props} shows the median metallicity of clusters with $\FeH > -3$ as a function of galactocentric radius for all simulated $L^\star$ galaxies. The left-hand panel shows the metallicity-radius relation for our fiducial simulations, whilst the right-hand panel shows the same for the simulations in which the model components governing the cluster formation properties have been disabled, i.e.~we assume that an environmentally-independent, fixed fraction of 10 per cent of all stars form in clusters and the maximum cluster mass is set to infinity. We focus here on a comparison of the radial trend exhibited by the simulation with that of observed clusters, rather than the precise normalisation of the metallicities, because nucleosynthetic yields are uncertain at the factor $\simeq 2$ level \citep[see e.g.][]{Wiersma_et_al_09} and hence the uncertainty budget is likely dominated by systematic effects.

In the fiducial case (left panel), the metallicity of clusters close to the centres of our simulated galaxies ($r < 1.5$ kpc), and at relatively large galactocentric distances ($r >20$ kpc), are compatible with the inferred metallicities of the MW's GCs. However, the radial trends are rather different, with the metallicity of clusters at intermediate distances (1.5-20 kpc) being significantly greater in the simulations than is observed. We attribute this to the (numerically) inefficient disruption of clusters born from gas at low-to-intermediate pressure. Such clusters typically form in the disc of the galaxy and are metal rich $\FeH > -1$; increasing their disruption rate would therefore significantly suppress the characteristic metallicity of clusters within $\simeq 20$ kpc.

In the simulations omitting cluster formation physics (right panel), the metallicity of clusters broadly traces that of the galaxy as a whole (with some limited effect of cluster disruption). 
The cluster metallicities at radii $<1.5 \kpc$ are similar to the case of the fiducial runs. However, at all other radii, clusters are significantly more metal-rich than observed. The difference between the fiducial and `formation off' runs is caused by the differing cluster age distributions (clusters in the fiducial run being significantly older than in the cases without formation physics, see Fig.~\ref{fig:Mcfin}), since the mass-metallicity relation of the galaxies themselves evolves with time.

\begin{figure*}
  \includegraphics[width=84mm]{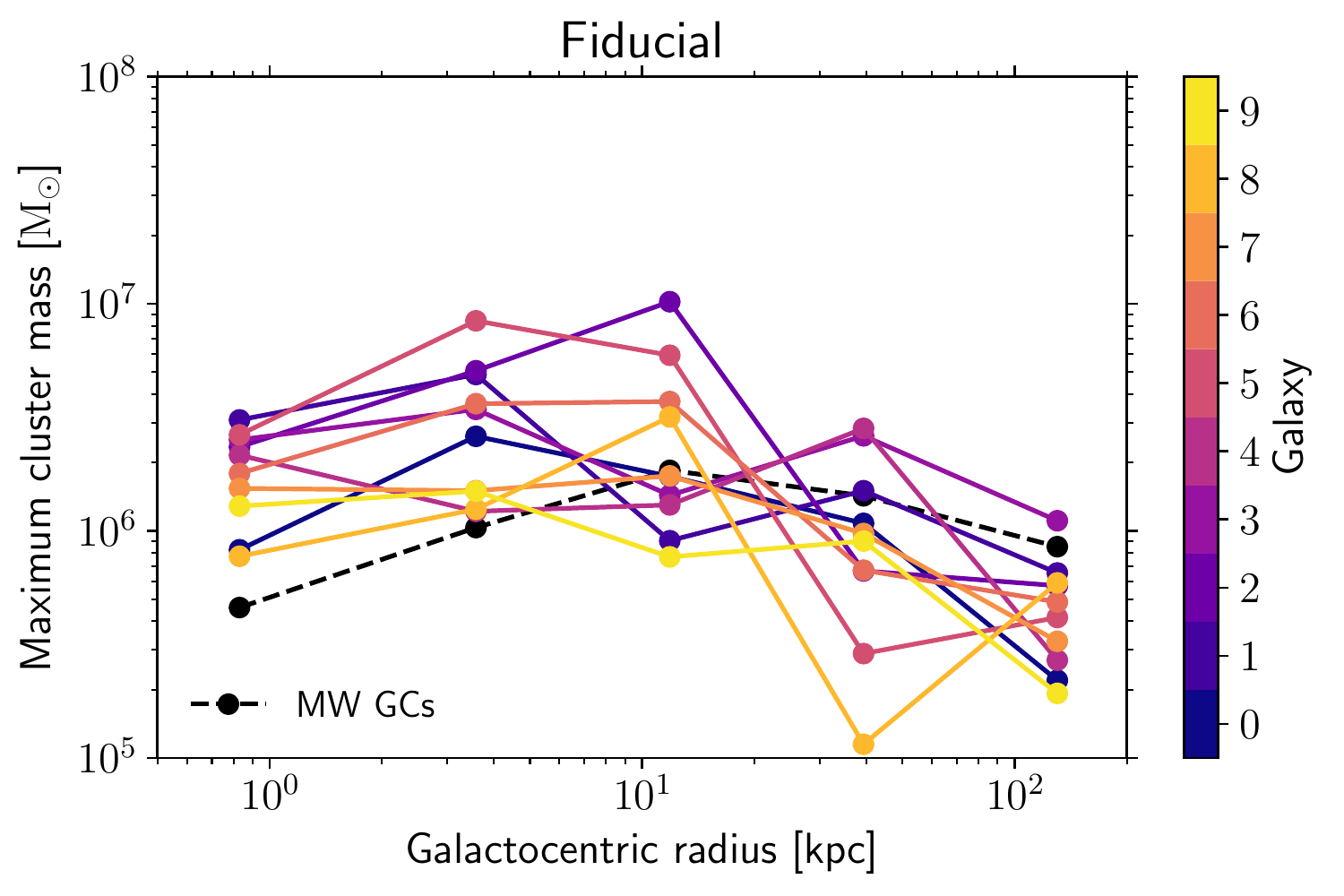}
  \includegraphics[width=84mm]{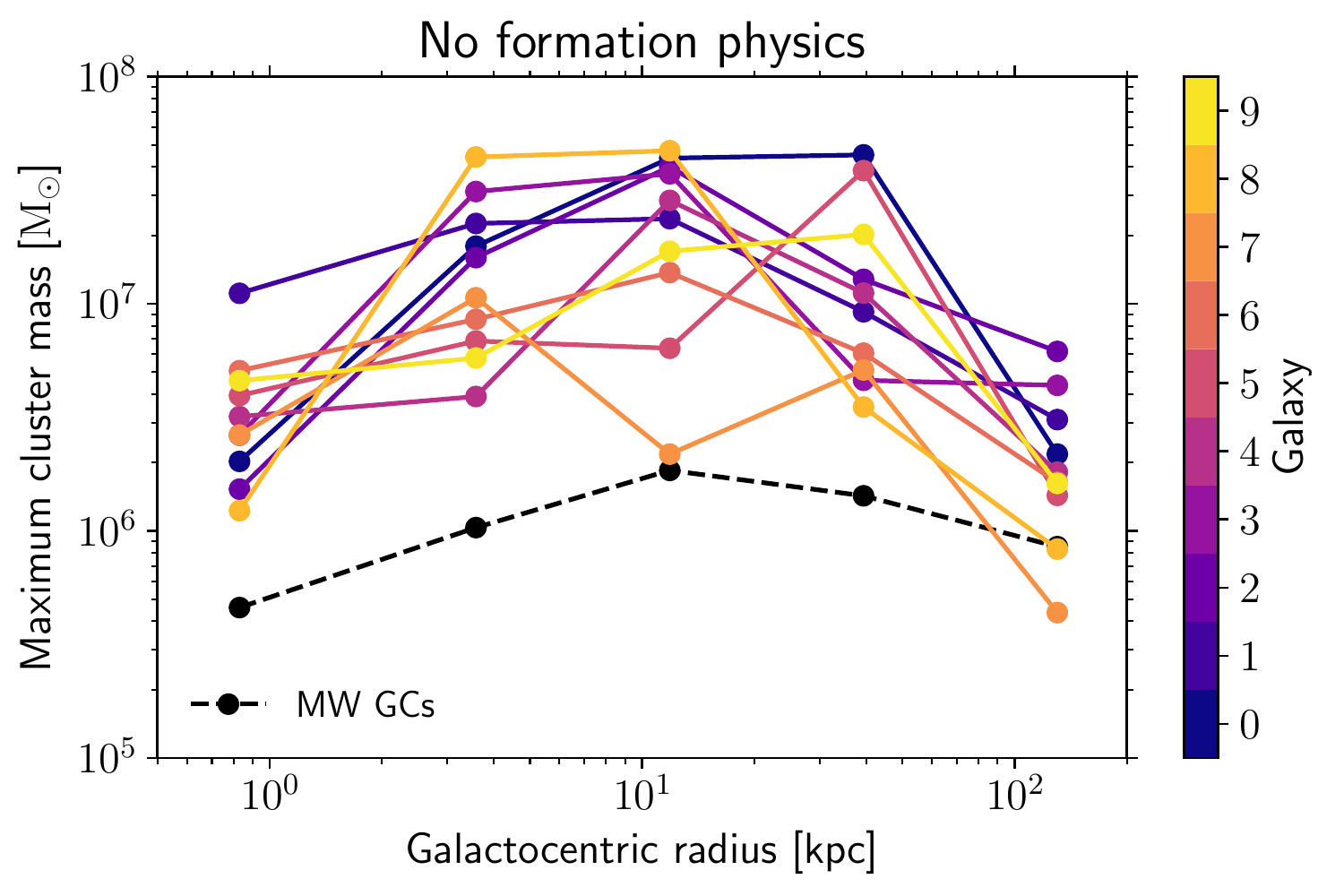}
  \caption{Most massive cluster as a function of galactocentric radius for all $\numgal$ galaxies and cluster ages greater than 6 Gyr. Dashed black lines show MW GCs \citep{Harris_96} with $M/L_V = 1.7 \, \rmn{M/L}_{\sun}$. The left panel shows the results for the fiducial runs, while the right panel shows the results for the `formation off' runs with a constant CFE $\Gamma=0.1$ and a power-law mass function. }
  \label{fig:radial-mass}
\end{figure*}

The maximum cluster mass in the fiducial runs, shown in the left panel of Fig.~\ref{fig:radial-mass}, is broadly similar to the maximum masses of MW clusters. At distances $\lesssim$5 kpc, massive clusters are removed by dynamical friction while at distances larger than 20 kpc the birth pressures limit the formation of massive clusters. This leads to the most massive clusters in the $L^\star$ galaxies typically being found at 3--10~kpc.
In the outer radial bin ($r \sim 100 \kpc$), the most massive GC observed in the MW is more massive than the GCs found at this distance in most of the simulations. However, the MW only has one cluster (NGC 2419) at this distance that is more massive than $10^5 \Msun$ and the next most massive cluster has a mass of $4 \times 10^4 \Msun$. In the other radial bins the second most massive cluster is of a similar mass to the most massive cluster. This illustrates that at large galactocentric distances the maximum cluster mass becomes highly stochastic. At radii less than 5~kpc, all simulations predict the existence of clusters more massive than is observed, which we attribute to the insufficient mass-loss of massive clusters in the simulations.

The simulations omitting cluster formation physics are shown in the right panel of Fig.~\ref{fig:radial-mass}. While the overall shape of the maximum cluster mass as a function of radius is similar between the simulations and the relation observed in the MW, with a peak at 3--10~kpc, the modelled maximum cluster masses are about an order of magnitude higher than the observed ones. This is a direct consequence of using an infinite ICMF truncation mass, showing that the combination of stochastic sampling and dynamical friction is insufficient to explain the absence of clusters $M>3\times10^6~\Msun$ in the MW. The fact that the maximum cluster mass is not a flat function of galactocentric radius is caused by two factors. At small radii, the most massive clusters are destroyed by dynamical friction. At large radii ($>20\kpc$), the mean maximum cluster mass does not necessarily change, but the scatter increases substantially compared to the galactic centre. This reflects the increased stochasticity of sampling high cluster masses in the absence of an upper truncation of the cluster mass function.

In summary, both Fig.~\ref{fig:radial-props} and~\ref{fig:radial-mass} confirm a key result obtained from the earlier sections in this paper, i.e.~that it is necessary to include a physically-motivated model for cluster formation physics when aiming to model the $z=0$ properties of the GC population. Omitting such a model, which is equivalent to particle tagging, leads to modelled cluster populations that are in qualitative disagreement with the properties of the MW GC population.


\section{Summary} \label{sec:summary}

We have introduced the E-MOSAICS project: a suite of cosmological, hydrodynamical simulations that couple the semi-analytic MOSAICS model of star cluster formation and disruption to the EAGLE galaxy formation model. We believe this is the first attempt to model the co-formation and co-evolution of galaxies and their star cluster populations over all of cosmic history in fully cosmological, hydrodynamical simulations.

Because the resolution of cosmological simulations of the galaxy population is generally insufficient to resolve star clusters, MOSAICS adopts a semi-analytic approach, in which the initial and evolving properties of clusters are governed by analytic expressions that depend on ambient quantities resolved by the numerical simulation to which it is coupled. MOSAICS includes models for star cluster formation, providing the fraction of star formation occurring in bound stellar clusters and the high-mass truncation for the cluster mass function, both of which are determined from the local physical properties of the natal gas at the site of star formation. Once formed, cluster populations undergo evolution and disruption via stellar mass-loss, tidally-limited two-body relaxation, tidal shocks and dynamical friction, where the dynamical mass-loss is determined by the local gravitational tidal field at the clusters' location in the numerical simulation. The advantage of this approach is that it requires fewer limiting approximations than fully analytic or semi-analytic approaches, whilst still being sufficiently computationally efficient to allow the populations of many galaxies to be followed from early cosmic epochs to the present day.

In this reference paper, we present the first set of cosmological zoom-in simulations of $\numgal$ MW-like, $L^\star$ galaxies in the E-MOSAICS project and discuss the co-formation and co-evolution of the galaxies and their star cluster populations. The principal findings of this work are as follows.
\begin{enumerate}
\item The clusters formed in the E-MOSAICS simulations in galaxies at low redshift are broadly compatible with observations of young clusters in nearby disc galaxies, demonstrating the ability of the model to predict star cluster properties from the properties of interstellar gas in the simulated galaxies. The mean CFE, which traces the gas pressure, decreases from $>10$ per cent at radii $<2 \kpc$ to a few per cent at $>4 \kpc$. The mean ICMF truncation mass, $\Mcstar$, is approximately constant with radius at $z=0$, exhibiting large scatter both between and within galaxies. This stems from the (sub-grid) molecular cloud masses being limited by Coriolis and centrifugal forces (through the epicyclic frequency $\kappa$) at the galactic centre ($<2\kpc$) and stellar feedback-limited at all other radii. We thus find that the maximum cluster mass does not simply follow from stochastic sampling statistics regulated by the SFR, but is set by environmentally-dependent, physical limits.

\item The simulations predict that GCs (i.e.~clusters with masses $>10^5~\Msun$) in $L^\star$ galaxies should be predominantly old, with formation redshifts $z \gtrsim 1$. This occurs due to the higher gas pressures and surface densities in the early Universe, which cause the CFE and $\Mcstar$ to increase with redshift and peak at $z=1$--$4$. Together, the evolution of the CFE and $\Mcstar$ with redshift impose limits on when massive clusters can (mostly) form during galaxy formation. This lends support to the hypothesis that GCs are the surviving population of clusters forming at early cosmic times of which the formation is reminiscent of YMCs observed in the local Universe.

\item The formation of massive star clusters at low redshift requires an elevation of the gas pressure, from its typical low-redshift values  ($P/k=10^3$--$10^5$ K cm$^{-3}$) to those more characteristic of the interstellar medium of galaxies at high redshift ($P/k=10^5$--$10^8$ K cm$^{-3}$). As a result, massive clusters rarely form in the local Universe, but their formation rates in $z=0$ galaxies are boosted during galaxy mergers. Few of the galaxies in our sample host young massive clusters at the present day.

\item We find a connection between the formation of star clusters and the overall stellar mass assembly of galaxies. Specifically, within a sample of galaxies with similar present-day stellar masses, those that form earlier and with lower metallicity form star clusters more efficiently, yielding populations that extend to higher cluster masses than galaxies that form later.

\item The strength of tidal heating, which governs cluster disruption by tidal shocks, varies strongly with environment and is greatest in star-forming regions. In general, this causes clusters to lose mass most rapidly in the gas-rich host galaxy disc in which they formed, which for GCs is at early cosmic times. Cluster disruption, as quantified by the evolution of the tidal strength and the tidal heating parameter, is most efficient during the peak of cluster formation from redshifts $1 \lesssim z \lesssim 4$, because the characteristic ambient gas density of clusters at birth peaks during this epoch.

\item The rate of cluster disruption by tidal shocks is a strong function of the ambient gas density in the immediate vicinity of the cluster. The greatest ambient density a cluster experiences is almost universally that of its natal gas. The mass function of clusters born at the highest pressures and densities realised by the simulations ($P/k \gtrsim 10^6 \K \cmcubed$; $n_\rmn{H} \gtrsim 20 \cmcubed$) evolves from the initial power-law form to a peaked log-normal distribution, similar to the observed GCMF of the MW GC population. However, mass functions of clusters formed from lower-pressure gas do not evolve so markedly, such that in general the simulations over-predict the number density of low-mass clusters, since tidal shocks are much less effective at disrupting clusters in this regime. We attribute this lack of disruption at low pressures to the absence of an explicit cold, dense interstellar gas phase in the EAGLE model. This numerical shortcoming of the current setup will need to be addressed in a future generation of models.

\item The high-mass end of the cluster mass functions realised by the simulations is generally compatible with that of the observed GCMF, with the most massive surviving cluster typically exhibiting a mass in the interval $10^6$-$10^7~\Msun$. The high-mass end of the cluster mass function is primarily shaped by dynamical friction, since the most massive clusters tend to form from high pressure gas within galactic centres. The observed mass of the most massive GC in the MW as a function of galactocentric radius is also broadly reproduced by the simulations. The most massive present-day cluster with age $>6~\Gyr$ is typically found 3--10~kpc from the galactic centre in each simulation.

\item At small ($<2 \kpc$) and large ($>20\kpc$) galactocentric radii at $z=0$, predicted GC (i.e.~clusters with masses $>10^5 \Msun$, ages $>6 \Gyr$) metallicities agree reasonably well with those in the MW. At intermediate radii the simulated cluster populations are too metal-rich, which we attribute to the insufficient disruption by tidal shocks in the simulations. For the same age and mass limits, clusters in simulations with the formation models off (i.e.~constant CFE and infinite $\Mcstar$) overpredict metallicities at all radii.
\end{enumerate}

We infer from these findings that it is not necessary to invoke separate mechanisms for the formation of star clusters at different epochs: the properties of both the populations of young clusters and old GCs can be reproduced by a model incorporating a single cluster formation mechanism. Differences between the young and old cluster populations are driven by the evolution of the characteristic pressure of star-forming interstellar gas as a function of cosmic time, with the conditions necessary for the formation of massive clusters being relatively common in the early Universe, whereas it is typically only realised in galaxy mergers at late times. This evolution of characteristic gas properties also acts to partially self-regulate the survival of high-mass clusters, since the conditions necessary for their formation are also those that are required to disrupt them through tidal shocks. Because more massive clusters are more likely to survive cluster disruption, the surviving population of old GCs has a higher characteristic mass scale than clusters forming at $z=0$. These findings also highlight the necessity of physically-motivated treatments of cluster formation and evolution when modelling the globular cluster population, since the cluster formation rate does not simply follow from the star formation rate.

We have shown that low-mass clusters are not disrupted in the E-MOSAICS simulations with the efficiency necessary to shape a power-law ICMF into the log-normal form exhibited by the MW GCMF. We demonstrate that the efficiency of disruption by tidal shocks is a strong function of the ambient density of star-forming gas, and infer that the true disruption rate of clusters is generally underestimated by current suite of E-MOSAICS simulations. The incorporation of an explicit model for the cold, dense interstellar gas phase into our hydrodynamical simulations would remedy this shortcoming. 

We intend to follow up this reference paper with several more studies that present a broad range of predictions of the E-MOSAICS simulations. These include a companion paper discussing how GCs can be used to trace the formation and assembly history of their host galaxy through the GC age-metallicity relation \citep{Kruijssen_et_al_18}, as well as targeted studies on e.g.~cluster formation histories (Reina-Campos et al., in prep.), the GC blue tilt (Usher et al., in prep.) and predictions for observations of galaxies at the epoch of GC formation (Bastian et al., in prep.). We will also extend the simulations to larger systems of galaxy groups and clusters such that the formation and co-evolution of galaxies and their star cluster populations can be explored in the most diverse possible range of cosmic environments.

\section*{Acknowledgements}
We thank the referee for a timely and constructive report and colleagues in the EAGLE team for helpful discussions.
JP and NB gratefully acknowledge funding from a European Research Council consolidator grant (ERC-CoG-646928-Multi-Pop). JMDK gratefully acknowledges funding from the German Research Foundation (DFG) in the form of an Emmy Noether Research Group (grant number KR4801/1-1, PI Kruijssen), from the European Research Council (ERC) under the European Union's Horizon 2020 research and innovation programme via the ERC Starting Grant MUSTANG (grant agreement number 714907, PI Kruijssen), and from Sonderforschungsbereich SFB 881 ``The Milky Way System'' (subproject P1) of the DFG. NB and RAC are Royal Society University Research Fellows.  
This work used the DiRAC Data Centric system at Durham University, operated by the Institute for Computational Cosmology on behalf of the STFC DiRAC HPC Facility (\url{www.dirac.ac.uk}). This equipment was funded by BIS National E-infrastructure capital grant ST/K00042X/1, STFC capital grants ST/H008519/1 and ST/K00087X/1, STFC DiRAC Operations grant ST/K003267/1 and Durham University. DiRAC is part of the National E-Infrastructure.
The study also made use of high performance computing facilities at Liverpool John Moores University, partly funded by the Royal Society and LJMU's Faculty of Engineering and Technology.



\bibliographystyle{mnras}
\bibliography{emosaics}



\appendix

\section{Tensor method for circular and epicyclic frequencies} \label{app:kappa}

\begin{figure*}
  \includegraphics[width=\textwidth]{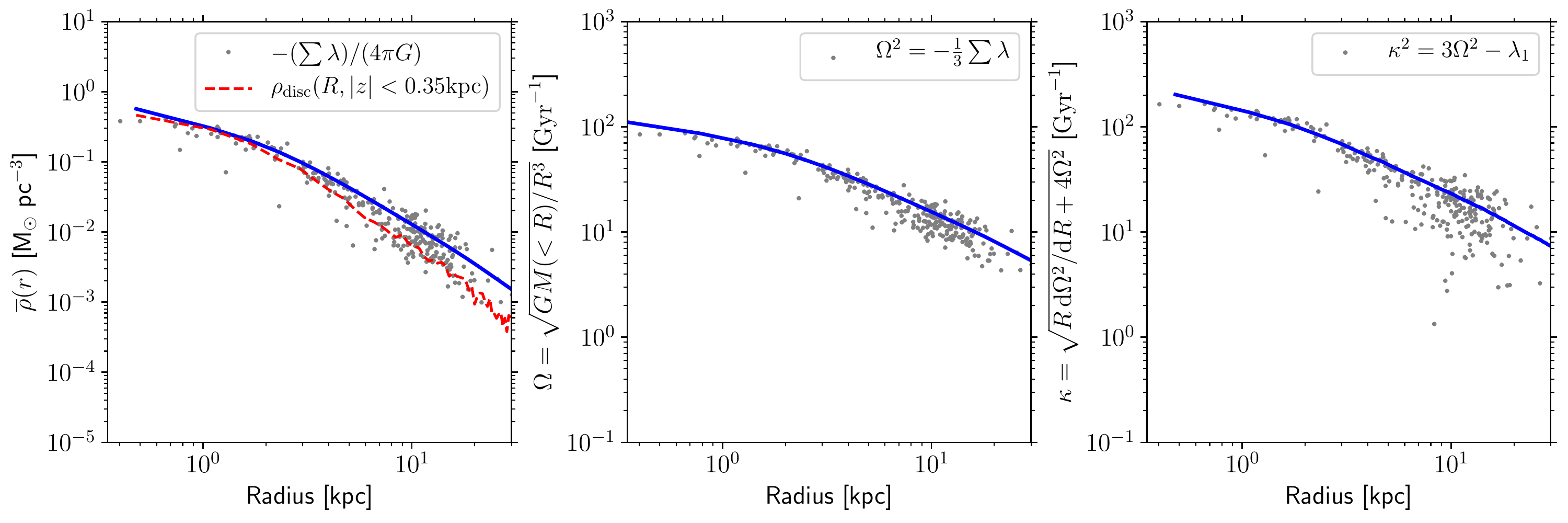}
  \caption{ 
Enclosed mean density $\mean{\rho}(r)$ (left panel), circular frequency $\Omega$ (middle) and epicyclic frequency $\kappa$ (right panel) calculated using our tidal tensor method for young stars ($<50$ Myr old; grey points), compared with that obtained by projecting the galaxy (solid line) for the simulated galaxy Gal009 at $z=0$. In the left panel we also show the mean density in the disc ($z<0.35$kpc) within cylindrical annuli about the galaxy (red dashed line). The agreement of the tidal tensor method for $\Omega$ and $\kappa$ with that obtained by projecting the galaxy is very good over the full radial range. }
  \label{fig:kappa}
\end{figure*}

In this section we derive equations for the circular ($\Omega$) and epicyclic frequencies ($\kappa$) from the tidal tensor at any given position within the simulations. These relations are used for calculating the Toomre mass (Appendix \ref{app:Sigma}) and tidal field strength (Appendix \ref{app:tidal_field}).

The circular frequency is defined as $\Omega \equiv v_c/r$ \citep{Binney_and_Tremaine_08}, where $v_c$ is the circular speed at radius $r$, and therefore may be written
\begin{equation}
\Omega^2 = \frac{v_c^2}{r^2} = \frac{G M(r)}{r^3} ,
\end{equation}
with $M(r)$ the mass enclosed within radius $r$.
This may also be written in terms of the mean density enclosed within the radius $r$, $\mean{\rho}(r)$, giving
\begin{equation} \label{eq:Omega_rho}
\Omega^2 = \frac{4}{3} \pi G \mean{\rho}(r).
\end{equation}

The eigenvalues of the tidal tensor are related to Poisson's equation by $\nabla^2 \Phi = 4\pi G \rho = -\sum_i \lambda_i$.
In Fig. \ref{fig:kappa} we compare the enclosed density (blue line), the density in the star-forming disc (red dashed line) and the sum of the tidal tensor eigenvalues for young stars ($<50$ Myr, grey points) for Gal009 (chosen simply because star formation spans the full radial range in the figures, particularly in the galactic centre). As the stars formed in overdensities relative to the mean density at a given radius, the points for young stars are elevated above $\rho_\rmn{disc}$. However the sum of the eigenvalues provides an very good fit to the mean enclosed density. We have verified this holds for all galaxies in our sample, as well as a range in galaxy stellar masses ($10^8-10^{10.5} \Msun$).
Therefore we simply calculate the enclosed density as
\begin{equation} \label{eq:enc_dens}
4 \pi G \mean{\rho}(r) = - \sum_{i} \lambda_i .
\end{equation}
When calculating tidal tensors for $\mean{\rho}$ we use a differentiation interval of the gas smoothing length. This is necessary to reduce local contributions to the tidal tensor that otherwise results in further deviations at large galactocentric distances ($\gtrsim 10$ kpc), while keeping the differentiation interval at the galactic centre small such that $\mean{\rho}$ is not decreased further than that caused by the softening length.

From our derived relation for $\mean{\rho}$, we finally derive expressions for $\Omega$ and $\kappa$.
Combining Eqs. \ref{eq:Omega_rho} and \ref{eq:enc_dens}, the expression for the circular frequency becomes
\begin{equation} \label{eq:Omega}
\Omega^2 = - \frac{1}{3} \sum_{i} \lambda_i .
\end{equation}
Near the equatorial plane in an axisymmetric potential the epicyclic frequency can be written as \citep{Binney_and_Tremaine_08}
\begin{equation}
\kappa^2 (R) = 3 \Omega^2 + \left( \pder[2]{\Phi}{R} \right) .
\end{equation}
In an axisymmetric system, the first (maximal) of the tidal tensor is given by $\lambda_1 = - \partial^2 \Phi/\partial R^2 $.
Therefore we obtain an expression for the epicyclic frequency in terms of the eigenvalues of the tidal tensor:
\begin{equation} \label{eq:kappa}
\kappa^2 = -\left( \sum_{i} \lambda_i \right) - \lambda_1.
\end{equation}
In the middle and right panels of Fig. \ref{fig:kappa} we show how our method for calculating $\Omega$ and $\kappa$ through the tidal tensor compares with the standard definition for Gal009.
As for the enclosed density, $\Omega$ and $\kappa$ are in good agreement with that calculated in post-processing.

\section{Local calculation of gas surface density and Toomre mass} \label{app:Sigma}

\begin{figure*}
  \includegraphics[width=\textwidth]{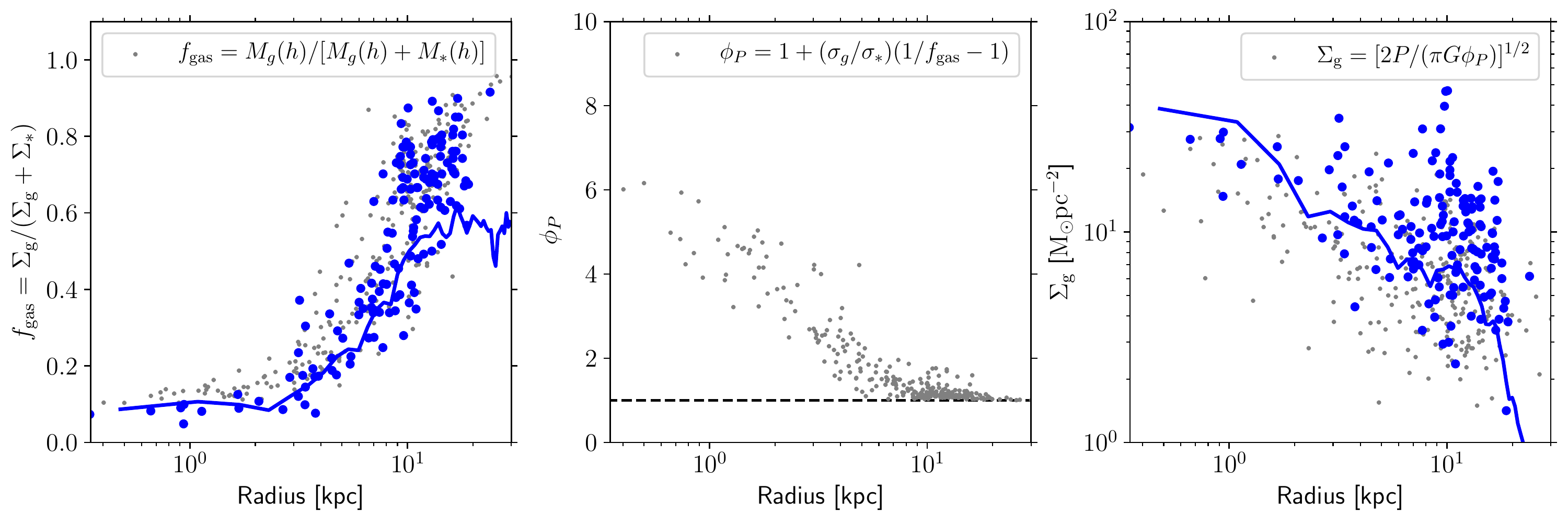}
  \caption{ Gas fraction, $\phi_P$ and gas surface density calculated locally (as indicated by the legends) for the simulated galaxy Gal009 at $z=0$ for stellar particles younger than 50 Myr (grey points). Solid lines show the results for linearly-spaced annuli of the projected galaxy and large filled circles show the values calculated for the 2000 baryonic particles projected closest around 150 randomly chosen star formation-eligible gas particles (i.e.~those within 0.5 dex of the temperature floor, see Section \ref{sec:eagle}). }
  \label{fig:Sigma}
\end{figure*}

\begin{figure*}
  \includegraphics[width=\textwidth]{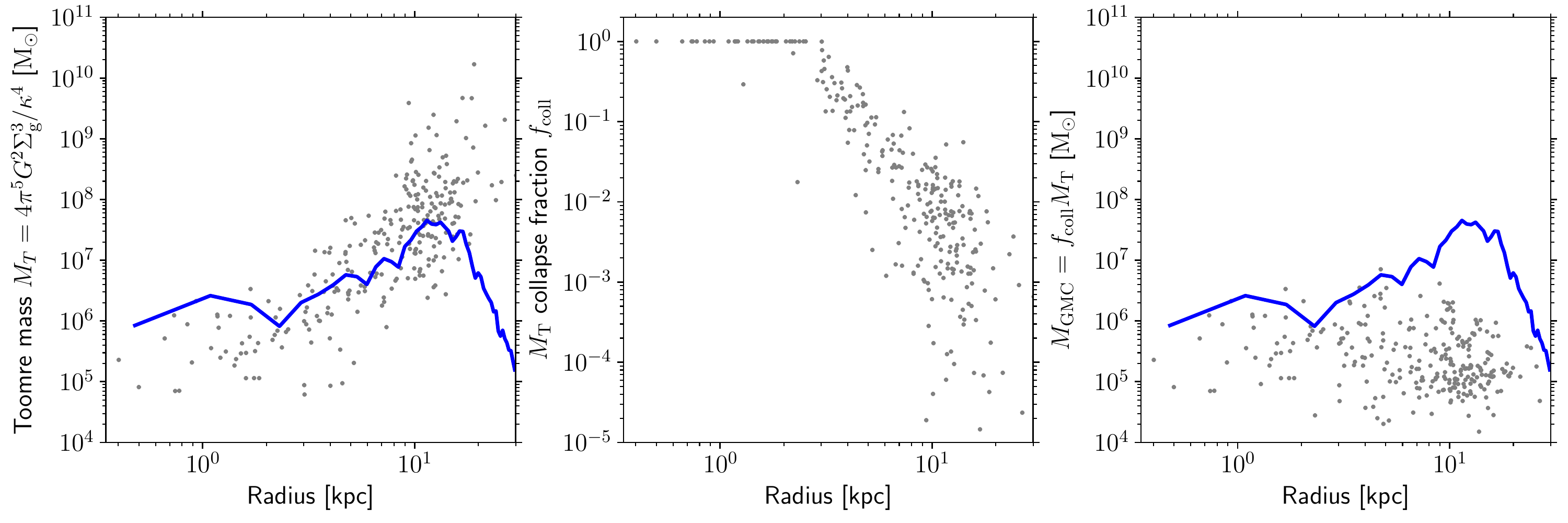}
  \caption{ Toomre mass, mass collapse fraction and molecular cloud mass for Gal009. The solid line shows the Toomre mass calculated in linearly-spaced annuli around the centre of the galaxy.}
  \label{fig:MToomre}
\end{figure*}

In this section we demonstrate the accuracy of our method (Section \ref{sec:clform}) for determining the gas surface density ($\Sigma_g$, Eq. \ref{eq:Sigma_g}) and Toomre mass ($\Mtoomre$, Eq. \ref{eq:Mtoomre}) from local variables in the simulation. 
Gas surface density is calculated from the local pressure assuming hydrostatic equilibrium (Eq. \ref{eq:Sigma_g}) and requires determination of the parameters $\phi_P$ and $f_\rmn{gas}$ (Eq. \ref{eq:phi_P}). 
The left panel of Fig. \ref{fig:Sigma} compares the local (i.e.~within a smoothing kernel as described in Section~\ref{sec:clform}) determination of $f_\rmn{gas}$ at the time of star formation (grey points; for stars with ages $<50$ Myr to limit any radial migration) with values calculated in projection about star-forming gas particles (blue filled circles) for Gal009 at $z=0$.
The local calculation of $f_\rmn{gas}$ agrees very well with the projected values, showing an similar trend with radius and with a comparable level of scatter. At small radii ($<3\kpc$) gas accounts for 10 per cent of the mass and beyond this the fraction increases with radius up to 90 per cent at $20\kpc$. At $<10\kpc$ the locally calculated values trace the mean values calculated in annuli (solid line). Beyond this, the gas fraction of star-forming regions increases while the mean value remains a constant $\sim55$ per cent.

The middle panel of Fig. \ref{fig:Sigma} shows the locally-determined $\phi_P$ as a function of radius in the galaxy. $\phi_P$ decreases from $\sim$6 at the galactic centre to $\sim$1 at radii $>10\kpc$. This range agrees well with the expected range determined by \citet{Krumholz_and_McKee_05}. Recently, \citet{Johnson_et_al_16} determined $\phi_P$ for the disc in M31, finding a range from 1.4 (outer disc) to 5.6 (inner disc), which is also in good agreement with our estimates.

In the final panel of Fig. \ref{fig:Sigma}, we compare the local calculation of $\Sigma_g$ with that calculated in projection. 
In comparison with the values calculated around star-forming gas particles, our local calculation underestimates $\Sigma_g$ by a factor $\sim$2. However, at 10 kpc the scatter in $\Sigma_g$ between both methods is of a similar order, suggesting the scatter in the local calculation represents a physical, rather than numerical, scatter. The underestimation may be caused by the assumption that the pressure of the SPH particle represents the mid-plane pressure, which will tend to underestimate the mid-plane pressure.
The locally-determined values provide a better approximation to the radially averaged value of $\Sigma_g$ (solid line), showing a very similar trend with radius. Only at radii $<2\kpc$, the local calculation underestimates the true value by a factor $\sim 1.5$.

Overall, our local method for locally determining $\Sigma_g$ provides a reasonable approximation, if slightly underestimating the true value. Therefore we now apply it in the local calculation of the Toomre mass. We also compare the Toomre mass, $\Mtoomre$, with the maximum molecular cloud mass, $\Mgmc$, used to set the maximum cluster mass scale (Eqs. \ref{eq:Mcstar} and \ref{eq:Mgmc}).

In the left panel of Fig. \ref{fig:MToomre}, we use the $z=0$ snapshot of galaxy Gal009 to compare the local calculation of the Toomre mass (grey points) with the azimuthally-averaged Toomre mass calculated in projection as a function of galactocentric radius (solid blue line; with the epicyclic frequency $\kappa^2 = R {\rm d}\Omega^2 / {\rm d}R + 4 \Omega^2$ and the circular frequency $\Omega^2 = G M(R) / R^3$).
The Toomre mass in Gal009 increases from $\sim10^6 \Msun$ at $<1\kpc$ to $5\times10^7 \Msun$ at 10 kpc. Beyond 15 kpc $\Mtoomre$ decreases rapidly as the gas surface density decreases (Fig. \ref{fig:Sigma}). 
The locally-calculated $\Mtoomre$ is typically 0.5 dex too low compared to the projected value at $<2$ kpc, but in good agreement at larger radii.  This stems from the underestimation of $\Sigma_g$ at small radii.
At large radii ($\sim 10 \kpc$) the locally calculated values for $\Mtoomre$ show a significant scatter to high masses, due to the scatter in the calculation of $\kappa$ (Section \ref{app:kappa}; recall $\Mtoomre$ scales with $\kappa^{-4}$, so a small error in $\kappa$ can result in a large error in $\Mtoomre$). 

In the middle and right panels of Fig. \ref{fig:MToomre}, we show the Toomre collapse fraction, $f_\rmn{coll}$, and maximum molecular cloud mass, $\Mgmc = f_\rmn{coll} \Mtoomre$. Recall that $f_\rmn{coll} = 1$ indicates masses limited by the epicyclic frequency $\kappa$ and $f_\rmn{coll} < 1$ indicates feedback-limited masses.
The collapse fraction shows a very strong trend with radius: At radii $<2\kpc$ $\Mgmc$ is limited by $\kappa$, while beyond this $\Mgmc$ is feedback-limited and $f_\rmn{coll}$ decreases rapidly with radius. 
This results in an $\Mgmc$ (right panel) that decreases mildly with radius in the galaxy from $10^6 \Msun$ at 1 kpc to $10^5 \Msun$ at 10 kpc.
In the case that $\Mgmc$ becomes feedback-limited, $\kappa$ cancels from the relation (since $f_\rmn{coll} \propto \kappa^4$ while $\Mtoomre \propto \kappa^{-4}$). Therefore the large scatter in $\Mtoomre$ at large radii is not propagated to $\Mgmc$, and the scatter in $\Mgmc$ mainly stems from $\Sigma_g$.

\section{Tidal field strength} \label{app:tidal_field}

\begin{figure}
  \includegraphics[width=84mm]{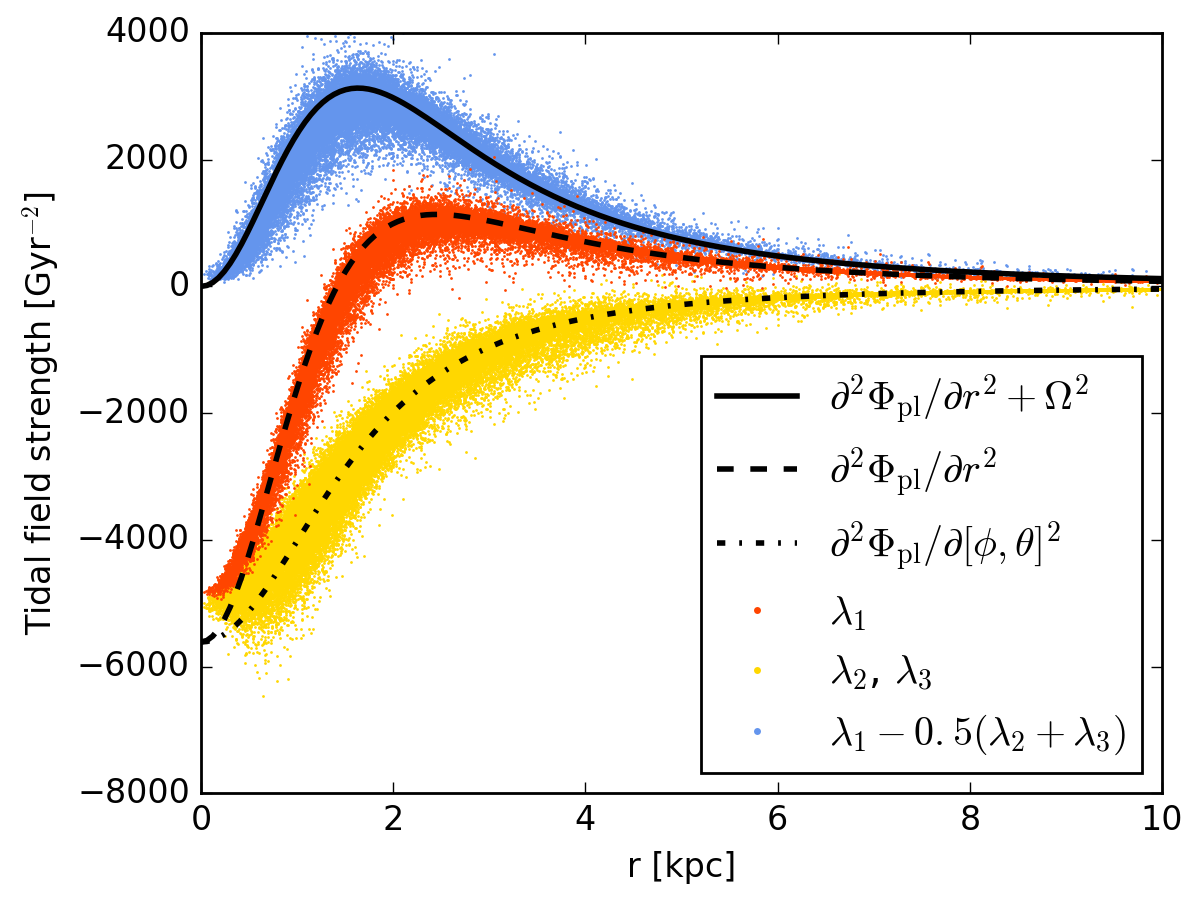}
  \caption{ Tidal field strength for a Plummer sphere. Red and yellow points show the radial ($\lambda_1$) and tangential ($\lambda_2$, $\lambda_3$) eigenvalues of the tidal tensor, which follow the second derivatives of the potential in the radial (dashed line) and tangential (dash-dotted line) directions, respectively. The tidal field strength in spherically symmetric systems $\partial^2\Phi/\partial r^2 + \Omega^2$ (solid line) can be determined from the eigenvalues of the tidal tensor as $\lambda_1 - \frac{1}{3}\sum\lambda = \lambda_1 - \frac{1}{3}(\lambda_1 + 2 \lambda_2)$ (blue points).}
  \label{fig:plummer_TFS}
\end{figure}

\begin{figure}
  \includegraphics[width=84mm]{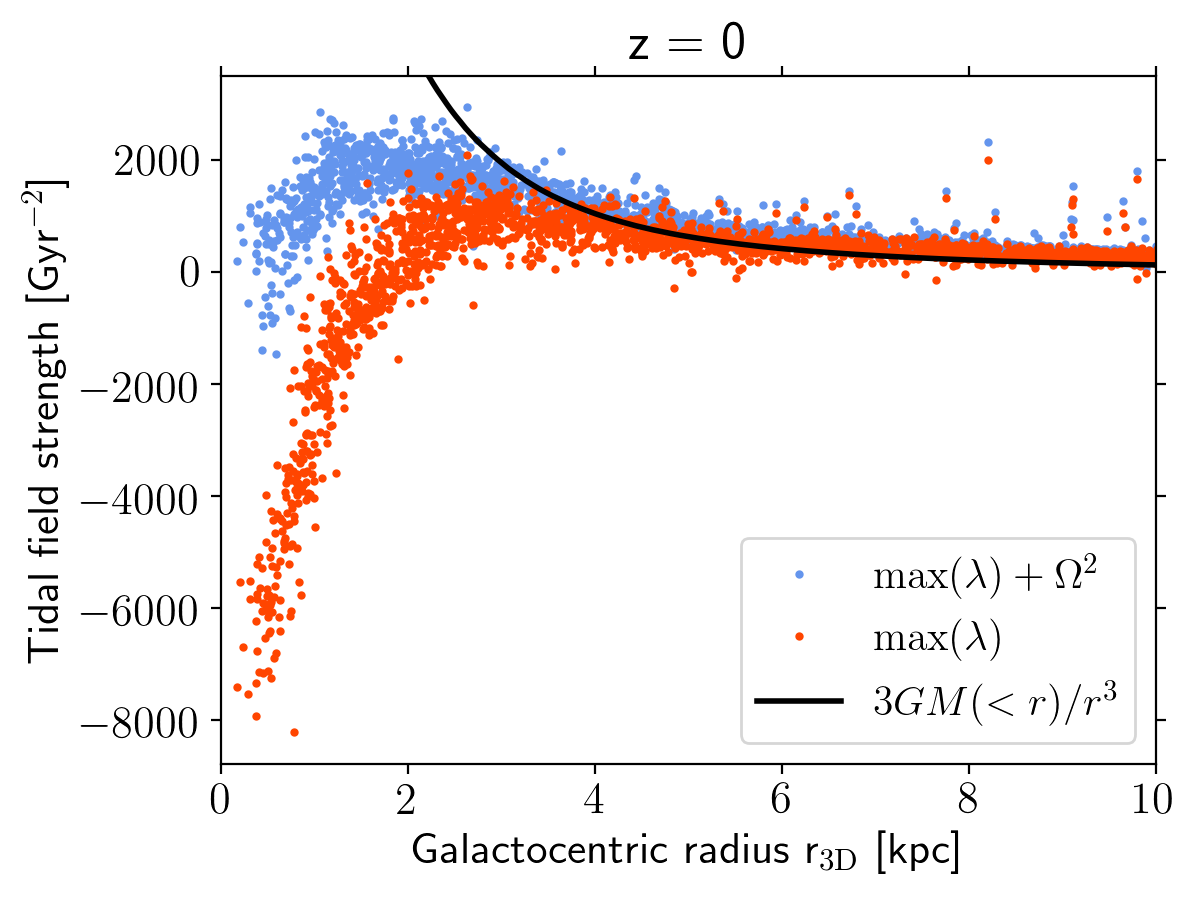}
  \caption{Tidal field strength with and without the inclusion of the circular frequency term $\Omega^2$ for Gal004 at $z=0$. For reference, the solid line shows the tidal field strength (including the $\Omega^2$ term) for a point mass with a mass equivalent to the enclosed mass at each radius. The factor $\Omega^2$ is mainly important within $5 \kpc$ of the galactic centre and has limited effect at larger radii.}
  \label{fig:tidal_field}
\end{figure}

\citet{King_62} found that the tidal field strength that sets the tidal radius of a cluster on a circular orbit is given by $\partial^2\Phi/\partial r^2 + \Omega^2$ \citep[see also][]{Renaud_Gieles_and_Boily_11}. In the previous work with the MOSAICS model \citep{Kruijssen_et_al_11, Kruijssen_et_al_12}, the tidal field strength $T$ (which sets cluster mass-loss by two-body relaxation) was taken to be the maximum eigenvalue of the tidal tensor $T = \max(\lambda)$. 
We first test this method for a Plummer model, given by the potential
\begin{equation}
\Phi_\rmn{pl} = -\frac{GM}{\sqrt{r^2+r_c^2}} ,
\end{equation}
where $M$ is the total mass and $r_c$ the scale radius. 
The tidal field strength for a Plummer model is given by
\begin{equation} \label{eq:Plummer_TFS}
\pder[2]{\Phi_\rmn{pl}}{r} + \Omega^2 = \frac{G M (r/r_c)^2}{r_c^3 \left[ 1+(r/r_c)^2 \right]^{5/2}} ,
\end{equation}
which we show as the solid line in Fig. \ref{fig:plummer_TFS}.
The Plummer model in this test has $10^5$ particles and was chosen to have a mass of $10^{10} \Msun$ and scale radius of $2 \kpc$. We used a gravitational softening length of $0.35 \kpc$, as in our fiducial cosmological simulations. 
The red ($\lambda_1$) and ($\lambda_2$, $\lambda_3$) yellow points in Fig. \ref{fig:plummer_TFS} show the eigenvalues of the tidal tensor calculated for each particle in the inertial frame. We find that the calculated eigenvalues correspond to the tidal tensor in a spherical coordinate system: $\lambda_1$ corresponds to the radial component $\partial^2\Phi_\rmn{pl}/\partial r^2$ (dashed line) while $\lambda_2$ and $\lambda_3$ correspond to the tangential components $\partial^2\Phi_\rmn{pl}/\partial [\phi,\theta]^2$ (dash-dotted line).
However, $\max(\lambda)$ (red points) does not correspond to the expected tidal field strength (solid line). At radii $r<r_c/\sqrt{2}$ the maximum eigenvalue $\partial^2\Phi_\rmn{pl}/\partial r^2 < 0$, in which case the cluster mass-loss rate by relaxation would be set to zero (Section \ref{sec:clevo}).

Therefore, in order to calculate the tidal field strength which sets the tidal radius of a cluster one must also account for the term $\Omega^2$ in addition to $\partial^2\Phi_\rmn{pl}/\partial r^2=\max(\lambda)$.
In Eq. \ref{eq:Omega} we derived the circular frequency in an axisymmetric system determined from the eigenvalues of the tidal tensor. In a spherically symmetric system it is given by $\Omega^2 = -\lambda_2 = -\lambda_3 $, which we calculate from the simulation as $\Omega^2 = -0.5(\lambda_2 + \lambda_3)$ for numerical reasons ($\lambda_2$ and $\lambda_3$ are not necessarily identical in a simulation). 
We show the tidal field strength calculated from the eigenvalues of the tidal tensor as the blue points in Fig. \ref{fig:plummer_TFS}, which show good agreement with the theoretical relation (solid line, Eq. \ref{eq:Plummer_TFS}). Therefore, for a Plummer model, the tidal field strength is always positive and the cluster mass-loss rate by relaxation always greater than zero.

In view of the above results, the updated version of the MOSAICS model includes the circular frequency term, setting $T = \max(\lambda) + \Omega^2$. The circular frequency was calculated via the tidal tensor according to Eq. \ref{eq:Omega}. We compare the two approaches in Fig. \ref{fig:tidal_field} for Gal004 at $z=0$. At radii $\gtrsim 5 \kpc$, the methods are nearly equivalent since $\max(\lambda) > \Omega^2$. However the methods diverge at $< 2 \kpc$ where $\max(\lambda) < 0$, while $\max(\lambda)+\Omega^2 > 0$.
This change only affects the mass-loss rate from two-body relaxation (Eq. \ref{eq:dmrlx}), making cluster evaporation most efficient at radii between 1--2 kpc in the galaxy at $z=0$, instead of 2--3 kpc in the previous formulation.

\section{Cluster disruption tests} \label{app:disruption}

\begin{figure*}
  \includegraphics[width=\textwidth]{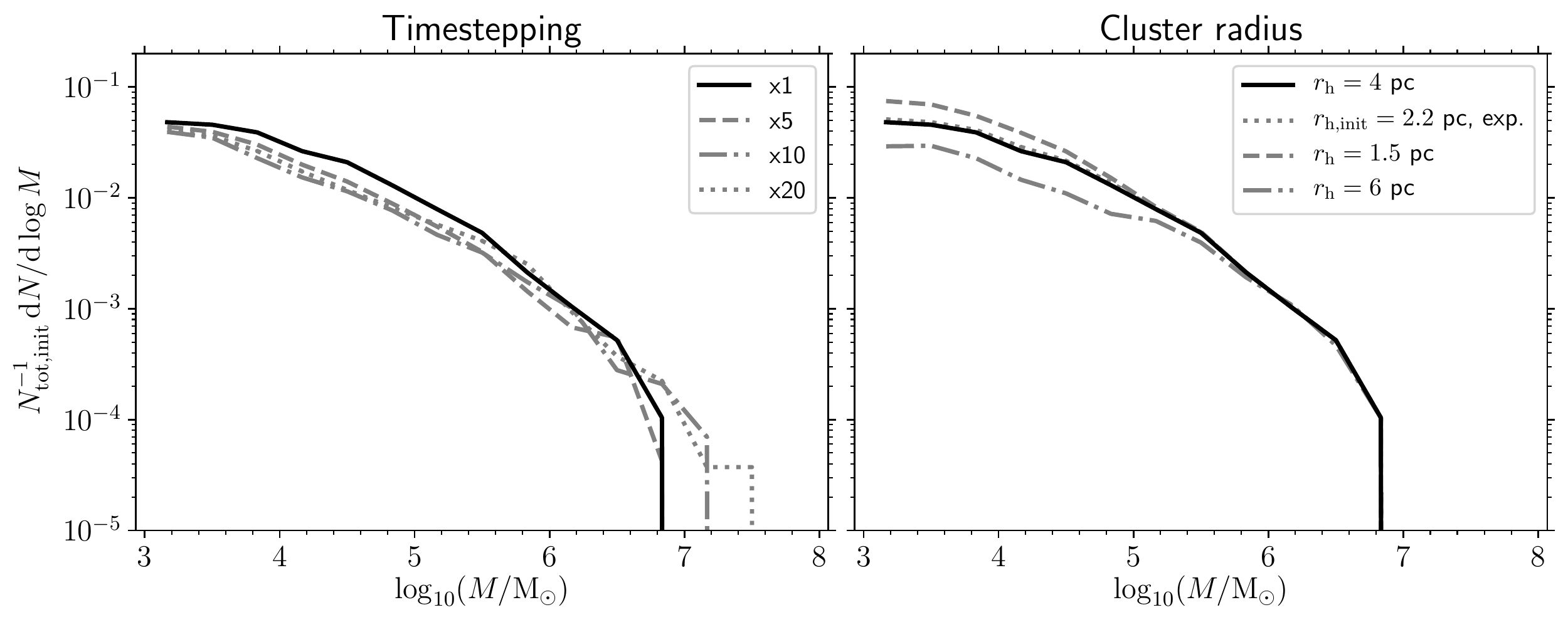}
  \caption{ Cluster mass functions at $z=0$ for clusters with ages $> 6 \Gyr$, normalized to the total initial population $N_\rmn{tot,init}$. Black solid lines show the fiducial simulations. The left panel shows particle timestepping tests. Timestepping mainly affects the mass-loss of clusters with final masses between $10^4$-$10^{5.5} \Msun$. Using more than 10 times the timesteps of the fiducial simulation ($\times$1) has no further effect on cluster mass-loss. The right panel shows cluster radii tests. Including cluster radius evolution due to stellar evolution mass-loss (dotted line; see Fig. \ref{fig:age-size} for the radius evolution as a function of cluster age) has little effect on the final cluster population. Due to the dependence of tidal shock mass-loss (Eq. \ref{eq:dmsh}) on cluster radius, compact clusters (dashed line) show less mass-loss than the fiducial simulation, while more extended clusters show higher mass-loss (dash-dotted line). }
  \label{fig:disrupt_tests}
\end{figure*}

\begin{figure}
  \includegraphics[width=84mm]{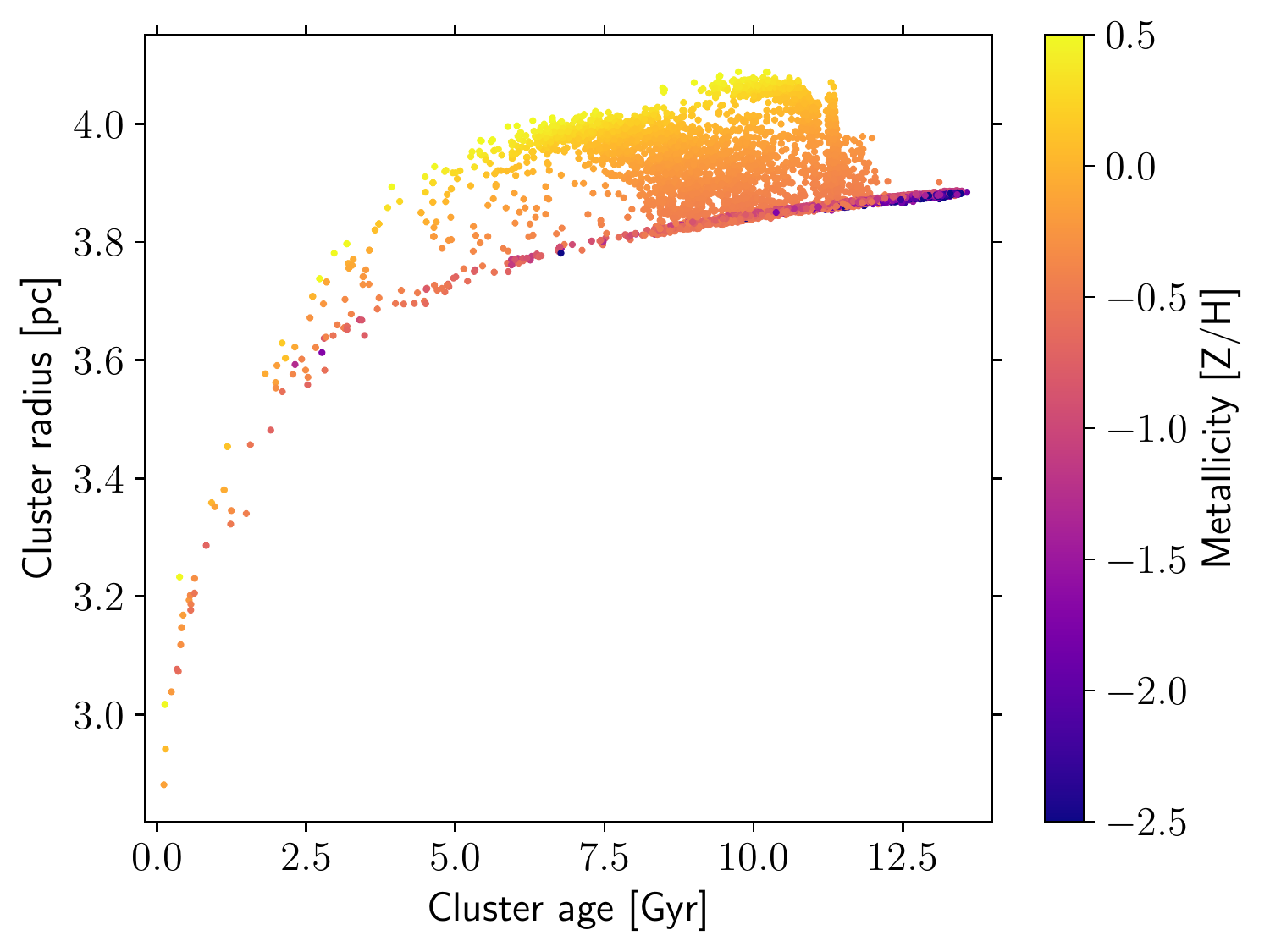}
  \caption{ The age-radius relation for clusters with initial radii of $2.2 \pc$ assuming adiabatic expansion due to mass-loss from stellar evolution. Higher metallicity clusters reach larger radii since stellar evolution mass-loss depends on metallicity in the EAGLE model. }
  \label{fig:age-size}
\end{figure}

In this section we test the effect of timestepping and cluster radius on cluster mass-loss.
Due to the adiabatic correction used when calculating tidal heating (Eqs. \ref{eq:Itid} and \ref{eq:AW}), both variables may affect mass-loss.
The timestepping provides a minimum absolute timescale for a tidal shock, comprising three timesteps. This timescale results in a maximum mass for clusters to be disrupted by tidal shocks, since, for a given cluster radius, $A_\rmn{w}$ decreases with increasing cluster mass, resulting in weaker shocks.

The effect of timestepping is shown in the left panel of Fig. \ref{fig:disrupt_tests} where we show the cluster mass function (scaled to the total initial cluster number) at $z=0$ for old clusters ($>6 \Gyr$) for different stellar particle timesteps. Dynamical friction was not included so that dynamical mass-loss at the high-mass end of the mass function can be compared. Here we only increase the number of timesteps for stellar particles while keeping gas and dark matter timesteps at the standard resolution. Note that the galaxy star formation history differs slightly between the runs, which mainly affects the very high mass end ($\gtrsim5\times10^5 \Msun$) of the mass function (particularly the $\times$10 run) due to stochasticity. 

The main effect of increasing the timestepping is to deplete the cluster mass function between $10^4$-$10^{5.5}\Msun$, relative to the standard run. This effect is maximal at $\approx3\times10^4$ and decreases the mass function by 0.35 dex between the $\times$1 and $\times$10 runs. Increasing the timestep resolution beyond the $\times10$ run has no further effect, since the $\times20$ run gives nearly identical results.
For masses $\gtrsim5 \times 10^5 \Msun$ increasing the timestepping does not significantly affect cluster mass loss.
At $z=2$, the typical stellar timesteps for the $\times$10 run are 0.05 Myr, which gives a maximum adiabatic correction of $A_{\rm w,max} \approx 0.5$ for a $10^7 \Msun$, $r_{\rm h} = 4 \pc$ cluster and $A_{\rm w,max} \approx 0.95$ for a $10^6 \Msun$ cluster. We conclude from these tests that cluster mass-loss at the high-mass end of the mass function $>10^{5.5}\Msun$ is not affected by the choice of timestepping.

In the right panel of Fig. \ref{fig:disrupt_tests} we show the effect of cluster radius on cluster mass-loss.  In the fiducial model we assume a constant radius of $r_{\rm h} = 4 \pc$ (black solid line) and we also tested constant radii of $1.5$ (dashed line) and $6 \pc$ (dash-dotted line). Here the galaxy star formation history is identical between all runs.
For the $r_{\rm h} = 1.5 \pc$ run, cluster mass-loss by tidal shocks decreases relative to the fiducial run. This is caused by the mass loss rate scaling of ${\rm d}M/{\rm d}t\propto r_{\rm h}^3$ for tidal shocks. In addition, the maximum cluster mass for which shock-driven mass loss is important decreases, because the adiabatic correction suppresses the mass loss at lower cluster masses (note that this is not significantly altered by also decreasing the timestep length). The inverse is true for the $r_{\rm h} = 6 \pc$ run, for which tidal shocks are generally more effective at disrupting clusters and also affect more massive ones, because the adiabatic correction only suppresses mass loss in the most massive clusters. The choice of cluster radius has an effect on cluster mass-loss at least as large as timestepping, particularly at the low mass end of the cluster mass function. 

We also evaluate a simple model to account for cluster expansion due to mass-loss from stellar evolution. Clusters are initially formed with radii $r_{\rmn{h,init}} = 2.2 \pc$, after which they expand assuming adiabatic expansion according to 
\begin{equation}
r_{\rm h} = r_{\rmn{h,init}} \frac{ m_{\ast,\rmn{init}} }{ m_\ast } , 
\end{equation}
where $m_{\ast,\rmn{init}}$ and $m_\ast$ are the initial and current stellar particle masses, which gives radii $\sim4 \pc$ for old clusters that have experienced most of their stellar evolutionary mass loss. The radii of clusters at $z=0$ as a function of the cluster age are shown in Fig. \ref{fig:age-size}. At a given age, the cluster radius increases with metallicity, because stellar mass-loss depends on metallicity in the EAGLE model. Cluster radius increases rapidly within the first few Gyr, reaching $3.1 \pc$ at $0.3 \Gyr$ ($\approx 40$ per cent of expansion) and $3.6 \pc$ at $2 \Gyr$ ($\approx 60$ per cent of expansion).
The effect of this simple model for cluster expansion on cluster mass-loss is shown as the dotted line in the right panel of Fig. \ref{fig:disrupt_tests}. The predicted mass function from this model is nearly indistinguishable from having a constant cluster radius of $r_{\rm h} = 4 \pc$. The main change compared to a constant cluster radius is that cluster disruption is very slightly delayed (typically by a few hundred Myr) until cluster radii increase. Therefore initial cluster expansion has little effect on the final cluster population properties.

\section{Constant star formation density threshold} \label{app:SFThresh}

\begin{figure*}
  \includegraphics[width=0.80\textwidth]{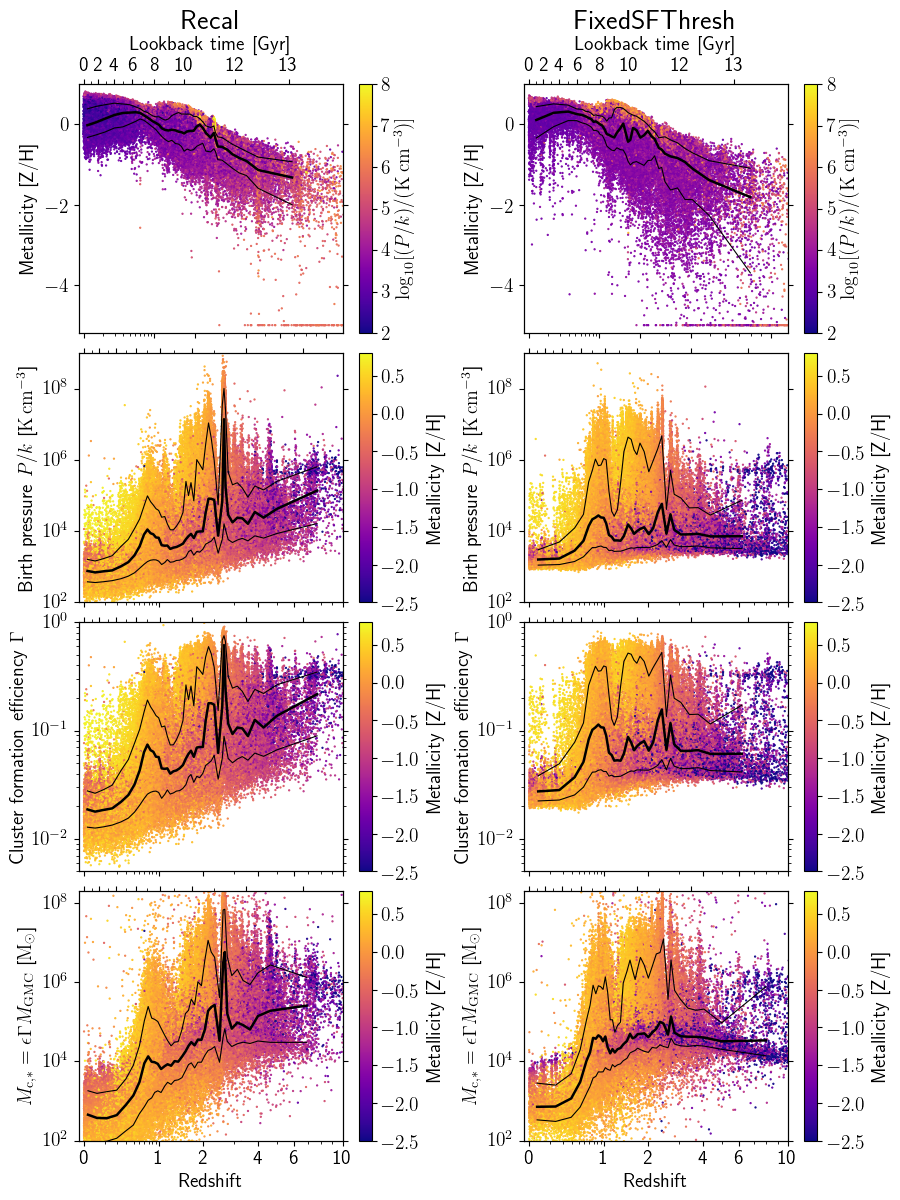}
  \caption{ Stellar particle and cluster formation properties (for particles within 100 kpc at $z=0$) for simulations of Gal004 with the standard metallicity-dependent star-formation threshold (Recal; left column) and a constant star formation density threshold of $n_H = 0.1 \cmcubed$ (SFThresh; right column). Very low metallicity particles are shown at $\ZH = -5$ dex. }
  \label{fig:SFThresh}
\end{figure*}

The standard EAGLE model adopts a metallicity-dependent gas density threshold for star formation. This threshold is motivated by the onset of the thermogravitational collapse of warm, photoionized interstellar gas into a cold, dense phase, which is expected to occur at lower densities and pressures in metal-rich gas \citep{Schaye_04}. 
In Fig. \ref{fig:SFThresh}, we compare cluster formation properties using the metallicity-dependent threshold (`Recal') with those for a constant density threshold of $n_H = 0.1 \cmcubed$ (`FixedSFThresh') for simulations of Gal004. At $z=0$ the galaxies in each simulation have very similar masses and global metallicities. A consequence of the metallicity-dependent threshold is that very few particles are formed with metallicities $\ZH < -3$ dex, while with the fixed threshold star particles are formed at very low metallicities over a much longer timescale.
Relative to a constant density threshold, star formation with the metallicity-dependent model occurs at higher pressures (through the EOS) in low metallicity gas ($\ZH < -1$) at high redshift ($z>3$). 
The major effect of changing the density threshold for star formation on cluster formation in the MOSAICS model is that the CFE ($\Gamma$) and mass function truncation mass ($\Mcstar$) are higher for low metallicity particles at $z>3$ in the Recal model than in the FixedSFThresh model.
From redshifts $z<2$, the evolution of the cluster formation properties is not significantly different between the models, since star formation in $\ZH \sim 0$ gas occurs at similar densities.

\section{ISM equation of state} \label{app:EOS}

\begin{figure*}
  \includegraphics[width=\textwidth]{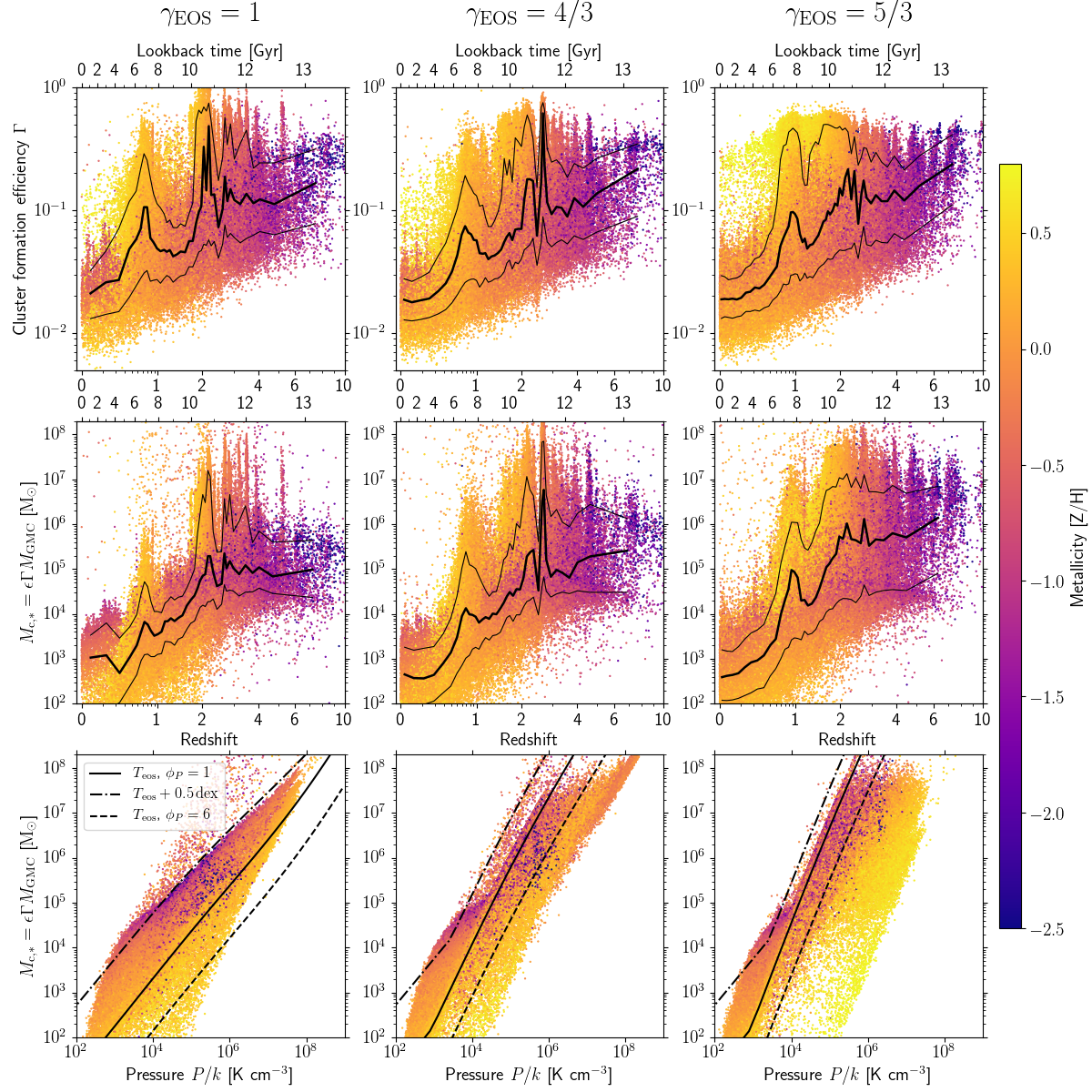}
  \caption{ Effect of the equation of state exponent, $\gamma_\rmn{EOS}$, on the star cluster formation properties CFE (top row) and MF truncation mass (middle row: redshift evolution; bottom row: compared with particle birth pressure) in Gal004. Columns from left to right show simulations with $\gamma_\rmn{EOS} = 1$ (isothermal), $\gamma_\rmn{EOS} = 4/3$ (fiducial simulation) and $\gamma_\rmn{EOS} = 5/3$ (adiabatic).}
  \label{fig:EOS_test}
\end{figure*}

In this section we discuss the effect of the equation of state (EOS) exponent, $\gamma_\rmn{EOS}$, on cluster formation properties. The pressure law scheme for star formation in EAGLE ensures that the model does not need to be recalibrated when changing $\gamma_\rmn{EOS}$ \citep[see][for discussion]{S15, C15}. Fig. \ref{fig:EOS_test} shows simulations of Gal004 with $\gamma_\rmn{EOS}=1$ (isothermal), $\gamma_\rmn{EOS}=4/3$ (standard Recal) and $\gamma_\rmn{EOS}=5/3$ (adiabatic). Note that the isothermal EOS enables higher gas densities than the standard EOS, meaning the Jeans scales are no longer resolved in this simulation.

For the CFE, the median evolution with redshift is relatively similar between all three simulations, which reflects the underlying similarity in the distribution of star formation pressures (recall that CFE depends almost entirely on gas pressure; Fig. \ref{fig:CFE}). However, the peaks in the CFE differ for each EOS exponent and tend to be higher towards low $\gamma_\rmn{EOS}$. In particular, for $\gamma=5/3$ the peak at $z\approx2.5$ is absent. This illustrates that the CFE is weakly dependent on the EOS exponent, again because it scales with pressure.

For $\Mcstar$, the median $\Mcstar$ evolution at $z<0.5$ is similar for all EOS exponents since star formation at this epoch mainly occurs at $n_\rmn{H}<10^{-1}\cmcubed$, i.e.~at densities below where the EOS takes effect. At $z>0.5$ the evolution of $\Mcstar$ differs in each case. For an isothermal EOS the peaks in $\Mcstar$ are significantly reduced relative to the fiducial simulation ($\gamma_\rmn{EOS}=4/3$), with the exception between redshifts 2--3.5 where merger-driven elevated gas pressures result in a peak of $\Mcstar$. For an adiabatic EOS, both the median and peak values of $\Mcstar$ at $z>2$ are nearly a factor of 10 higher than for the fiducial EOS. This happens because $\Mcstar$ increases steeply with the turbulent velocity dispersion ($\sigma = \sqrt{P/\rho}$) in the feedback-limited regime through $f_\rmn{coll}$ (see Eqs. \ref{eq:f_coll} and \ref{eq:t_fb}), which means that for higher values of $\gamma_{\rm EOS}$, a given variation of the density drives larger excursions in $\Mcstar$. For an isothermal EOS, at a given pressure $P$, $\rho$ is increased resulting in a $\sigma$ that is smaller relative to the fiducial runs (and conversely for the adiabatic EOS). In view of the above experiments, we conclude that the maximum cluster masses depend on the chosen EOS, particularly for metallicities $\ZH<-1$ at $z>3$

In the bottom row of Fig. \ref{fig:EOS_test}, we compare $\Mcstar$ with the particle birth pressure. The three lines in each panel show the expected relation from Eq. \ref{eq:Mcstar} for feedback-limited masses (i.e.~$f_\rmn{coll} < 1$, in which case the epicyclic frequency $\kappa$ cancels from the relation). The solid and dash-dotted lines show the relations for particles on the polytropic temperature floor ($T_\rmn{eos}$) and at 0.5 dex above the temperature floor (the limit for star formation), respectively, assuming $\phi_P=1$. The dashed line shows the relation for $T_\rmn{eos}$ with $\phi_P = 6$. For feedback-limited masses and for particles on the EOS ($n_\rmn{H}>10^{-1} \cmcubed$), the truncation mass scales with pressure as approximately $\log_{10}(\Mcstar) \propto \gamma_\rmn{EOS}^{3/2} \log_{10}(P)$.
Particles in the panels below the $\phi_P=6$ line generally correspond to masses limited by $\kappa$ ($f_\rmn{coll} = 1$), which, in the $\kappa$-limited regime, lowers $\Mcstar$ at a given pressure relative to the feedback-limited case \citep[recall the method selects the minimum of feedback-limited and $\kappa$-limited masses; e.g. see Fig. 7 in][]{Reina-Campos_and_Kruijssen_17}.

\section{Environment density} \label{app:env_density}

\begin{figure*}
  \includegraphics[width=\textwidth]{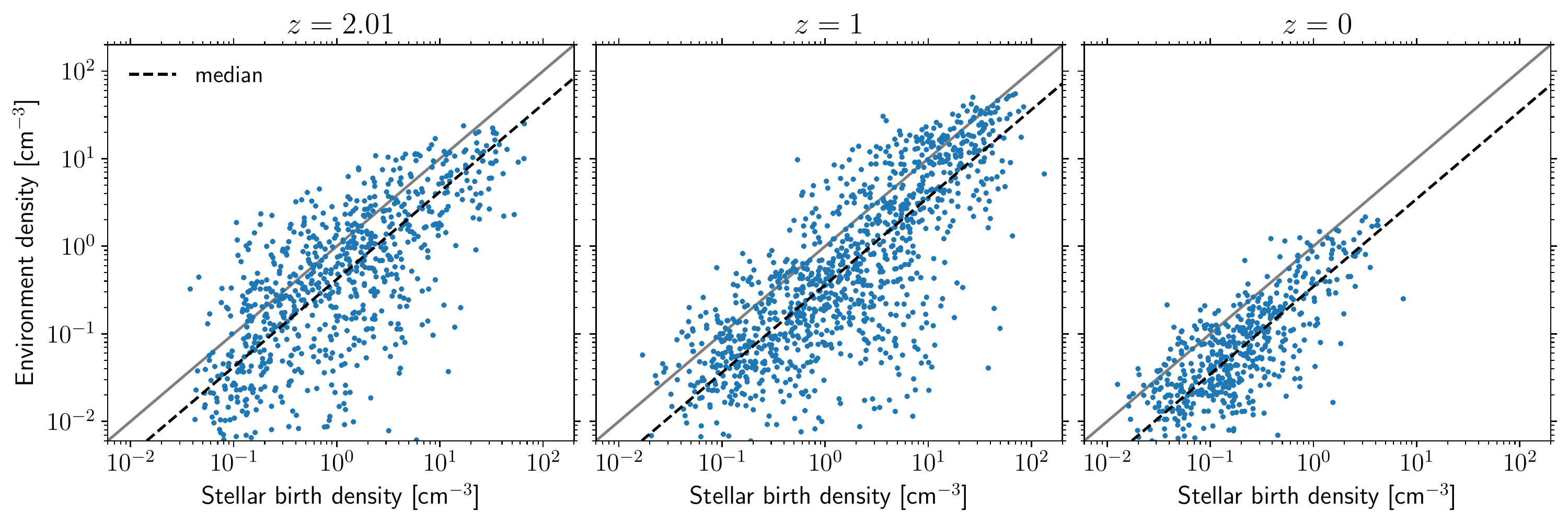}
  \caption{ Stellar particle gas birth density compared with surrounding environment gas density for young stars           ($<100\Myr$) in Gal009 at redshifts 2 (left), 1 (middle) and 0 (right panel). The density of the surrounding environment scales with birth density, but is typically a factor of $\approx3$ lower.}
  \label{fig:env_density}
\end{figure*}

In this section we test whether the natal gas density of stellar particles is related to the gas density of the surrounding environment, which young star clusters experience shortly after formation and has sets the mass-loss rate due to tidal shocks (see Section \ref{sec:GCMF}).
We estimate the typical distance young stars may be expected to travel from their formation site in a star-forming disc as $r_\rmn{env} = \pi \sigma_\rmn{g,1D} / 2 \kappa \approx \sigma_{\rmn g} / \kappa$, where $\sigma_\rmn{g,1D} \approx \sigma_{\rmn g} / \sqrt{3}$ and $\sigma_{\rmn g}$ are the 1D and 3D velocity dispersion of gas particles within the SPH kernel at the time of star formation and $\kappa$ is the epicyclic frequency (i.e. the maximum distance from the mean radius of orbit is reached after a time $\pi/2\kappa$). The gas density of the environment of the particle is then calculated within the radius $r_\rmn{env}$.

We apply this method to young ($<100 \Myr$) star particles in Gal009 at redshifts 2, 1 and 0. At $z=0$, the method gives $r_\rmn{env} \approx 0.5 \kpc$ within galactocentric radii of $2 \kpc$, and $r_\rmn{env} \approx 2 \kpc$ (with large scatter) at radii $>5 \kpc$. In practice, we impose a maximum $r_\rmn{env}$ of $5 \kpc$ (the exact value does not affect the results) to limit the effect of low values of $\kappa$.
Fig. \ref{fig:env_density} shows the comparison of the gas birth density with the gas density of the surrounding environment using this method. At all redshifts there is a strong relationship between the quantities (though with significant scatter), with the environment density being approximately one-third of the birth density.


\bsp	
\label{lastpage}
\end{document}